\documentclass[aps,prd,preprintnumbers,showpacs,nofootinbib]{revtex4}
\usepackage{amsmath}
\usepackage{amssymb}
\usepackage{enumerate}

\usepackage{graphicx}
\usepackage{color}
\usepackage{dsfont}
\usepackage[latin1]{inputenc}
\usepackage{pdfpages}

\usepackage{amsmath}
\usepackage{amsfonts}
\usepackage{amssymb}
\usepackage{color}
\usepackage{slashed}
\usepackage{graphicx}
\usepackage{bm}
\usepackage{braket} 
\usepackage[usenames,dvipsnames,svgnames]{xcolor}
\usepackage[colorlinks=true,
            linkcolor=red,
            urlcolor=gray,
            citecolor=blue]{hyperref}
\PassOptionsToPackage{unicode}{hyperref}
\PassOptionsToPackage{naturalnames}{hyperref}

\newcommand{\destacar}[1]{{\slshape \color{blue} #1}}
\definecolor{amarillo}{cmyk}{0,0.5,1,0}

\begin{document} 

\title{\destacar{Phenomenological aspects from vacuum stability and
metastability in models with soft breaking  of a \texorpdfstring{$U\left( 1\right)$}{lg} global
symmetry}}
\author{Andr\'es Castillo}
\email{afcastillor@unal.edu.co} 
\affiliation{Departamento de F\'isica, Universidad Nacional de Colombia, Sede Bogot\'a}
\author{Rodolfo A. Diaz}
\email{radiazs@unal.edu.co}
\affiliation{Departamento de F\'isica, Universidad Nacional de Colombia, Sede Bogot\'a} 
\author{John Morales}\email{jmoralesa@unal.edu.co}\affiliation{Departamento de F\'isica, Universidad Nacional de Colombia, Sede Bogot\'a} \author{Carlos G. Tarazona}\email{caragomezt@unal.edu.co}
\affiliation{Departamento de F\'isica, Universidad Nacional de Colombia, Sede Bogot\'a}

\begin{abstract}
We study the vacuum behavior of an extended Higgs sector with two doublets in a scenario with a softly broken \texorpdfstring{$U(1)$}{lg} global symmetry. The soft-violation term is introduced to avoid massless-axion particles arising when the global symmetry becomes spontaneously broken. This model has metastable states through the possible presence of multiple non-degenerate minima, which is unwanted from the phenomenological point of view if the metastable state is not long-lived enough. The analysis of this fact leads to find possible exclusion limits over parameter space of quartic couplings. Results improve the individual behavior of initial conditions for renormalization group equations; also determining unstable zones for the effective Higgs potential at one loop level. Besides vacuum stability analyses, the influence of absence of charge violation minima is considered as a limiting case excluding zones in the parameter space. Extremal cases for the model as well as criticality phenomena are discussed with the aid of relation among Higgs masses or splittings among them. From vacuum behavior and LHC results, phenomenological aspects in the searching of charged and heavier Higgs bosons are considered to evaluate the scalar alignment regimen of the two Higgs doublet model.
\end{abstract}

\maketitle

\section{Introduction}

The most successful discovery in high-energy physics since fermions in the third family -top quark and
tau neutrino- is the scalar boson founded in LHC by CMS and ATLAS collaborations in July of 2012. 
The properties of this scalar particle, as its couplings with gauge bosons and fermions, are compatible so far with those expected for a Standard Model (SM) 
Higgs boson at a mass about 125 GeV \cite{PDG2,Atlas,CMS}. New data and phenomenological studies also establish that this scalar particle resembles the Higgs boson in a broad sense since it is consistent with a particle of spin zero and positive parity, expected by attributes of SM with Spontaneous Symmetry Breaking (SSB) \cite%
{CMSspin}. This experimental scenario open new frames to study
the theoretical and phenomenological behavior of the SM itself (via precision tests) and models of
physics beyond, e.g. Extended Higgs Sectors as the Two Higgs Doublet Model. We consider these new possibilities from a theoretical point of view by involving the following phenomenological scenario: SM Higgs boson properties, as couplings and mass, are inherited by one of the new Higgs bosons (the lighter one with positive parity) and the remaining scalars are settled in any energy scale. This limit is commonly called as \emph{the alignment regime} \cite{Carena2}.

One significant benchmark in particle physics is the fact that in this mass region for Higgs boson, the SM-effective Higgs potential is \emph{metastable} at high energy scales according to
the most precise computations carry out in different papers with NNLO corrections \cite{Degrassifin,Elias,ArkaniHamed,Bezrukov,Burgess}. Indeed, those critical effects might be a consequence of symmetry itself, a fine tuning or a dynamical
effect among new parameters from a physical completion for SM \cite{JCasas}.

Vacuum stability condition is one of the most relevant
features for Spontaneous Symmetry Breaking (SSB), due to it determine
natural solutions for stationary and minimization equations for a well-defined ground state. For instance in the SM, this condition
is such that the Higgs potential, in their part of the fourth dimension, should always be positive in the field space; translating it into that the quartic
coupling would be $\lambda >0,$ even for asymptotically values of $%
\left\vert \Phi \right\vert \rightarrow \infty.$ This condition at tree
level can be extrapolated to the effective Higgs potential at, e.g., one
loop level or at Next to Next Leading Order (NNLO). Then, using Renormalization
Group Equations (RGEs) for quartic coupling, gauge and Yukawa couplings and
their simultaneous solutions, it is possible to find out the stability
regimes for the effective Higgs potential concerning the energy scales $%
\mu .$ In those regimes, the Higgs potential shape might change, and it will
drive out to instabilities in fact. The instability energies can be seen as
cut scales where the theory (as a bottom-up approach) is only one effective description up to those energy values, and hence new physics or new freedom degrees are expected to enter in the foundations to tackle these problems. In our treatment, the new physics effects
come from a well motivated extended scalar sector due to two Higgs doublet
model (2HDM) with a Higgs potential that softly breaks a $U(1)$ global
symmetry. This symmetry implementation encourages a Higgs potential without explicit CP
violation and a Yukawa sector with an absence of Flavor Changing Neutral Currents (FCNCs) \cite{Report}.

Despite the general procedure to find positivity relations could be more sophisticated in models beyond SM, there are many elegant methods to find out three level vacuum behavior in 2HDMs. Variational calculus and hidden symmetries in
the Higgs potential belong to establish relatively manageable formalism which becomes useful in this searching \cite{IvanovSimetrias}. As was pointed out in \cite{FerreiraPer}, most elaborated difficulties appear when the relations gotten at tree level are elevated in higher order corrections, because of the introduction of new fields and couplings; translating finally in more RGEs to solve simultaneously. For example, in the two Higgs
doublet model, the vacuum stability conditions rely on the asymptotical behavior of the extended field space.
Furthermore, RGEs at one loop level for scalar and general Yukawa
couplings are intricate to control, since they need additional assumptions, e.g., the initial conditions over the respective parameters building all combinations for field space. Hence,
the consequences over parameter space are tricky, since initial conditions of
parameters are highly unknown. Therefore, the last fact is a motivation to introduce other theoretical and phenomenological constraints in the general
study of vacuum stability regions for extended models, like the behavior of oblique parameters and perturbative unitarity.

In addition to the standard issues in the bounded from below conditions, the 2HDM has a richer vacuum structure wherein metastable states could be present. This effect is a consequence of relating the number of critical points in the Higgs potential, which particularly in its minimum consists on the
composition of two vacuum expectation values, also depending on stationary points nature. The scalar potential of 2HDM might have simultaneously two neutral minima, two $CP$ conserving or two $CP$ violating \cite%
{Ferreira1}\footnote{In 2HDMs exist the possibility of having charge violating critical points. Despite these stationary points must be avoided and coexistence with normal ones is forbidden, we consider limiting where this vacuum is possibly generated as one assumption to describe hierarchical structures in scalar masses computed in a neutral vacuum.}. In those cases, from a vacuum state belonging in a metastable local minimum, there would be the possibility of decaying later, for large enough times, into the global minimum. Descriptions of metastability with underlying two Higgs doublet dynamics focused on formal aspects have been broadly studied  in \cite%
{Ivanov,8Ferreira1,Barroso2,Ivanovnuevo,Barroso:2007rr}.
Phenomenological aspects of the possibility of two global minima have been treated comprehensively in \cite{Met1,Met2}. Nonetheless, studies about vacuum metastability in softly $U(1)$ models have been carried out, comparison among one loop behavior of the effective Higgs potential and the presence of two neutral minima is still an issued to be addressed. In this direction, recently in \cite{Dipankar} studied the impact of considering a softly breaking term to get stable zones in energies $\mu>10^{10}$ GeV.  For Inert models, in \cite{Ferreira2015} has been shown as the parameter space compatible with the coexistence of both possible neutral vacua is larger than the predicted by tree-level analyses; also demonstrating how the nature of vacuum can change at one loop level with established at tree level. Therefore, potential regions investigated by vacuum stability can be constrained by the presence of an inert-like vacuum (where fermions are massless) at one loop level.

In the moment of building a model containing a Higgs potential with several distinct neutral minimums breaking the same symmetries, possibility of taking a metastable minimum as a physically acceptable vacuum state is allowed if this effect might be suppressed by the fact of having a long lived enough minimum state. This lead to describe the
masses for particles in a realistic approach, without being concerned about the tunneling effects changing this phenomenological scenario. Nevertheless, from a cosmological point of view, a metastable regime of 2HDM could bring many consequences and issues in baryogenesis mechanisms. For instance, the critical temperature for a (strong) first order EW phase transition might have an ambiguous definition if possible minima of the effective Higgs potential are not considered properly \cite{Cline}.
However, LHC-phenomenology analyses as decays ratios of new physics could give some information about suppression metastable states \cite{Met2} in 2HDMs. Moreover, LHC collaborations have given contours in the measured golden-decays with a good confidence level \cite{Lilith}. Therefore, experimentally constrained regions can be used to extract plausible frameworks in the vacuum realization for 2HDMs.   
 
With all these ideas in mind, our study is organized as follows: In section \ref{sec:TL} and by symmetries and basis invariant transformations in 2HDMs, we use formalism presented in \cite{IvanovSimetrias} to find out positivity constraints in this simple $U(1)$ scalar sector with a soft-violation term. The systematic begins with a general way to write a more compact form for the Higgs potential which is invariant under a parameterization of the Lorentz group in the future light cone. Under impositions over
doublets, and its extrapolation into Minkowskian space is possible to find out vacuum stability constraints associated with the requirement at tree level for a bounded from below Higgs potential in regimes of the field space associated with points or orbits defined in the future light cone. Moreover, that covariant structure of the
Higgs potential yields a viable frame to analyze the possibility to reach metastable states in vacua with two normal-minima (where VEVs are real). On the other hand, bilinears form of the Higgs potential carries out information about structures of stationary conditions compatible with a complete SSB and one neutral vacuum.  These conditions can be considered from a precise point of
view; studying when possible critical points are minimum indeed. For these purposes, we first review eigenstates masses for scalars in \ref{sec:masseigenstates}. The phenomenological starting point to interpret a possible model realization, the alignment regimen is studied in \ref{sec:align}. Then, in section \ref{sec:MetI}, we review the problem for critical conditions of a Higgs potential with a softly broken $U(1)$-symmetry. Section \ref{sec:phenU(1)} dedicates to describe some phenomenological and theoretical aspects of these models. In \ref{sec:LO}, we compute the vacuum behavior at tree level and for different combinations of the parameter space. Meanwhile, NLO calculations have been studied in section \ref{sec:NLO}. Likelihood proofs for two photons decay of Higgs boson and oblique parameters analyses are established in \ref{sec:Likelihood} with the aim to see the compatibility of vacuum studies with models in the alignment scenario for 2HDMs. Finally, in conclusions and remarks, we describe the relevance of our treatment in the interpretation of
vacuum and metastability analyses and the compatibility with Electroweak precision tests and likelihood proofs in the two photons decay channel.

 \section{Vacuum behavior and positivity constraints in a 2HDM with a softly
violation for \texorpdfstring{$U\left( 1\right)$}{lg} symmetry \label{sec:TL}}

This section is devoted to introducing generalities of the Real $U(1)$-2HDM, as well as its theoretical constraints. The 2HDM potential with a soft breaking of a $U(1)$ global symmetry is

\begin{align}
V_{H}&=m_{11}^{2}\Phi _{1}^{\dagger }\Phi _{1}+m_{22}^{2}\Phi _{2}^{\dagger
}\Phi _{2}-m_{12}^{2}\left( \Phi _{1}^{\dagger }\Phi _{2}+\Phi _{2}^{\dagger
}\Phi _{1}\right)\notag\\ 
&+\frac{1}{2}\lambda _{1}\left( \Phi _{1}^{\dagger }\Phi
_{1}\right) ^{2}+\frac{1}{2}\lambda _{2}\left( \Phi _{2}^{\dagger }\Phi
_{2}\right) ^{2}+\lambda _{3}\left( \Phi _{1}^{\dagger }\Phi _{1}\right)
\left( \Phi _{2}^{\dagger }\Phi _{2}\right) +\lambda _{4}\left( \Phi
_{1}^{\dagger }\Phi _{2}\right) \left( \Phi _{2}^{\dagger }\Phi _{1}\right) .
\label{POTH1}
\end{align}

By using re-parametrization invariance \cite{IvanovSimetrias} given by $SO(1,3)$ group transformations, Higgs potential regarding gauge orbit vectors $r^{\mu}$ is

\begin{align}
 V_{H}=-M_{\mu}r^{\mu}+\frac{1}{2}\Lambda_{\mu\nu}r^{\mu}r^{\nu},\label{Higgspot}
\end{align}

with $r^{\mu}=(r^{0},r^{i})=(\boldsymbol{\Phi}^{\dagger}\boldsymbol{\Phi},\boldsymbol{\Phi}\sigma_{i}\boldsymbol{\Phi})$ and where

\begin{align}
 \boldsymbol{\Phi}=\begin{pmatrix}
                    \Phi_{1}\\
                    \Phi_{2}
                   \end{pmatrix}.
\end{align}

Here $\boldsymbol{\Phi}$ is a 2-dimensional vector and $\sigma_{i}$ are the Pauli matrices. In this particular case, cuadrivector of bilinears couplings  is

\begin{align}
M_{\mu}=\left(\frac{m_{11}^{2}+m_{22}^{2}}{2},\text{Re}\left(m_{12}^{2}\right),0,\frac{m_{11}^{2}-m_{22}^{2}}{2}\right). 
\end{align}
In the scenario of a $U(1)-$ Higgs potential, diagonal $\Lambda_{\mu\nu}$ tensor of quartic couplings has the following form 

\begin{equation}
\Lambda =\frac{1}{2}\left( 
\begin{array}{cccc}
\lambda _{3}+\sqrt{\lambda _{1}\lambda _{2}} & 0 & 0 & 0 \\ 
0 & -\lambda _{4}& 0 & 0 \\ 
0 & 0 &-\lambda _{4}  & 0 \\ 
0 & 0 & 0 & \lambda _{3}-\sqrt{\lambda _{1}\lambda _{2}}
\end{array}%
\right) .
\end{equation}

A bounded from below Higgs potential demands $\Lambda_{\mu\nu}$ must be positive definite in the future light cone $LC^{+}$, i.e., $r_{\mu}r^{\mu}\geq0$. Through this formalism for re-parametrization, it is possible to find out the Higgs potential positivity constraints in the fourth dimension terms:

\begin{equation}
\lambda _{1}+\lambda _{2}>|\lambda _{1}-\lambda _{2}|.
\end{equation}

which is equivalent to $\lambda _{1}>0$ and $\lambda _{2}>0$. And%

\begin{subequations}
\begin{eqnarray}
\lambda _{3} &>&-\sqrt{\lambda _{1}\lambda _{2}}. \\
\lambda _{4}+\lambda _{3} &>&-\sqrt{\lambda _{1}\lambda _{2}}.
\end{eqnarray}
\end{subequations}

In addition to the traditional relations for vacuum stability at tree-level, the absence of charge violation vacua yields to one possible condition $\lambda _{4}<0$ \cite{Ginzburg2,Kanishev2,Kanishev3}. Despite in 2HDMs at tree level two minima that break different symmetries cannot coexist, this condition can be rendered as a limiting hypothesis for our assumption of neutral vacua. 
This fact has significant phenomenological consequences and which will be treated exhaustively in the following sections, for example to
describe possible hierarchical structures in the masses for scalar states.

On the other hand, unitarity constraints can be obtained from the following eigenvalues $%
\boldsymbol{\Lambda }_{Y,\sigma \pm }^{Z_{2}}$\footnote{%
Matrices are constructed on  the Isospin basis, $\sigma $ and Hypercharge $%
Y,$ which are conserved quantities of scalar scattering at high energies.} of
scattering matrices \cite{Ivanov}:

\begin{subequations}
\begin{align}
\boldsymbol{\Lambda }_{2,1\pm }^{even}& =\frac{1}{2}\left( \lambda
_{1}+\lambda _{2}\pm \sqrt{(\lambda _{1}-\lambda _{2})^{2}}%
\right), \label{eigen1} \\
\boldsymbol{\Lambda }_{2,0\pm }^{even}& =\lambda _{3}-\lambda _{4}. \\
\boldsymbol{\Lambda }_{0,1\pm }^{even}& =\frac{1}{2}\left( \lambda
_{1}+\lambda _{2}\pm \sqrt{(\lambda _{1}-\lambda _{2})^{2}+4\lambda _{4}^{2}}%
\right) , \\
\boldsymbol{\Lambda }_{0,0\pm }^{even}& =\frac{1}{2}\left[ 3(\lambda
_{1}+\lambda _{2})\pm \sqrt{9(\lambda _{1}-\lambda _{2})^{2}+4\left(
2\lambda _{3}+\lambda _{4}\right) ^{2}}\right].
\label{eigen4}
\end{align}
\end{subequations}

with the perturbative unitarity bound

\begin{equation}
\left\vert \boldsymbol{\Lambda }\right\vert <\frac{1}{8\xi \pi }
\end{equation}

$\xi=2$ is a factor for indistinguishable particles present in the initial or final states.

\section{Softly broken \texorpdfstring{$U\left( 1\right) -$}{lg}Higgs Potential: Mass Eigenstates}
\label{sec:masseigenstates}

In the following, mass eigenstates and respective relations among Higgs potential couplings are considered. To that end, we describe the standard parametrization of Higgs doublets concerning physical mass eigenstates%
\begin{align}
\Phi _{1}& =\frac{1}{\sqrt{2}}\left( 
\begin{array}{c}
\sqrt{2}\left( G^{\pm }\cos \beta -H^{+}\sin \beta \right) \\ 
v\cos \beta -h^{0}\sin \alpha +H^{0}\cos \alpha +i\left( G^{0}\cos \beta
-A^{0}\sin \beta \right)%
\end{array}%
\right) . \\
\Phi _{2}& =\frac{1}{\sqrt{2}}\left( 
\begin{array}{c}
\sqrt{2}\left( G^{\pm }\sin \beta +H^{+}\cos \beta \right) \\ 
v\sin \beta +h^{0}\cos \alpha +H^{0}\sin \alpha +i\left( G^{0}\sin \beta
+A^{0}\cos \beta \right)%
\end{array}%
\right) .
\end{align}

Here $-\pi/2\leq\alpha\leq\pi/2$ and $0<\beta<\pi/2$. The Higgs masses and Higgs eigenstates are defined with the parameters $%
m_{ij}^{2}$ and $\lambda _{i}$ from the potential (\ref{POTH1}), and
consequently, depend on the symmetries to write the Lagrangian indeed. 
Moreover, the mass matrix depends on the neutral vacuum structure selected in the above parameterization, 

\begin{equation}
\label{VEVs}
\langle \Phi _{1}\rangle _{0}=\frac{1}{\sqrt{2}}\begin{pmatrix}
                           0\\
                           v_{1}
                          \end{pmatrix}
 \text{ and }\langle \Phi _{2}\rangle _{0}=\frac{1}{\sqrt{2}}\begin{pmatrix}
                           0\\
                           v_{2}
                          \end{pmatrix}
\end{equation}

where $v_{1}=v\cos\beta$ and $v_{2}=v\sin\beta$. In the case of a Higgs
potential with soft breaking of a $U\left( 1\right) -symmetry$, relations among quartic couplings and masses are given by%

\begin{subequations}
\begin{eqnarray}
\lambda _{1} &=&\frac{1}{v^{2}\cos ^{2}\beta }\left( \cos ^{2}\alpha
m_{H^{0}}^{2}+\sin ^{2}\alpha m_{h^{0}}^{2}-m_{12}^{2}\tan \beta \right)
\equiv \Delta S_{1}^{2}, \\
\lambda _{2} &=& \frac{1}{v^{ 2}\sin^{ 2}\beta }\left( \sin ^{2}\alpha
m_{H^{0}}^{2}+\cos ^{2}\alpha m_{h^{0}}^{2}-m_{12}^{2}\cot \beta \right)
\equiv \Delta S_{2}^{2}, \\
\lambda _{3} &=&\frac{2m_{H^{\pm }}^{2}-m_{A^{0}}^{2}}{v^{2}}+\frac{\sin
2\alpha \left( m_{H^{0}}^{2}-m_{h^{0}}^{2}\right) }{v^{2}}\equiv \Delta
S_{3}^{2}, \\
\lambda _{4} &=&\frac{2m_{A^{0}}^{2}-2m_{H^{\pm }}^{2}}{v^{2}}\equiv \Delta
S_{4}^{2}.
\end{eqnarray}

and 
\begin{equation}
m_{A^{0}}^{2}=\frac{m_{12}^{2}}{\sin \beta \cos \beta }=\frac{2m_{12}^{2}}{%
\sin \left( 2\beta \right) }.\label{A0}
\end{equation}
\end{subequations}

These arrays of equations are valid for $0<\beta <\pi /2$, excluding
inert models for 2HDM since impossibility of diagonalizing mass eigenstates at the same time that critical conditions are preserved.

\section{Alignment Regime}
\label{sec:align}
The scalar \emph{alignment regime}, where the lighter Higgs CP-even behaves
as SM Higgs, independently of masses of remaining scalars is aimed to
establish compatibility between theoretical analysis and precision searches
for beyond SM physics. Despite phenomenologically it seems likely that
alignment will only be realized approximately, rather than exactly, it can
be translated in ground studies to interpret scalar signal at 125 GeV results from extended models like 2HDMs. As first glance, we present a study where the exact alignment is
achieved for our model. The decoupling limit, where the low-energy spectrum
contains only the SM Higgs and no new light scalars, is only a subset of one
more general alignment limit. For Higgs potential described in (\ref{POTH1})
with, the alignment scenario has the following conditions \cite%
{Carena2}:

\begin{subequations}
\begin{eqnarray}
m_{h}^{2} &=&v^{2}\left( \lambda _{1}\cos ^{2}\beta +\left( \lambda
_{3}+\lambda _{4}\right) \sin ^{2}\beta \right) , \\
m_{h}^{2} &=&v^{2}\left( \lambda _{2}\sin ^{2}\beta +\left( \lambda
_{3}+\lambda _{4}\right) \cos ^{2}\beta \right) .
\end{eqnarray}
\end{subequations}
If there is a $\tan \beta $ satisfying the above equations, then the
alignment limit would occur for arbitrary values of $m_{A^{0}}$ and does not
require non-SM-like scalars to be heavy. We scan out the parameter space in such way that both conditions are satisfied simultaneously:

\begin{equation}
\tan ^{2}\beta =\frac{\lambda _{1}-\lambda _{3}-\lambda _{4}}{%
\lambda _{2}-\lambda _{3}-\lambda _{4}}.  \label{alignment}
\end{equation}

In the limit of $\beta \rightarrow 0,$ both conditions have a natural
solution only if $\lambda _{1}=\lambda _{3}+\lambda _{4}.$  On the other hand, $\tan\beta\to 1$, implies $\lambda_{1}=\lambda_{2}$.

\section{Metastability theorems: Particular cases}
\label{sec:MetI}

We briefly discuss the origin of multiple stationary points in the 2HDM by
considering tadpoles at tree level equations for the Higgs potential and through of a
revision of systematics developed comprehensively in \cite{Ivanov2, Ivanovnuevo}. Firstly, we find the critical points equations (i.e. non-trivial tadpoles at tree level) for Higgs potential (\ref{POTH1}) that give rise to the different stationary points (based on a neutral vacuum): 

\begin{subequations}
\begin{eqnarray}
T_{1} &\equiv&2m_{11}^{2}v_{1}-2m_{12}^{2}v_{2}+%
\lambda _{1}v_{1}^{3}+\lambda _{4}v_{1}v_{2}^{2}+\lambda
_{3}v_{1}v_{2}^{2}=0.  \label{min1} \\
T_{2} &\equiv&2m_{22}^{2}v_{2}-2m_{12}^{2}v_{1}+%
\lambda _{2}v_{2}^{3}+\lambda _{4}v_{1}^{2}v_{2}+\lambda
_{3}v_{1}^{2}v_{2}=0.  \label{min2}
\end{eqnarray}
\end{subequations}

Notice that one cannot have solutions of the form $\left\{ v_{1}=0,v_{2}\neq
0\right\} $ or $\left\{ v_{1}\neq 0,v_{2}=0\right\} $, unless that $m_{12}^{2}=0$. Those are the natural inert models discussed broadly in \cite{Kanishev2,Kanishev3,Borah}. 
A trivial solution of these equations is clearly $v=0,$ equivalent to one theory without EW symmetry
breaking. Excluding that case, the stationarity conditions (\ref{min1}) and (\ref{min2}) become

\begin{subequations}
\begin{eqnarray}
v^{2}+\frac{2m_{11}^{2}-2m_{12}^{2}\tan \beta }{\lambda _{1}\cos ^{2}\beta
+\lambda _{4}\sin ^{2}\beta +\lambda _{3}\sin ^{2}\beta }&=&0 .
\label{min11} \\
\left( 2m_{11}^{2}-2m_{12}^{2}\tan \beta \right) \left( \lambda _{2}\tan
^{2}\beta +\lambda _{4}+\lambda _{3}\right) -\left(
2m_{11}^{2}-2m_{12}^{2}\cot \beta \right) \left( \lambda _{1}\cot ^{2}\beta
+\lambda _{4}+\lambda _{3}\right) \tan \beta &=&0  \label{min12}
\end{eqnarray}
\end{subequations}

We have also been following discussion and formalism presented in \cite%
{Barroso:2007rr}. Eq. (\ref{min11}) tells us that, other than its sign, the
value of VEV $v$ is given unequivocally by $\tan \beta $ function. Eq. (\ref{min12})
is an equation of fifth order on $\tan \beta $, having at most five possible real solutions. These two equations describe
therefore ten possible solutions $\left\{ v_{1},v_{2}\right\} $, due to
the ambiguity on the sign of $v$; since 2HDM potential in Eq. (\ref{POTH1})
is also invariant under the transformation $\Phi _{1}\rightarrow -\Phi _{1}$
and $\Phi _{2}\rightarrow -\Phi _{2}$. These ten solutions
correspond to only four different physical scenarios. Adding the trivial
solution $v_{1}=v_{2}=0$ (no EW symmetry), we have a total of eleven solutions.

There are at most two different values of $\tan \beta $ what satisfy both equations;  implying exist a maximum of six stationary points. Indeed, it could lead more than one normal minimum, with different depths. To establish it formally is necessary to make use of Morse's systematics \cite%
{MorseT,Barroso:2007rr}: For a given real function of two variables, let $%
\eta _{0}$, $\eta _{1}$ and $\eta _{2}$ be the number of its minima, saddle points, and maximums inside of the Higgs potential, respectively. For a polynomial function in $v_{1}$ and $%
v_{2}$, bounded from below, such as the one we are dealing with, Morse's inequalities state that:
\begin{itemize}
\item $\eta _{0}\geq 1.$

\item $\eta _{1}\geq \eta _{0}-1.$

\item $\eta _{0}-\eta _{1}+\eta _{2}=1.$
\end{itemize}

The foundations of Morse's inequalities give bounds over critical points, in particular over minimum behavior in the Higgs potential and its influence in extremum conditions and mass matrices. The 2HDM potential in Eq (\ref{POTH1}) has $\eta _{0}+\eta _{1}+\eta _{2}=2n +1$ stationary solutions, $n=0,...,5$: at most $2n$ real roots of eqs. (\ref%
{min11}), (\ref{min12}) plus the trivial solution $v_{1}=v_{2}=0$ (No EW symmetry breaking). We can use Morse's inequalities to get $\eta _{0}+\eta
_{2}=n+1.$ We analyze several possibilities for the number of minima $\eta_{0}$, depending on the number of real solutions $n$. By counting extremal cases, all the different combinations of stationary points leads to find the following general aspects: There are critical points without symmetry breaking associated to every case of maximums, minima or saddle points combinations. Thus, typical situations of SSB with a global minimum are given for $1\leq n\leq2$. Meanwhile, for $n=3$ and using Morse's inequalities:  $\eta _{0}+\eta _{2}=4$ yields a SSB scenario plus a trivial minimum located at the origin. The $n=4$ case translates in two pairs of degenerate minima away from the origin. It is not mandatory that these two pairs of minima have the same depth. Therefore as first glance, we might have one normal minimum deeper than another. Finally, a higher order in $n$ solutions yields trivial plus global and non-global minima.

Hence if there are more than two solutions for $v_{2}/v_{1}$ ratio, which
said the 2HDM might have more than one normal minimum away from the origin
with different depths. However, no more than two of such minima can
exist by physical grounds (for a non-long-lived enough minimum state). The analysis of non-global minimum structures must see as potential exclusion regions in the parameter
spaces for these particular cases of 2HDM.

From the form of critical points and Morse's inequalities, we can see as multiple non-degenerate minima can be present in the Higgs potential. It is worthwhile now to analyze where can be ensured the existence of one and only one global minimum. For this purpose, we restrict the following phenomenological study to avoid two minima with different depths. Taking the case of four real solutions, in \cite{Ferreira1,8Ferreira1} have shown that the difference in the values of the potential in those two normal vacuum structures $N_{1}$ and $N_{2}$ is given by

\begin{equation}
V_{N_{2}}-V_{N_{1}}=\frac{1}{4}\left[ \left( \frac{m_{H^{\pm }}^{2}}{v^{2}}%
\right) _{N_{1}}-\left( \frac{m_{H^{\pm }}^{2}}{w^{2}}\right) _{N_{2}}\right]
\left( v_{1} w_{2} -w_{1} v_{2}  \right) ^{2}.  \label{diffmet}
\end{equation}

with $v^{2}=v_{1}^{2}+v_{2}^{2}$ in $N_{1}$ and $w^{2}=w_{1}^{2}+w_{2}^{2}$
in $N_{2}.$ $N_{2}$ structure can be seen interchanging $v_{i}\rightarrow
w_{i}$ in the VEVs for respective doublets of (\ref{VEVs}). Nothing
establishes how to carry out the computations to determine which is the overall
sign in the difference. Another aspect of this result is that starting with the same Higgs potential, the demonstration depends on only of vacuum structure
in both minima. The following discriminant, written regarding tensor matrix
of the Higgs potential (\ref{Higgspot}), ensures the existence of one and
only one global minimum in the theory \cite{Ivanovnuevo}
\footnote{$\mathcal{D}$-discriminant, encouraging a global minimum in the Higgs potential, has been computed for 2HDMs from Hessian of the Higgs potential in the gauge orbit field using the reparameterization group $SO(1,3)$.}

\begin{equation}
\mathcal{D}\equiv -\det (\Lambda _{E}-\zeta I).
\end{equation}

$\zeta$ is an auxiliary function introduced in the Higgs potential to determine stationary conditions. This Lagrange multipliers is related with $\sum_{i=1}^{2}\Phi^{\dagger}_{i}\Phi_{i}\geq0$ constraint. $\Lambda_{E}$ is $\Lambda_{\mu\nu}$ of Higgs potential (\ref{Higgspot}) expressed by an Euclidean-metric and $I$ is the four dimensional identity matrix. In the diagonal basis ($\Lambda_{E}\to \text{diag}(\Lambda_{0},\Lambda_{1},\Lambda_{2},\Lambda_{3})$), the global minimum discriminant reads

\begin{equation}
\mathcal{D}\equiv (\Lambda _{0}-\zeta )(\zeta -\Lambda _{1})(\zeta -\Lambda
_{2})(\zeta -\Lambda _{3}).
\end{equation}

From the parameters inside of the Higgs potential with a softly breaking of $U(1)$-symmetry (\ref{POTH1}), 

\begin{equation}
\mathcal{D}=\left( m_{11}^{2}-\kappa ^{2}m_{22}^{2}\right) \left( \tan \beta
-\kappa \right) >0.  \label{MetastabilityDiscriminant}
\end{equation}

with $\kappa =\left( \lambda _{1}/\lambda _{2}\right) ^{1/4}.$ By the
implications of stationary conditions (\ref{min1}) and (\ref{min2}), we
exclude $\beta = 0$ and $\beta=\pi/2 $
values in the parameter space. Only we approach to them by means of their limit values, which bring out to some couplings in non-perturbative regions. In terms of scalar masses and splitting parameter $k_{S}=m_{A^{0}}/m_{H^{0}}$,  discriminant takes the form

\begin{align}
k_{S} <\frac{1}{\left( \frac{m_{H^{0}}^{2}}{m_{h^{0}}^{2}}\left( 1-\sin
^{2}(\beta+\alpha )\right) +\sin ^{2}(\beta+\alpha) \right) }. \label{MetasDiscr2}
\end{align}

This bound has been evaluated in the plane $m_{H^{0}}-\sin(\beta+\alpha)$ as is shown in Fig. (\ref{fig:sin}). For our purposes, both structures for metastability discriminant are useful in determining regions of parameter space compatible with vacuum stability. The second one give us direct information about scalar mass, while the
former give us information about Higgs mass splittings. $\mathcal{D}>0$ ensure a global minimum in the theory. If $\mathcal{D}<0,$ additional computations are necessary to discriminate between both vacuum structures.

\section{Phenomenological aspects of theories with softly breaking of \texorpdfstring{$U(1)$}{lg}}
\label{sec:phenU(1)}
The 2HDM model with Abelian global symmetries has been used as a ground basis to explain CP violation phases in strong interactions using the Peccei Quinn mechanism \cite{Peccei1,Peccei2}. When the global symmetry is spontaneously broken, the new scalar spectrum should contain an axion with zero mass, which is not wanted by theoretical facts \cite{Peccei4,Kim}.  If a massless (o with a mass of small size) scalar particle exist its detection and precision measurements are a real challenge; the phenomenological compatibility to explain strong-CP phases is still no accurate so far \cite{Kim2}. Furthermore, topological defects as vortices are generated in this regime when the global symmetry has been broken. To avoid those issues and to improve the experimental level of accuracy of the
Peccei-Quinn models, a dimension two term of $U(1)$ symmetry violation, is introduced in the Higgs potential. This term has a small impact on the evolution of Renormalization Group Equations\footnote{%
In the softly broken $U(1)$-model, $\bar{m}_{12}^{2}$ has the
following RGE 
\begin{equation*}
16\pi ^{2}\frac{d}{d\log \mu ^{2}}\bar{m}_{12}^{2}=\left( 2\lambda
_{3}+\lambda _{4}\right) \bar{m}_{12}^{2}.
\end{equation*}
It is possible to see as radiative corrections to $%
m_{12}^{2}$ are proportional to $m_{12}^{2}$ itself and are only logarithmically sensitive to the cutoff $\mu$.} and so on computations of stability and metastability regimes. $m_{12}^{2}$ term also yields a non-zero mass term for pseudoscalar particle $A^{0}$ (see Eq. \ref{A0}).

It is worth to say that the presence of dimension two-term, as well as the critical conditions, exclude the presence of an inert vacuum in some doublet. Hence inert scenarios are just accomplished in an approximate manner for fractions of $v_{2}/v_{1}\rightarrow 0$ (quasi-inert regimen); having many phenomenological consequences in dark matter searches \cite{Kanishev2,Kanishev3} and for the description of viable mass terms for neutrinos \cite{Logan1,Logan2}.

As was pointed out above, an outstanding aspect of general 2HDM is the presence of metastable states with two normal vacuum structures; what is a consequence of the solutions of stationary conditions combinations of Eqs. (\ref{min11})-(\ref{min12}). The simplest case is such
where a soft violation term appears in the Higgs potential invariant under a continuous global symmetry. Nevertheless, 2HDM nature restricts facts as the
coexistence of minima of different depths and different origins (CP
breaking and charge breaking -CB- vacuum structures) \cite{Ferreira1}. Besides, whenever a normal minimum exists in the 2HDM, the global minimum of the
potential is a normal one, and no tunneling to a deeper CB or CP minimum is
allowed \cite{Ferreira2}. On the other hand, if a CP (CB) violating minimum exist in the 2HDM,
it is the global minimum of the theory, and thoroughly stable and no
tunneling to a deeper normal or CB (CP) minimum can occur \cite{Report}.
Hereafter, we are only focused on normal behavior of EW vacuum and its
phenomenological consequences.

With this phenomenological approach, our primary goal is to compute the regions where the metastability due to two normal
vacua in EW scale arise; determining allowed scenarios in experimental data (e.g. diphotonic decays for SM like-Higgs)
and thus improving vacuum analysis carried out at NLO level.

\section{Global minimum behavior in \texorpdfstring{$m_{H^{0}}-\sin\left(\alpha+\beta\right)$}{lg} plane}
\label{sec:global}

\begin{figure}[tph]
\centering\includegraphics[scale=0.18]{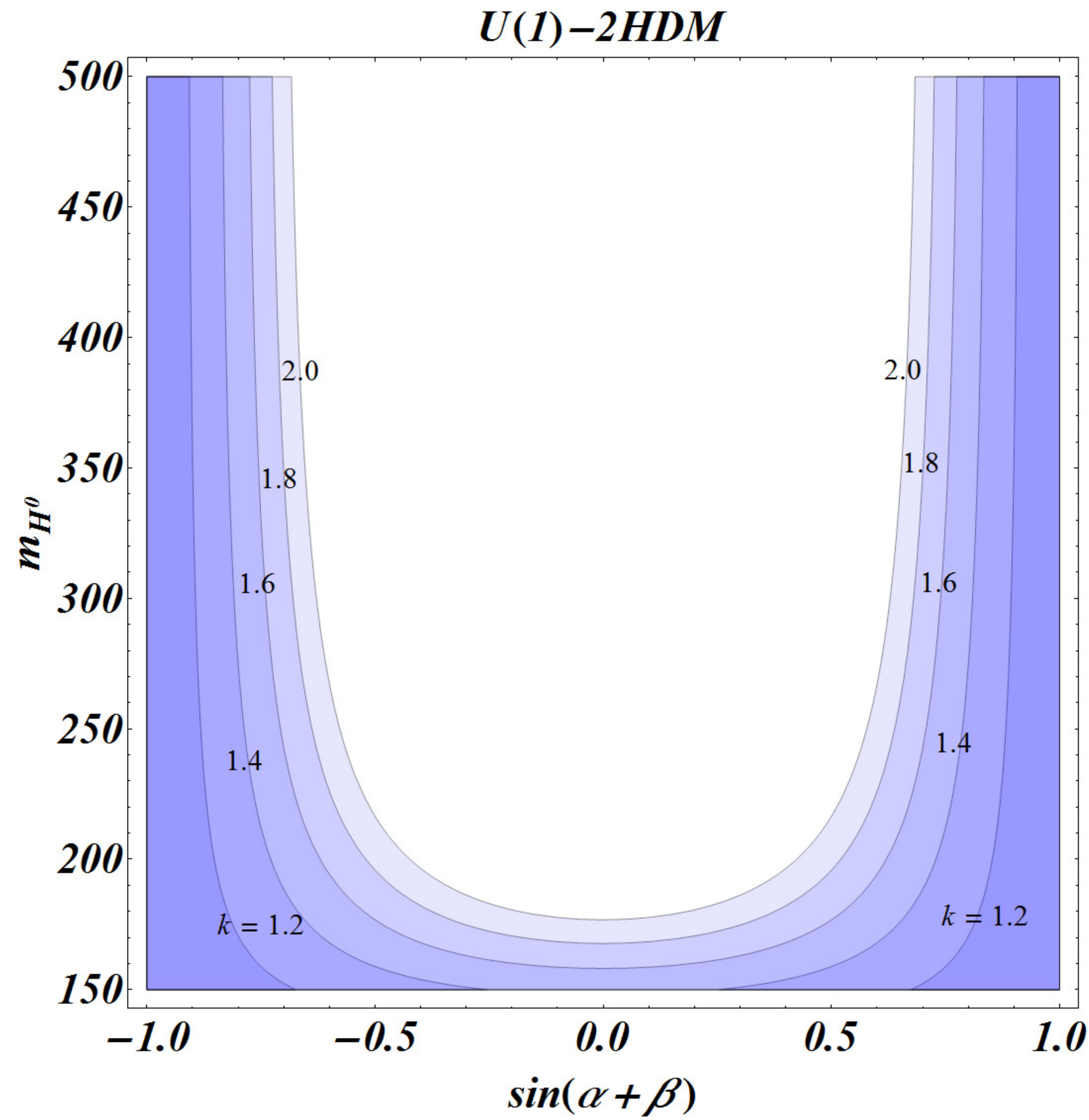}
\vspace{0.4cm}
\caption{Metastability behavior for $m_{H^{0}}-\sin(\protect\alpha+\protect%
\beta)$ plane represented by shadowed zones. Each scenario is characterized
by ratio $k_{S}=m_{A^{0}}^{2}/m_{H^{0}}^{2}$.}
\label{fig:sin}
\end{figure}

Considering Eq. (\ref{MetasDiscr2}) is possible to establish the zones where
absolute stability could be in conflict with the presence of a second minimum.
Ratio between Higgs eigenstates $k_{S}=m_{A^{0}}^{2}/m_{H^{0}}^{2}$ has been
used to compare different regions in the plane $m_{H^{0}}-\sin \left( \alpha
+\beta \right) ,$ as is despited in Fig. (\ref{fig:sin}) 

More generally and to avoid multiple minima at tree level, the lower limit on $m_{H^{0}}$ becomes weaker as $\sin(\alpha+\beta)\to 0$. Moreover, metastable states appear in a wide zone of $m_{H^{0}}-\sin(\beta+\alpha)$ when $k_{s}>1$ increases. In addition when $k_{s}<1$, the parameter space does not show exclusion zones. Finally, metastable states dominate all values of $m_{H^{0}}$ when $\vert\sin(\beta+\alpha)\vert=1$.

\section{Exclusion regions by stability and metastability analyses: \texorpdfstring{$0\leq\alpha\leq\pi/2$}{lg}}
\label{sec:LO}

Discrimination of exclusion zones from metastability in particular regions
of parameter space is relevant to estimate vacuum behavior for the theory
at NLO. Typical constraints found out involve many parameters of the Higgs
potential. Thus, to extract phenomenological information, we analyze
particular models. To that end, we have listed in Tab. \ref{tab:Models} some limiting models for
specific values of $\alpha $ and $\beta $ angles by using expressions of
Higgs eigenstates given in section (\ref{sec:masseigenstates}).

\begin{table}
\centering
\begin{tabular}{|l|l|l|l|l|l|}
\hline
Model & $\tan \beta $ & $\alpha $ & $\Delta S_{1}^{2}v^{2}$ & $\Delta S_{2}^{2}v^{2}$
& $\Delta S_{3}^{2}v^{2}$ \\ \hline\hline
$A_{I}$ & $10^{-3}$ & $0$ & $m_{H^{0}}^{2}$ & $\xi
_{0}\left( m_{h^{0}}^{2}-m_{A^{0}}^{2}\right) $ & $2m_{H^{\pm
}}^{2}-m_{A^{0}}^{2}$ \\ \hline
$A_{II}$ & $10^{-3}$ & $\rightarrow \pi /4$ & $\frac{1}{2}\left(
m_{H^{0}}^{2}+m_{h^{0}}^{2}\right) $ & $\frac{\xi _{0}}{2}\left(
m_{H^{0}}^{2}+m_{h^{0}}^{2}-2m_{A^{0}}^{2}\right) $ & $2m_{H^{\pm
}}^{2}+m_{H^{0}}^{2}-m_{A^{0}}^{2}-m_{h^{0}}^{2}$ \\ \hline
$A_{III}$ & $10^{-3}$ & $\pi /2$ & $m_{h^{0}}^{2}$ & $%
\xi _{0}\left( m_{H^{0}}^{2}-m_{A^{0}}^{2}\right) $ & $%
2m_{H^{\pm }}^{2}-m_{A^{0}}^{2}$ \\ \hline
$B_{I}$ & $1$ & $0$ & $\left(
2m_{H^{0}}^{2}-m_{A^{0}}^{2}\right) $ & $\left(
2m_{h^{0}}^{2}-m_{A^{0}}^{2}\right) $ & $2m_{H^{\pm
}}^{2}-m_{A^{0}}^{2}$ \\ \hline
$B_{II}$ & $1$ & $\rightarrow \pi /4$ & $\left(
m_{H^{0}}^{2}+m_{h^{0}}^{2}-m_{A^{0}}^{2}\right) $ & $\frac{1}{2}\left(
m_{H^{0}}^{2}+m_{h^{0}}^{2}-m_{A^{0}}^{2}\right) $ & $2m_{H^{\pm
}}^{2}+m_{H^{0}}^{2}-m_{A^{0}}^{2}-m_{h^{0}}^{2}$ \\ \hline
$B_{III}$ & $1$ & $\pi /2$ & $\left(
2m_{h^{0}}^{2}-m_{A^{0}}^{2}\right) $ & $\left(
2m_{H^{0}}^{2}-m_{A^{0}}^{2}\right) $ & $2m_{H^{\pm
}}^{2}-m_{A^{0}}^{2}$ \\ \hline
$C_{I}$ & $10^{2}$ & $0$ & $\simeq \xi _{1}%
(m_{H^{0}}^{2}-m_{A^{0}}^{2})$ & $\simeq m_{h^{0}}^{2}$ & $%
2m_{H^{\pm }}^{2}-m_{A^{0}}^{2}$ \\ \hline
$C_{II}$ & $10^{2}$ & $\rightarrow \pi /4$ & $\simeq \frac{\xi _{1}}{2}%
\left( m_{H^{0}}^{2}+m_{h^{0}}^{2}-2m_{A^{0}}^{2}\right) $ & $\simeq \frac{1%
}{2}\left( m_{H^{0}}^{2}+m_{h^{0}}^{2}\right) $ & $2m_{H^{\pm
}}^{2}+m_{H^{0}}^{2}-m_{A^{0}}^{2}-m_{h^{0}}^{2}$ \\ \hline
$C_{III}$ & $10^{2}$ & $\pi /2$ & $\simeq \xi _{1}\left(
m_{h^{0}}^{2}-m_{A^{0}}^{2}\right) $ & $\simeq m_{H^{0}}^{2}$
& $2m_{H^{\pm }}^{2}-m_{A^{0}}^{2}$ \\ \hline
\end{tabular}
\caption{Splittings among Higgs mass eigenstates for different models, which are
varying mixing angles $\alpha $ and $\beta $ ($\tan \beta $)$;$ being $%
\lambda _{4}={2\left( m_{A^{0}}^{2}-m_{H^{\pm }}^{2}\right) }/{v^{2}}\equiv
\Delta S_{4}^{2}$ independent of those parameters. $\xi _{0}$ and $\xi _{1}$
are related to values of $\cot \beta $ and $\tan \beta $ respectively.
Moreover, they are introduced to conserve perturbative behavior of $%
\lambda $'s couplings, i.e., $\lambda_{i}\sim O(1)$. From stability conditions of $\lambda _{1}>0$ and $%
\lambda _{2}>0$, it is also possible to infer a set of features of each model.}
\label{tab:Models}
\end{table}

\begin{figure}[tph]
\centering\includegraphics[scale=0.15]{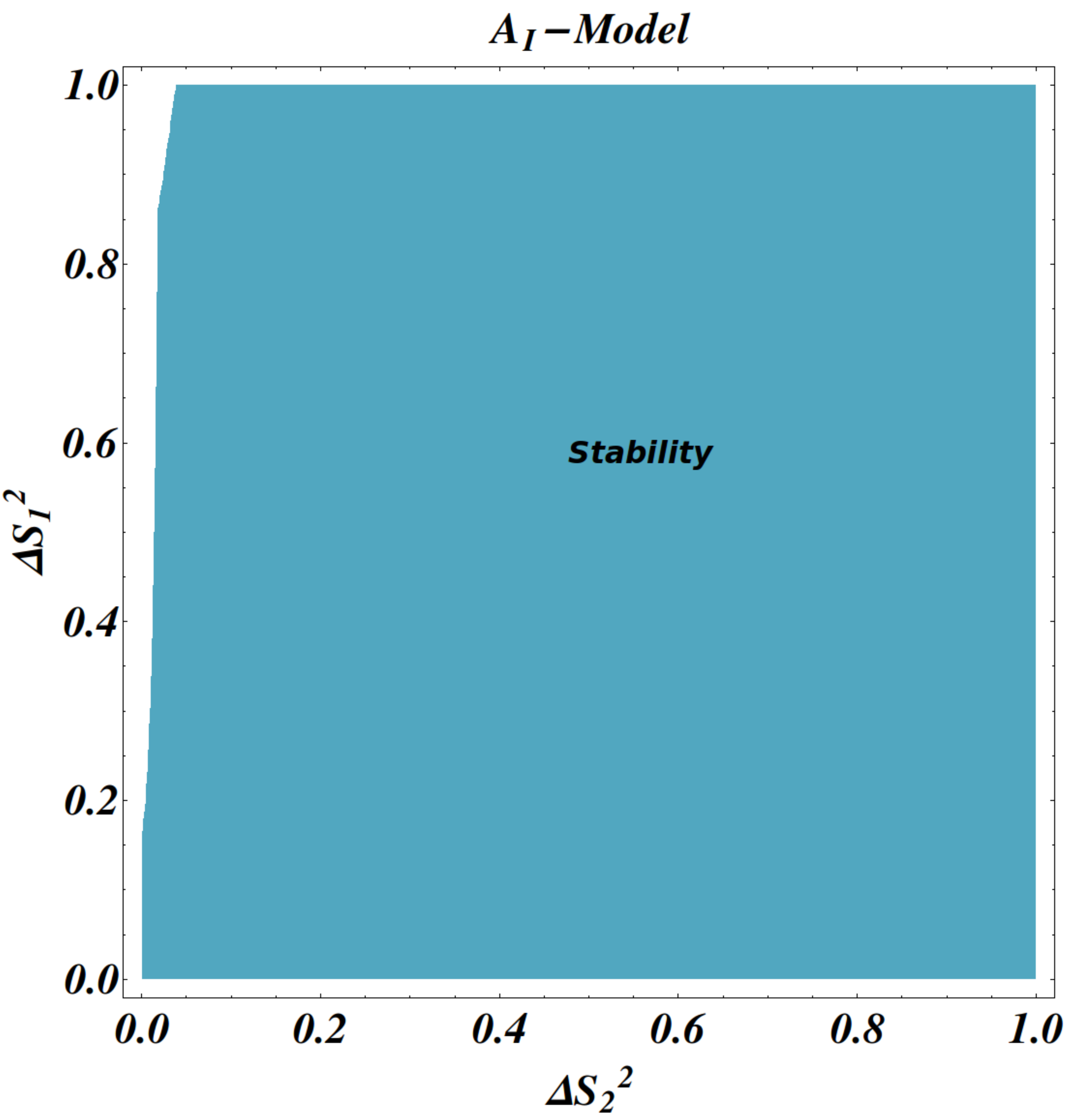} %
\includegraphics[scale=0.15]{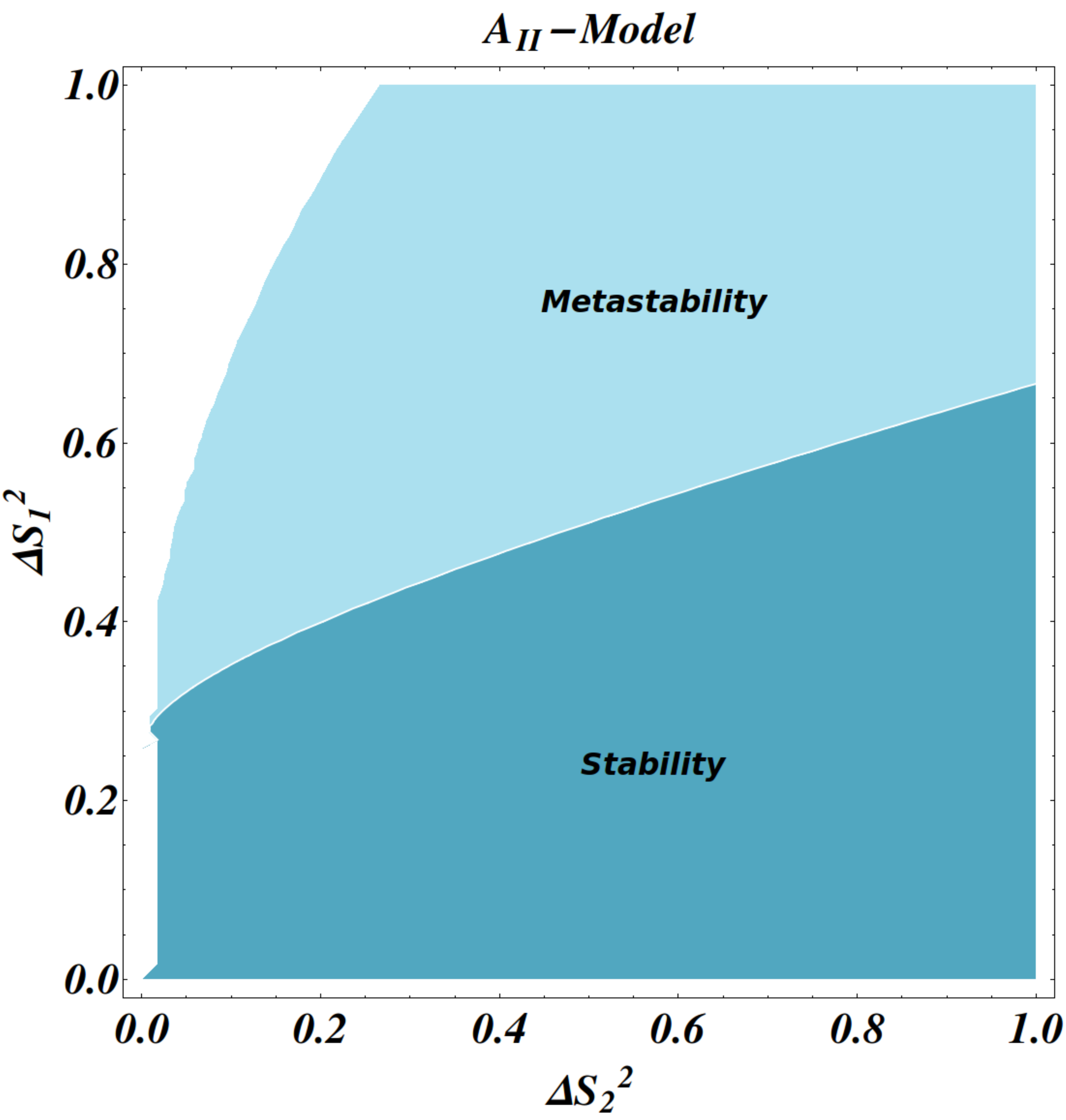} %
\includegraphics[scale=0.15]{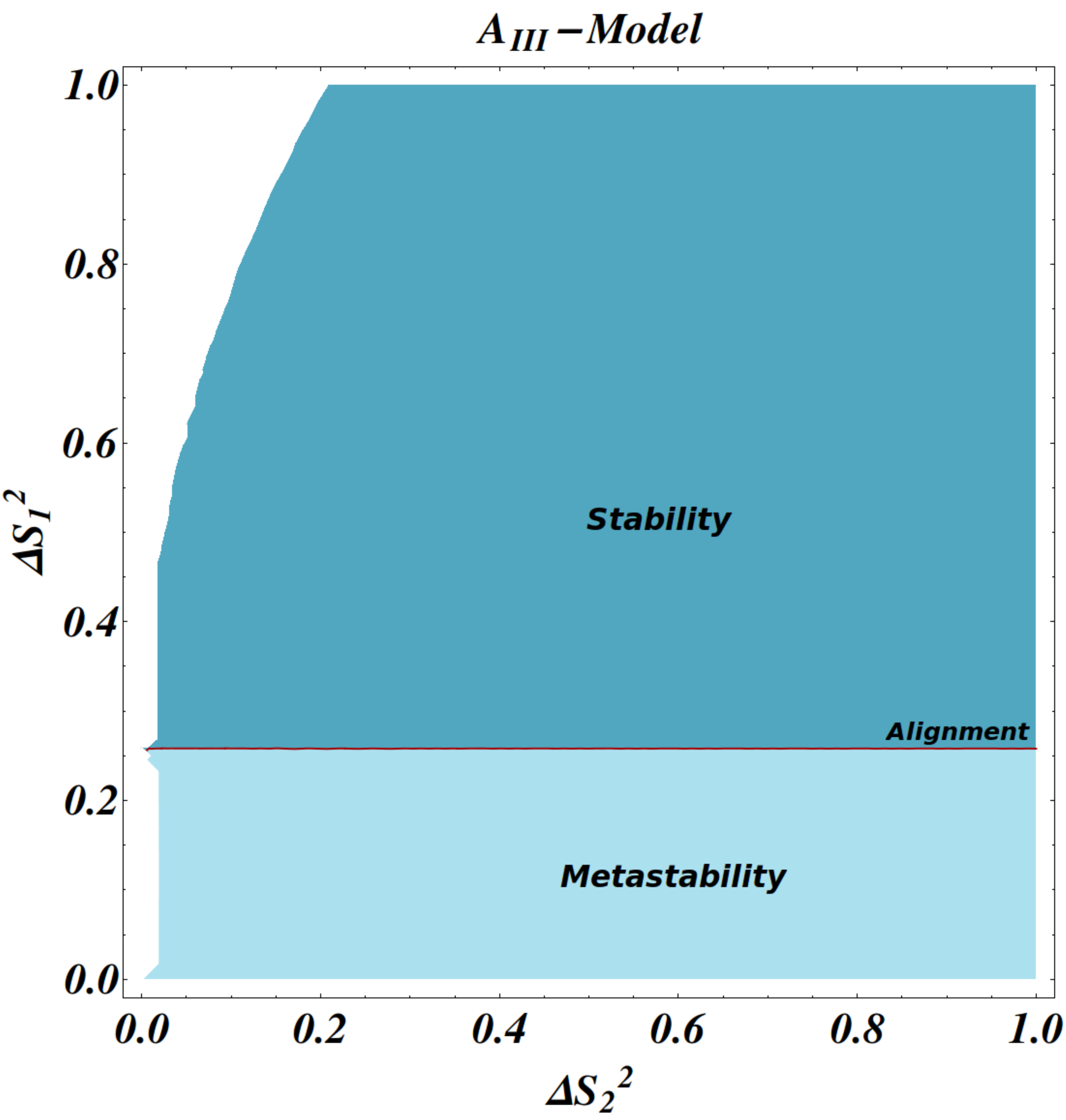} \includegraphics[scale=0.15]{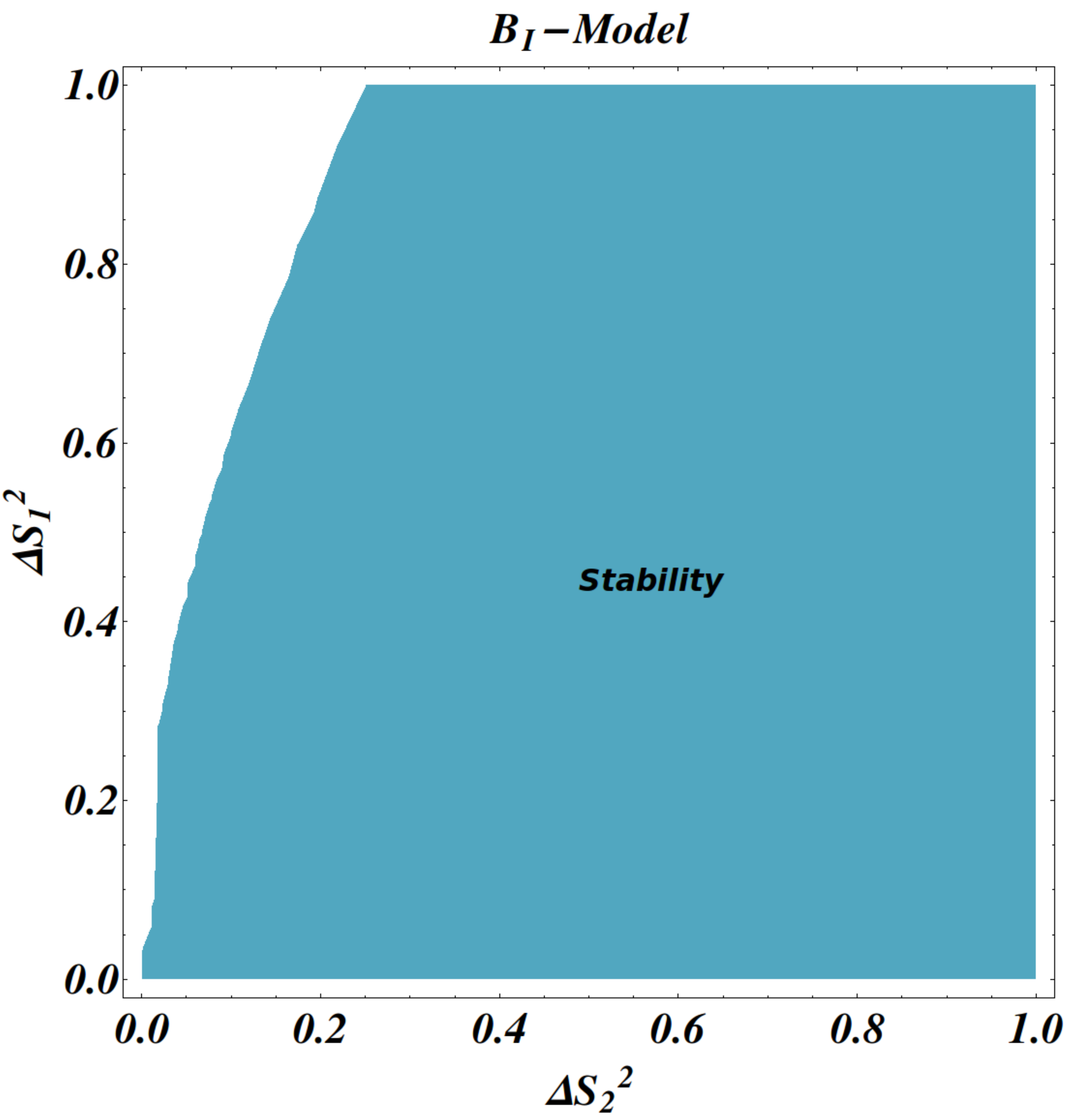}
\includegraphics[scale=0.15]{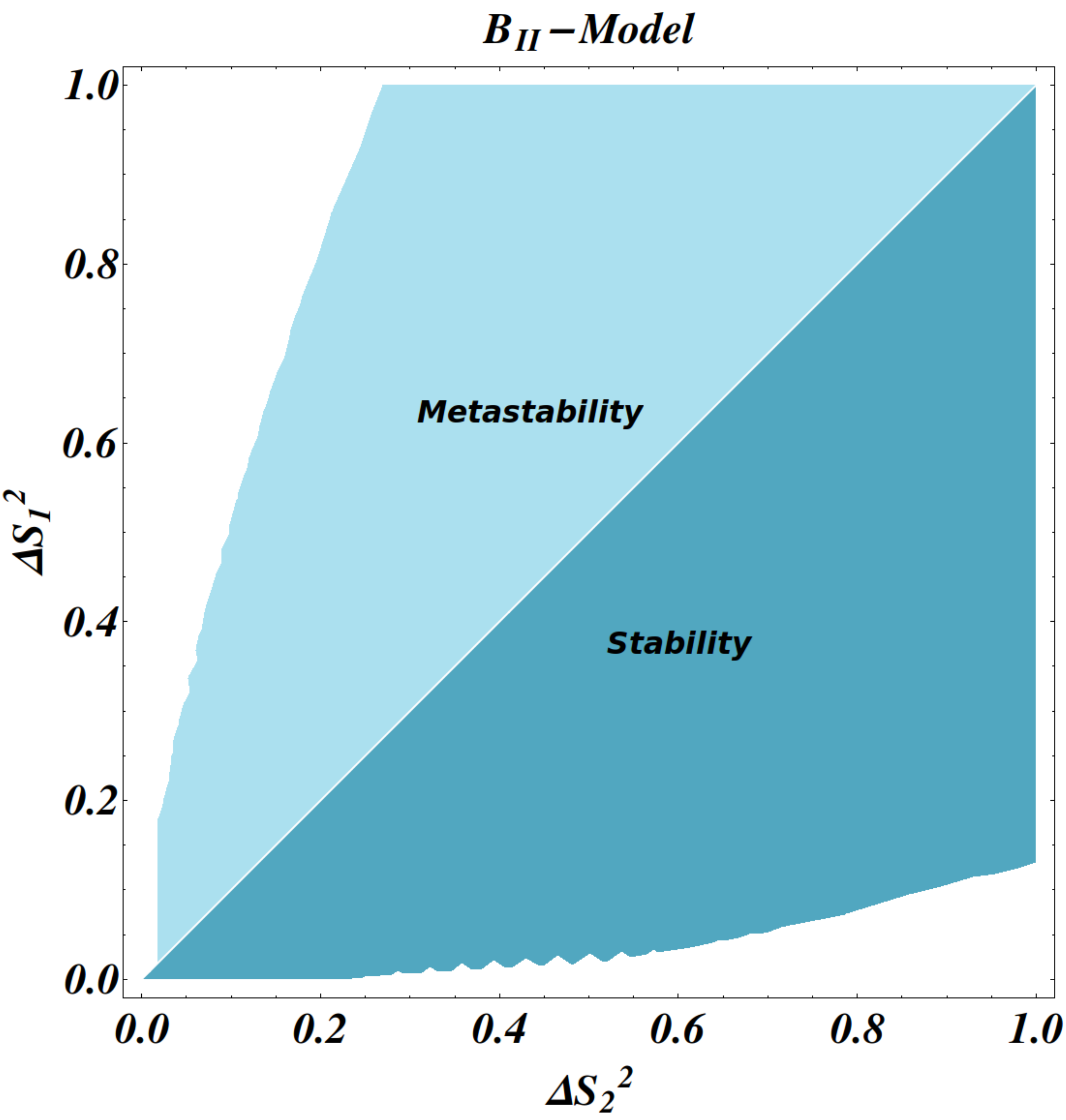} %
\includegraphics[scale=0.15]{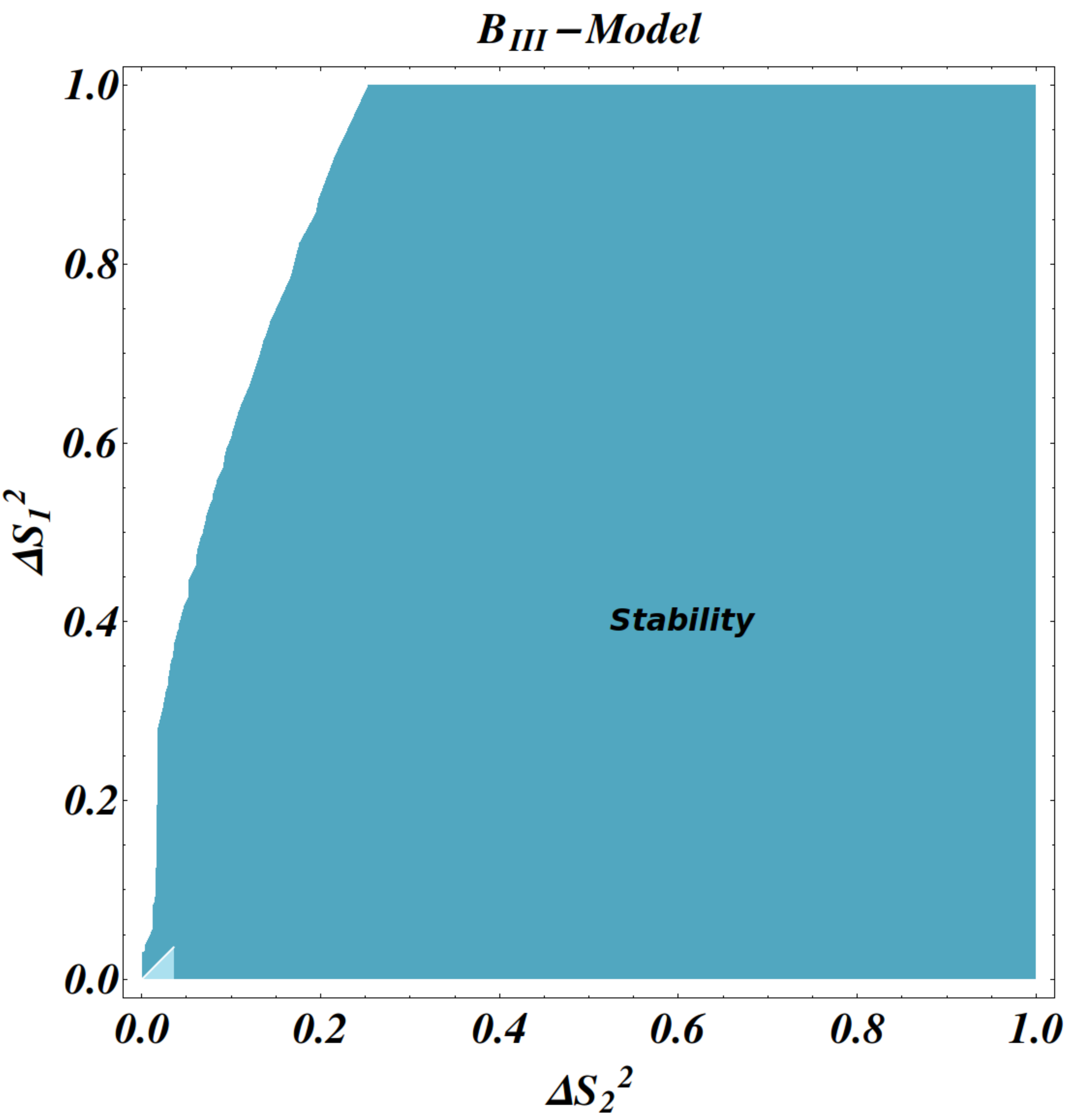} \includegraphics[scale=0.15]{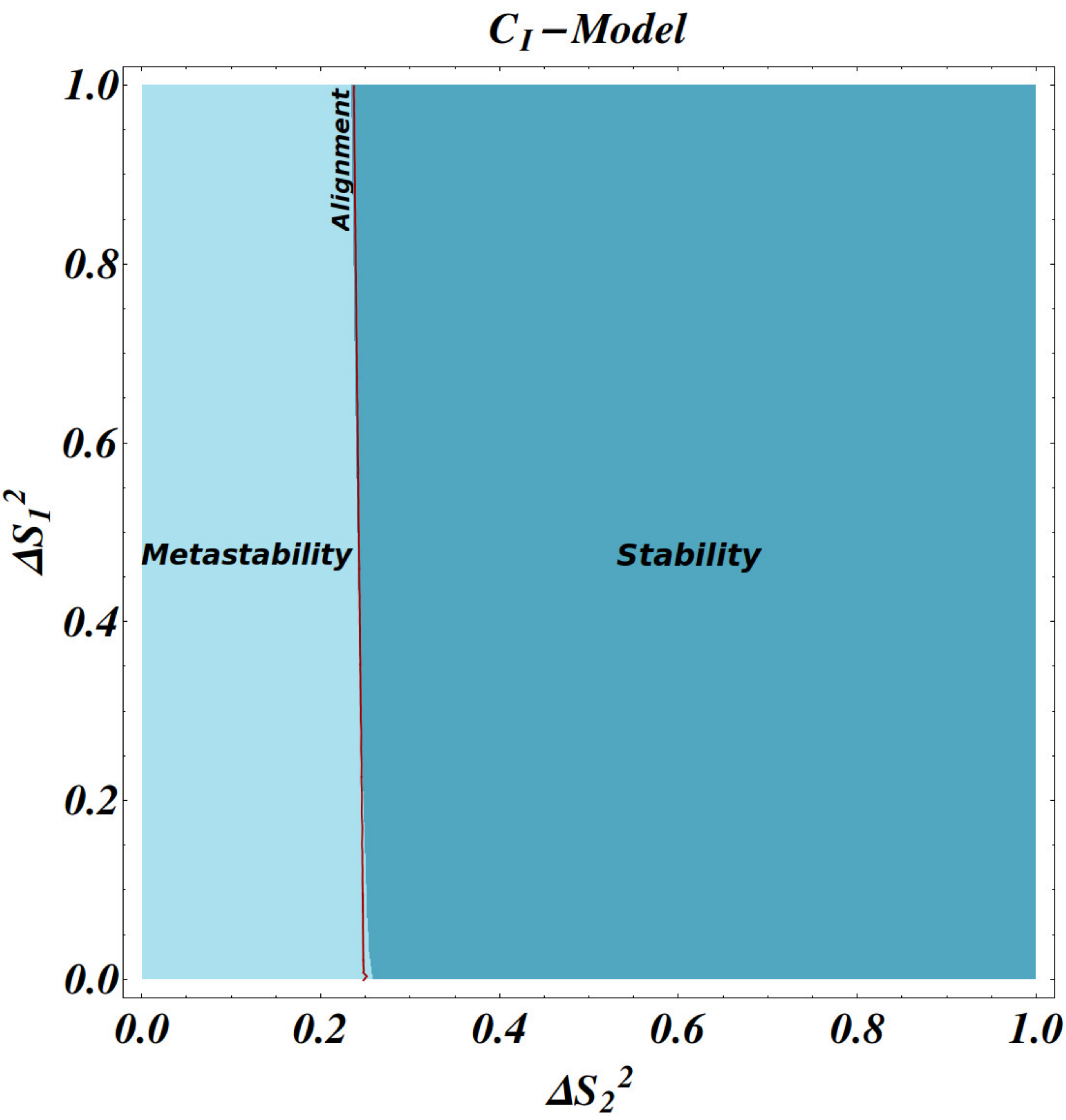}
\includegraphics[scale=0.15]{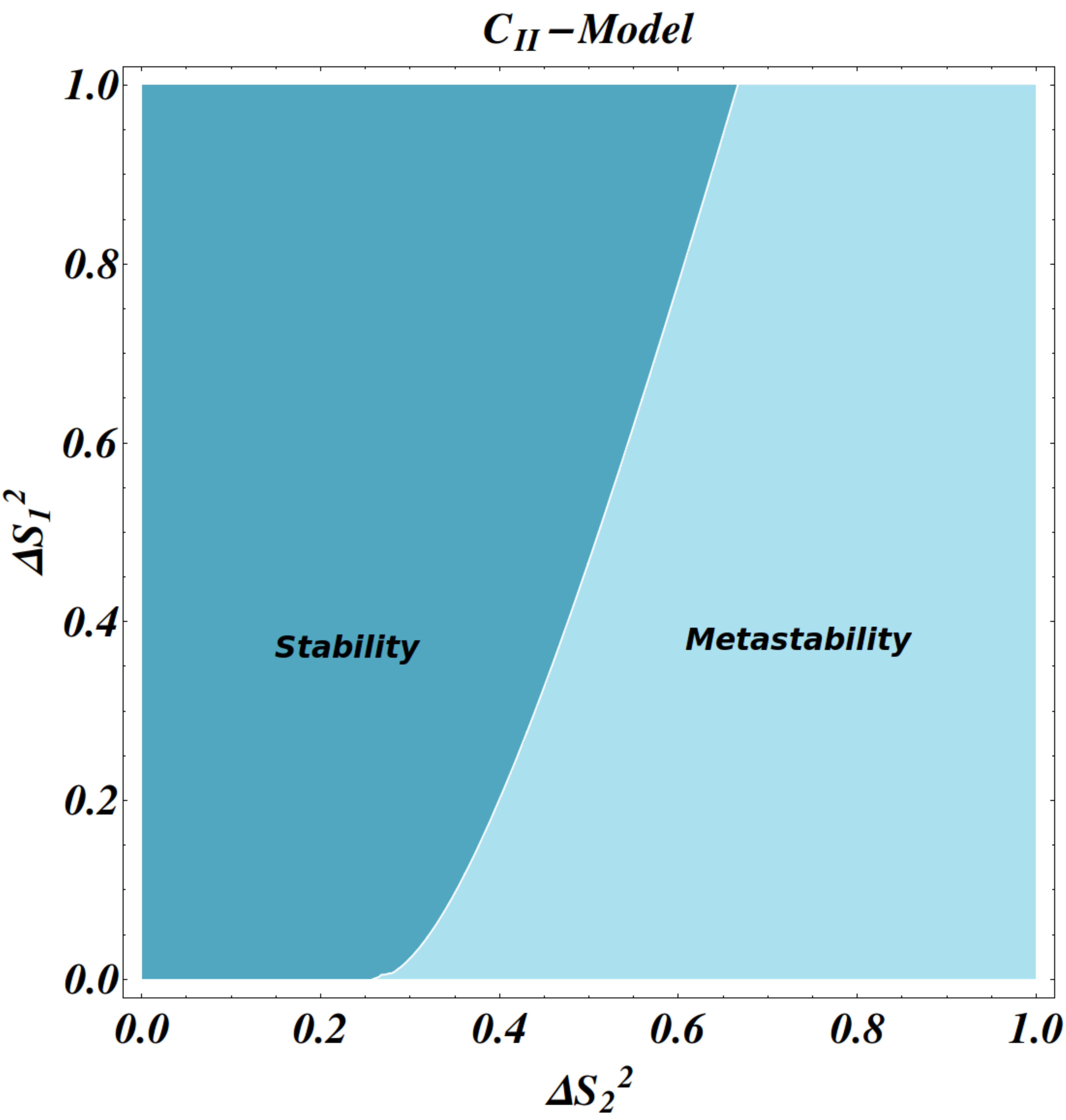} %
\includegraphics[scale=0.15]{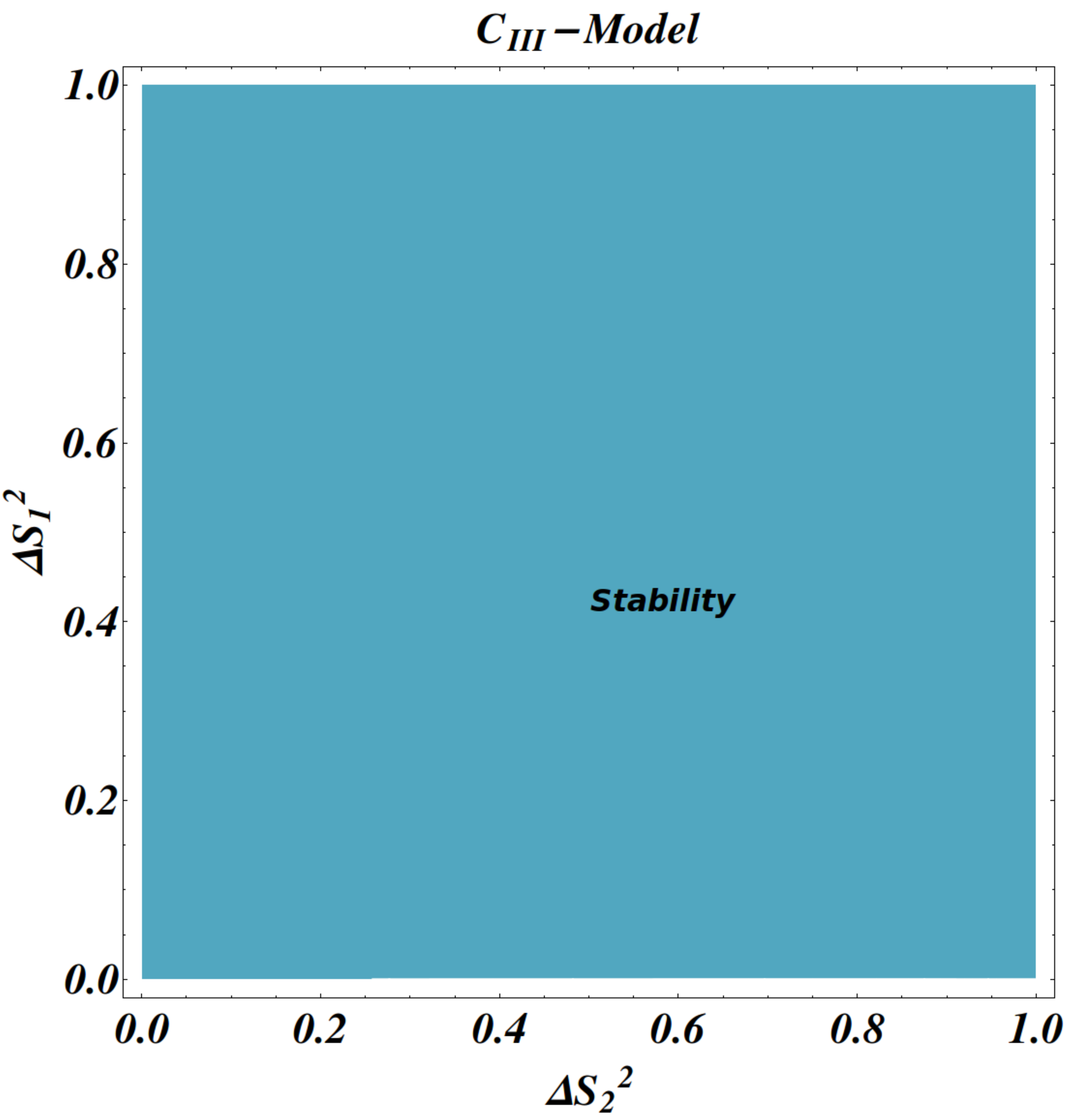}
\vspace{0.4cm}
\caption{Metastability and absolute stability region in $\Delta
S_{1}^{2}-\Delta S_{2}^{2}$ plane for $A_{i},B_{i}$ and $C_{i}$ models with $%
i=I,II,III$ described in Table \protect\ref{tab:Models}. Red lines in $A_{III}$ and $C_{I}$ models give information about as alignment regime behaves in those parameter spaces where $\cos(\beta-\alpha)\approx0$.}
\label{fig:conABC}
\end{figure}

Particularly, alignment regime for couplings ($\cos (\beta -\alpha
)\rightarrow 0$) is present in $A_{III}$ and $C_{I}$ models. For model $%
A_{III}$, condition (\ref{alignment}) implies besides $\lambda _{1}=\lambda
_{3}+\lambda _{4}$. Finally, fulfillment of alignment regime in model $C_{I}$ is achieved when $\lambda _{2}=\lambda _{3}+\lambda _{4}$. Other models
taking into account $h^{0}$ with the mass of $125$ GeV, but they do not emulate the same Higgs couplings as in the SM case.

Figure \ref{fig:conABC} shows a set of contours to see metastability states in the parameter
space and for particular cases given in Tab. \ref{tab:Models}. With this in mind, the
plane $\Delta S_{1}^{2}-\Delta S_{2}^{2}$ is the first candidate to observe
exclusion zones using discriminant \ref{MetastabilityDiscriminant}. By parameters counting, metastability zones ends up depending on mixing angles $\alpha,$ $\beta,$ quartic couplings $\lambda _{1}$
and $\lambda _{2},$ and from $m_{h^{0}}$ values. Given this context, we fixed the mass value for a lighter Higgs in $125$ GeV. We have been appealing to a search of a complete alignment scenario for 2HDM, where the lighter Higgs is identified with the SM Higgs boson, and the remaining scalars could be settled at any energy scale, even in the EW regime. Having this exclusion analysis, we proceed to
evaluate the RGE's evolution to describe initial conditions influence in the global solution in the face of vacuum behavior. Here the contours are interpreted by the minimum values obtained from the
discriminant in such a way that the presence of another minimum has been identified as a possible appearance of metastable states.

\begin{itemize}
\item $A_{I}$ model. Metastable states are suppressed in the plane $\Delta
S_{1}^{2}-\Delta S_{2}^{2}$. Model has only instabilities for the effective
Higgs potential. To maintain perturbativity in this framework $%
m_{A^{0}}\sim m_{h^{0}}$ ($\Phi _{2}$-direction) and to avoid minima with
charge violation, scalar spectrum behaves as $%
m_{h^{0},A^{0}}<m_{H^{0}}$ and $m_{A^{0}}<m_{H^{\pm }}$. This scenario
embodies a non-alignment case, where $H^{0}$ saturates couplings with SM
bosons and fermions.
\item $A_{II}$ model (limit case). Stable zones dominate over broad regions
in the respective plane. Metastable zones start to appear in values of $\Delta
S_{1}^{2}>0.258$ and $\Delta S_{2}^{2}$ approaching to zero. Non-perturbative zones are excluded when $2m_{A^{2}}\sim m_{H^{0}}^{2}+m_{h^{0}}^{2}$, which
is not compatible with stable zones at tree level (presence of one global
minimum in the EW theory).

\item $A_{III}$ model. Metastable zones dominate both directions $%
\Phi _{1}-\Phi _{2}$. In the limit, a small stable zone is located $\Delta
S_{1}^{2}> 0.258$. The lower bound corresponds to the observed  Higgs mass of $%
m_{h^{0}}\approx 125$ GeV and that has been identified with parameters
associated with $h^{0}$. The phenomenological scenario is the alignment regimen defining a boundary between stable and metastable scenarios. Non-perturbative zones are present unless $%
m_{H^{0}}\sim m_{A^{0}}$. This limit model emulates important features of a
$U(1)$-Inert 2HDM, like the hierarchy in the scalar spectrum: $m_{H^{\pm
}}>m_{A^{0},H^{0}}$. In the same way, that happens in the inert model and because of fulfilling of an alignment regime, $H_{0}$ and $A_{0}$ would behave as possible dark matter candidates. It is relevant to point out that all these Abelian models with softly breaking terms can not be reduced down at all to an inert 2HDMs since stationary conditions (\ref{min1}) do not satisfy the choice of $v_{2}=0$ simultaneously.

\item $B_{I}$ model. Metastability is absent in the plane $\Delta
S_{1}^{2}-\Delta S_{2}^{2}$ for zones compatible with SSB. In addition, and
in the same space parameter, vacuum stability at tree level imposes $2
m_{H^{0}}^{2}>m_{A^{0}}^{2}$ and $2 m_{h^{0}}^{2}>m_{A^{0}}^{2}$. Thus $A^{0}\to
h^{0}h^{0}$ and $A^{0}\to H^{0}H^{0}$ decays are suppressed.

\item $B_{II}$ model. Metastability appears in this parameter space significantly,
with a scenario where $\Delta S_{1}^{2}>\Delta S_{2}^{2}$. From vacuum
stability conditions, it is inferred that $%
m_{H^{0}}^{2}+m_{h^{0}}^{2}>m_{A^{0}}^{2}$. Non alignment scenario is present since $\cos(\beta-\alpha)\simeq 1$, thus $H^{0}$ impersonates to couplings SM Higgs with fermions and gauge bosons.

\item $B_{III}$ model. Stability through just one global minimum is ensured in almost all parameter space. Tiny zones of non-stability are
encoded for values of $\lambda_{1}\to0$. Here the scalar spectrum inherits an analog behavior for that obtained in the $B_{I}$ model, i.e., $2 m_{H^{0}}^{2}>m_{A^{0}}^{2}$
and $2 m_{h^{0}}^{2}>m_{A^{0}}^{2}$. The hierarchy structure for scalars
spectrum saves a similar pattern that in the $B_{I}$ model.

\item Model $C_{I}$: Parameter space is broadly dominated by one stable minimum. The presence of one and an only global minimum is ensured for $\Delta S_{1}^{2}\geq 0.258$. By providing perturbativity in the scalar sector,  spectrum should satisfy $%
m_{H^{0}}\sim m_{A^{0}}$; being both scalars plausible dark matter candidates. Once again, this model can emulate a pseudo inert scenario with $v_{1}=0$. It also contains an alignment scenario delimiting the boundary between stable and possible metastable states.

\item Model $C_{II}$: Metastable zones are present in values starting in $%
\Delta S_{2}^{2}\approx 0.258$ for small values of $\Delta S_{1}^{2}$, but
other metastable zones are also present in lower values of $\Delta S_{1}^{2}$.  Together with avoiding non perturbative scenarios, structure of mass eigenstates
implies $m_{h^{0}}^{2}+m_{H^{0}}^{2}\sim 2m_{A^{0}}^{2}$.

\item Model $C_{III}$: Large areas of stability are present in this
particular parameter space. Nonperturbative scenarios appear unless that $%
m_{h^{0}}\sim m_{A^{0}}$, with a small splitting preferring $%
m_{A^{0}}^{2}-m_{h^{0}}^{2}<0$. Here $H^{0}$ can be identified with possible
values taken by $\lambda _{2}$. In comparison to cases $A_{III},B_{II}$ and 
$C_{I}$, $C_{III}$ model ensures a non-alignment regime where couplings to
SM bosons and fermions are dominated by $H^{0}$, meanwhile $h^{0}$ is approximately decoupled of them. Perhaps, this model has many strong
constraints from a phenomenological point of view \cite{Carena2}.
\end{itemize}

\section{Exclusion regions by stability and metastability analyses: \texorpdfstring{$-\pi/2<\alpha<0$}{lg}}
 
By construction of mass eigenstates, there is another relevant phenomenological zone where $-\pi/2<\alpha<0$, with limit values of $\alpha\to-\pi/2$ and $\alpha\to-\pi/4$.
\begin{table}
\centering
\begin{tabular}{|l|l|l|l|l|l|}
\hline
Model & $\tan \beta $ & $\alpha $ & $\Delta S_{1}^{2}v^{2}$ & $\Delta S_{2}^{2}v^{2}$
& $\Delta S_{3}^{2}v^{2}$ \\ \hline\hline
$D_{I}$ & $10^{-3}$ & $\rightarrow -\pi /4$ & $\frac{1}{2}\left(
m_{H^{0}}^{2}+m_{h^{0}}^{2}\right) $ & $\frac{\xi _{1}}{2}\left(
m_{H^{0}}^{2}+m_{h^{0}}^{2}-2m_{A^{0}}^{2}\right) $ & $2m_{H^{\pm
}}^{2}+m_{h^{0}}^{2}-m_{A^{0}}^{2}-m_{H^{0}}^{2}$ \\ \hline
$D_{II}$ & $10^{-3}$ & $\rightarrow -\pi /2$ & $%
m_{h^{0}}^{2}$ & $\xi _{1}\left(
m_{H^{0}}^{2}-m_{A^{0}}^{2}\right) $ & $2m_{H^{\pm }}^{2}-m_{A^{0}}^{2}%
$ \\ \hline
$E_{I}$ & $1$ & $\rightarrow -\pi /4$ & $\frac{1}{2}\left(
m_{H^{0}}^{2}+m_{h^{0}}^{2}-m_{A^{0}}^{2}\right) $ & $\frac{1}{2}\left(
m_{H^{0}}^{2}+m_{h^{0}}^{2}-m_{A^{0}}^{2}\right) $ & $2m_{H^{\pm
}}^{2}+m_{h^{0}}^{2}-m_{A^{0}}^{2}-m_{H^{0}}^{2}$ \\ \hline
$E_{II}$ & $1$ & $\rightarrow -\pi /2$ & $\left(
2m_{h^{0}}^{2}-m_{A^{0}}^{2}\right) $ & $\left(
2m_{H^{0}}^{2}-m_{A^{0}}^{2}\right) $ & $2m_{H^{\pm
}}^{2}-m_{A^{0}}^{2}$ \\ \hline
$F_{I}$ & $10^{2}$ & $\rightarrow -\pi /4$ & $\simeq \frac{\xi _{0}}{2}%
\left( m_{H^{0}}^{2}+m_{h^{0}}^{2}-2m_{A^{0}}^{2}\right) $ & $\simeq \frac{1%
}{2v^{2}}\left( m_{H^{0}}^{2}+m_{h^{0}}^{2}\right) $ & $2m_{H^{\pm
}}^{2}+m_{h^{0}}^{2}-m_{A^{0}}^{2}-m_{H^{0}}^{2}$ \\ \hline
$F_{II}$ & $10^{2}$ & $\rightarrow -\pi /2$ & $\simeq \xi _{0}%
\left( m_{h^{0}}^{2}-m_{A^{0}}^{2}\right) $ & $\simeq %
m_{H^{0}}^{2}$ & $2m_{H^{\pm }}^{2}-m_{A^{0}}^{2}$ \\ \hline
\end{tabular}%
\vspace{0.4cm}
\caption{Splittings among Higgs mass eigenstates for different models, which are
varying mixing angles $\alpha $ and $\beta $ ($\tan \beta $)$;$ being $%
\lambda _{4}={2\left( m_{A^{0}}^{2}-m_{H^{\pm }}^{2}\right) }/{v^{2}}\equiv
\Delta S_{4}^{2}$ independent of those parameters. $\xi _{0}$ and $\xi _{1}$
are related to values of $\cot \beta $ and $\tan \beta $ respectively.
Moreover, they are introduced in order to conserve perturbative behavior of $%
\lambda $'s couplings, i.e., $\lambda_{i}\sim O(1)$. From stability conditions of $\lambda _{1}>0$ and $%
\lambda _{2}>0,$ it is possible to infer a set of features of each model}
\label{tab:Modelsan}
\end{table}

\begin{figure}[tph]
\centering\includegraphics[scale=0.15]{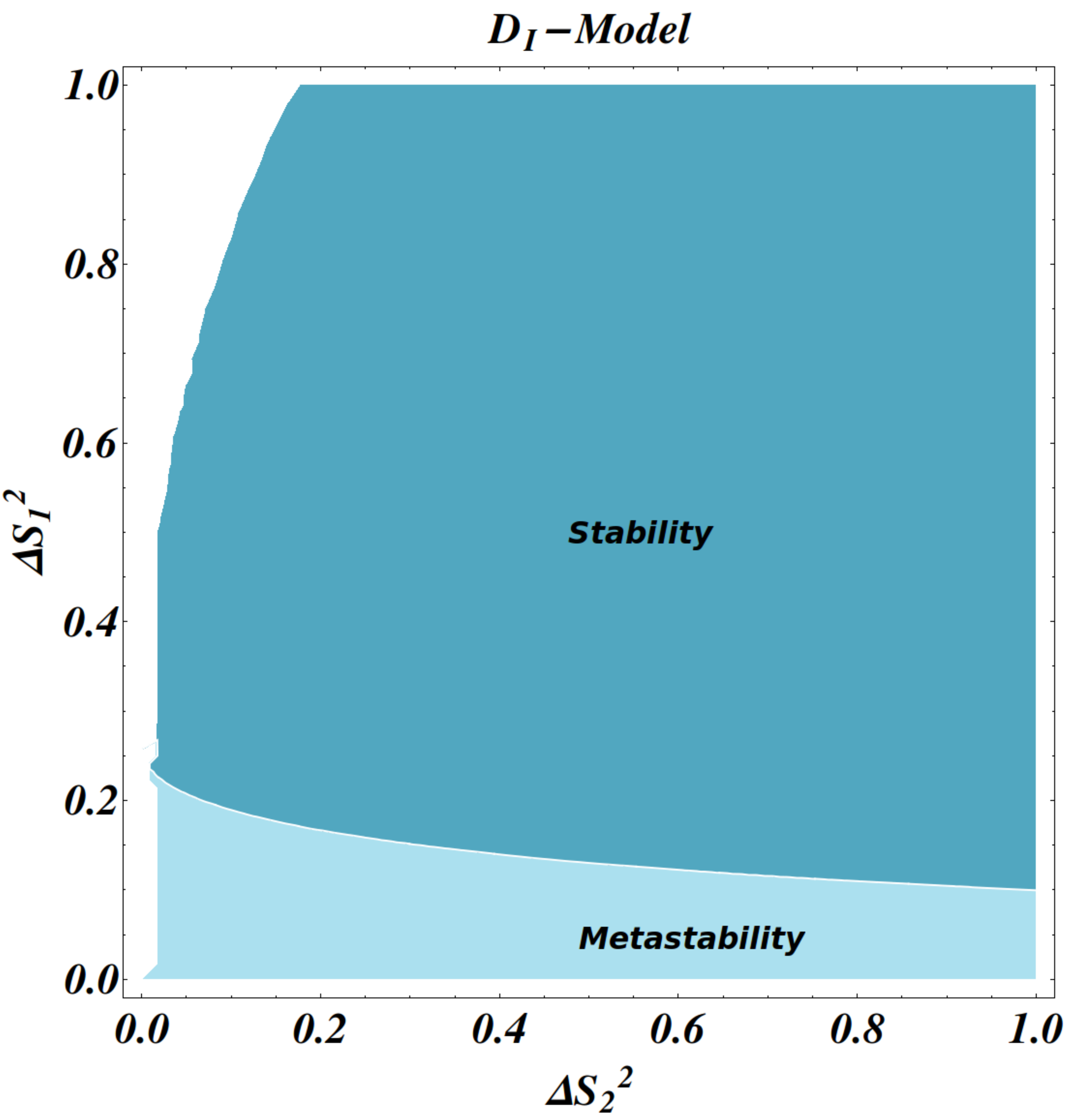} %
\includegraphics[scale=0.15]{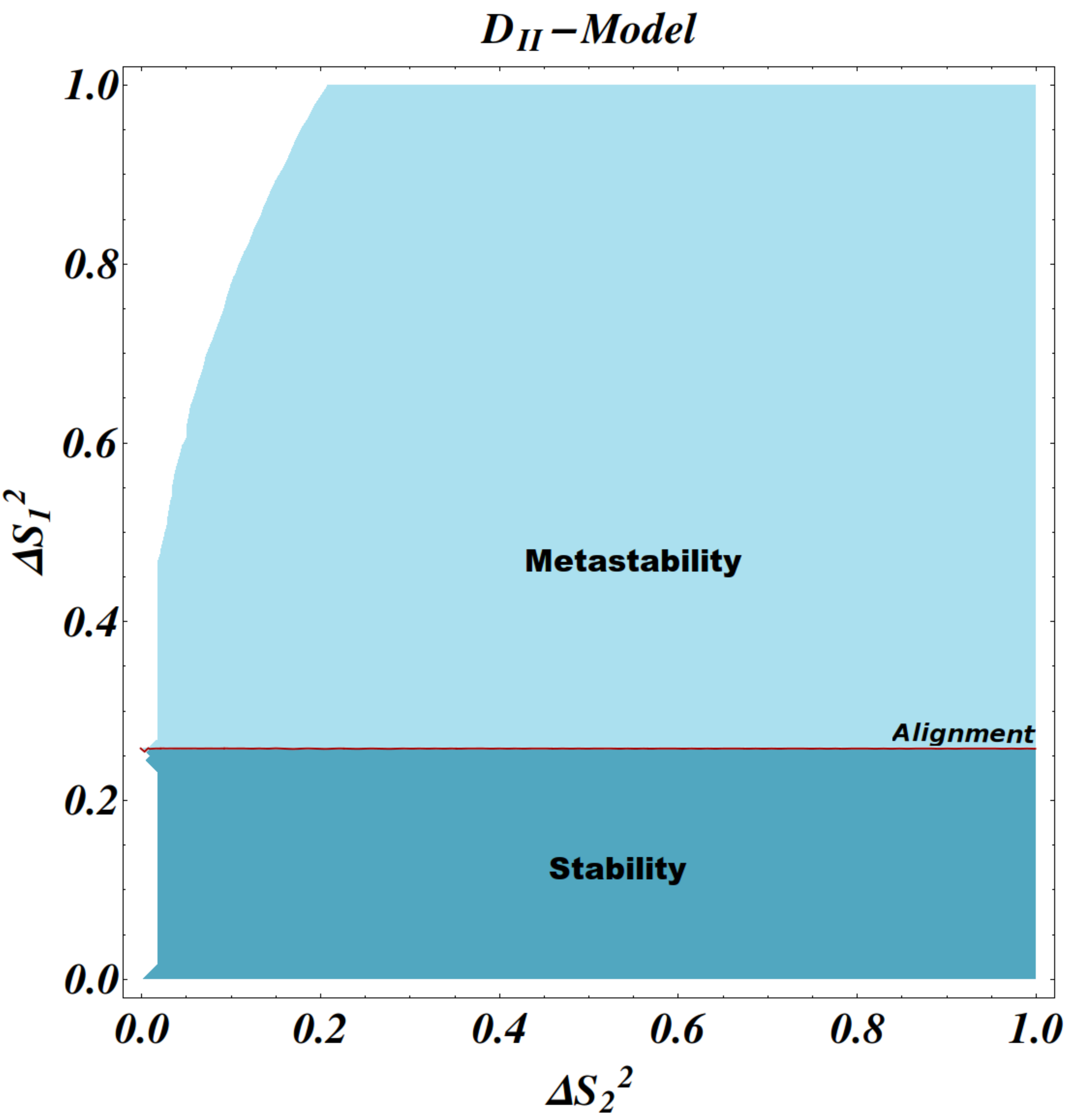} %
\includegraphics[scale=0.15]{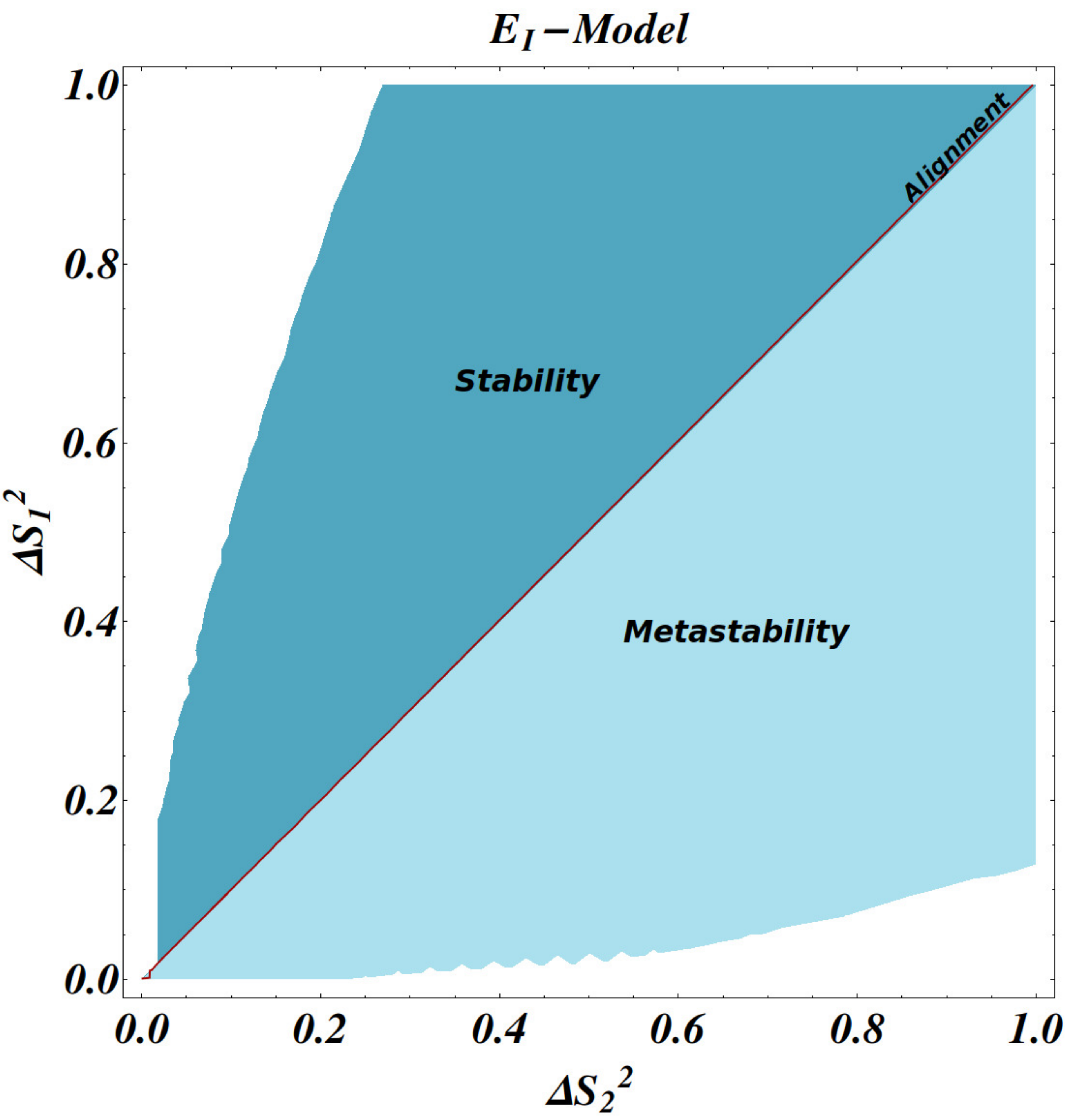} 
\includegraphics[scale=0.15]{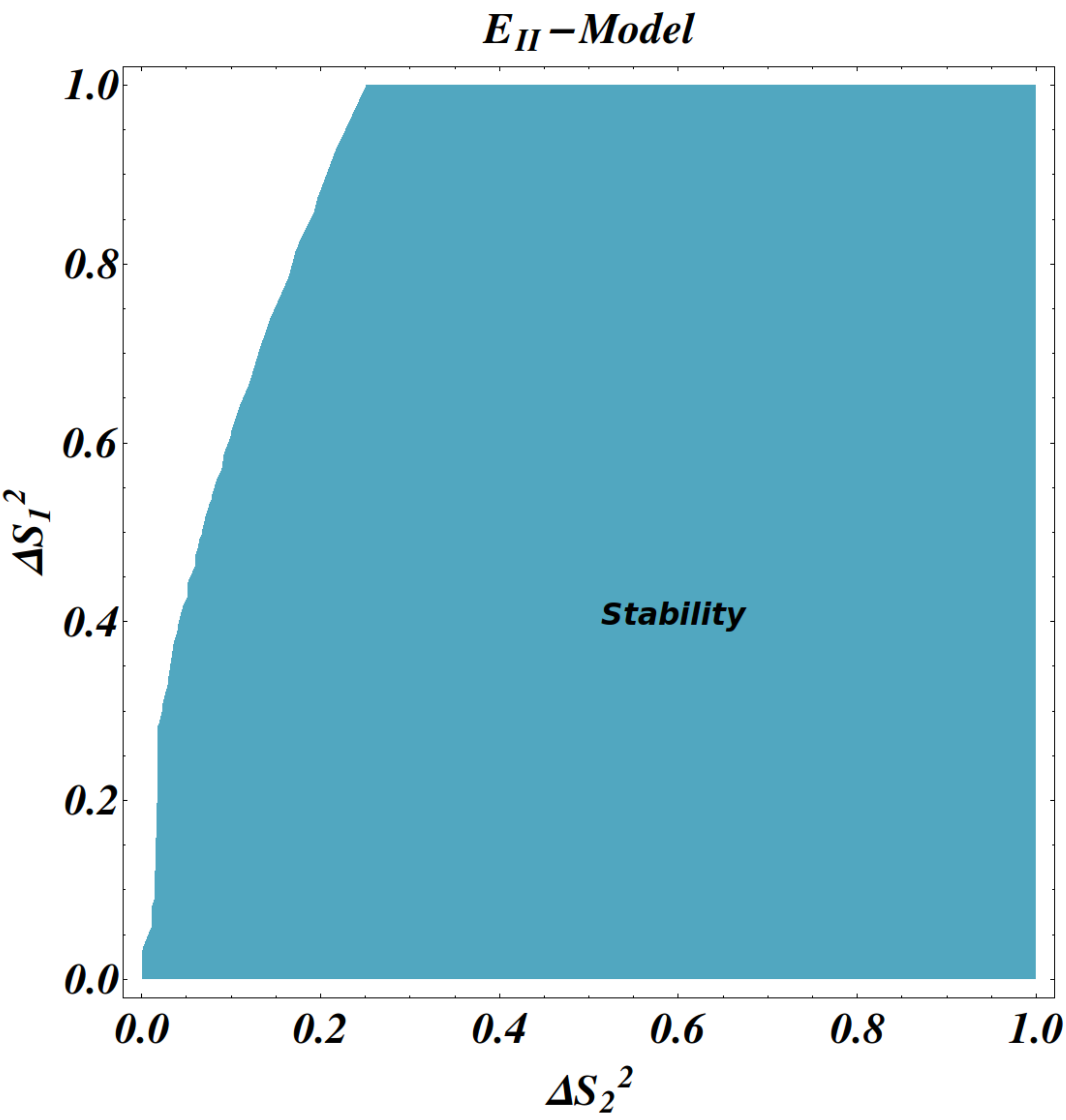} %
\includegraphics[scale=0.15]{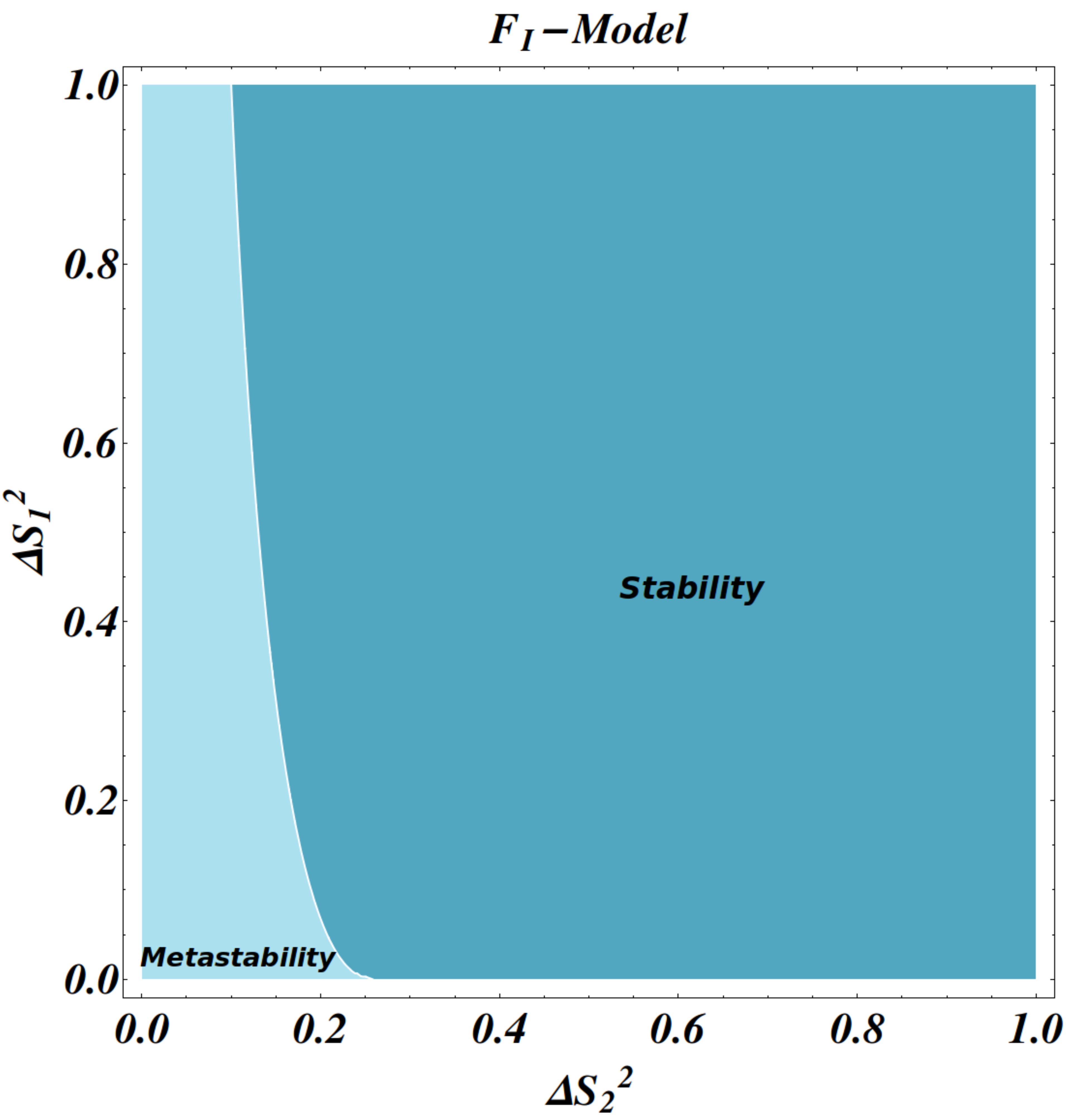} 
\includegraphics[scale=0.15]{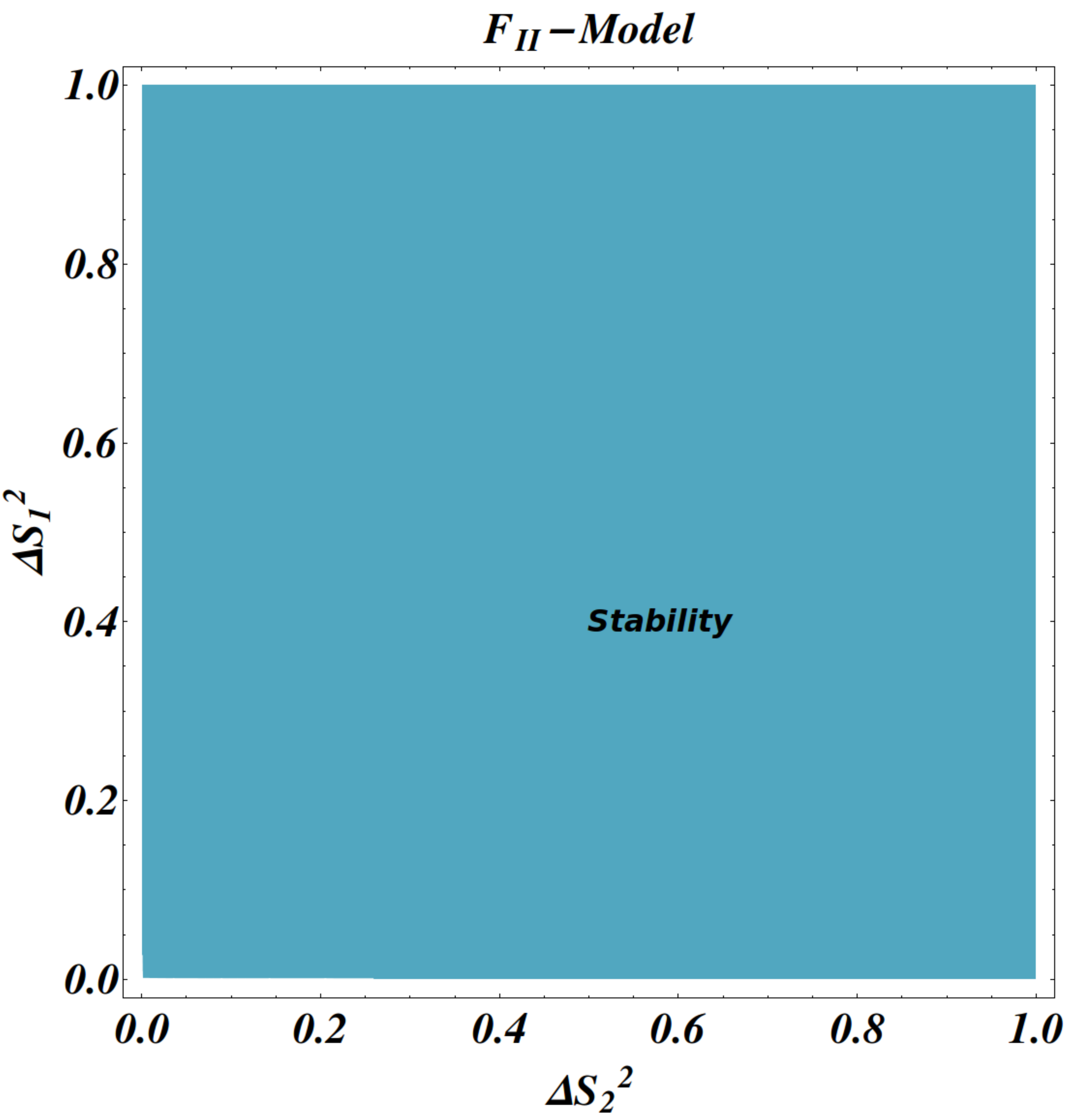} 
\vspace{0.4cm}
\caption{Metastability and absolute stability region in $\Delta
S_{1}^{2}-\Delta S_{2}^{2}$ plane for $D_{i},E_{i}$ and $F_{i}$ models with $%
i=I,II$ described in Table \protect\ref{tab:Modelsan}. Red lines in $D_{II}$ and  $E_{I}$ models give information about as alignment regime behaves in those parameter spaces where $\cos(\beta-\alpha)\approx0$.}
\label{fig:metastabilityn}
\end{figure}

There are two realizations of alignment regime given by the $D_{II}$ and $E_{I}$ models. From Fig. \ref{fig:metastabilityn} and mass eigenstates behavior described in Table \ref{tab:Modelsan}, we can extract the following features:

\begin{itemize}
 \item $D_{I}$-model: Despite this choice contains a metastable zone for lower values of $\Delta S_{1}^{2}<0.258$, the presence of a global minimum dominates this parameter space. The most stringent bound from metastable behavior is also in lower values of $\Delta S_{2}^{2}$. Moreover by perturbativity $m_{h^{0}}^{2}+m_{H^{0}}^{2}\sim 2m_{A^{0}}^{2}$; being slightly greater the sum of CP even states.
 \item $D_{II}$-model: A metastable zone arises in $\Delta S_{1}^{2}\geq0.258$. By virtue of identification of $h^{0}$ with SM Higgs, alignment regime is developed over this value separating global minima and metastability behavior. In addition $m_{H^{0}}\sim m_{A^{0}}$ condition coming from perturbativity request. In this model $A^{0}\to H^{0}H^{0}$ decays are forbidden.
 \item $E_{I}$ model: A broad zone of metastability is present for $\Delta S_{2}^{2}\geq\Delta S_{1}^{2}$, developing an alignment regime in the limit where both parameters are equal. Hence alignment scenario defines a boundary between a theory with a global minimum and one model with more of two minima. By the form of the couplings, alignment seems to be the best limit in the parameter space. Stability in the $\Phi_{1}$ and $\Phi_{2}$ directions leads to $m_{H}^{0}+m_{h^{0}}>m_{A^{0}}$.
 \item $E_{II}$-model: Parameter space compatible with SSB does not contain metastable states for any combination of couplings. In this scenario $2m_{h^{0}}^{2}>m_{A^{0}}^{2}$ and $2m_{H^{0}}^{2}>m_{A^{0}}^{2}$; avoiding $A^{0}\to h^{0}h^{0}$ and $A^{0}\to H^{0}H^{0}$ decays. Thus $m_{A^{0}}<176$ GeV. This model shares some features of metastability and phenomenology with the $B_{I}$ and $B_{III}$-models.
 \item $F_{I}$-model: Metastability zones are developed in lower values of $\Delta S_{2}^{2}<0.258$ extending roughly in all values of $\Delta S_{1}^{2}$. In this parameter space, perturbativity demands $m_{H^{0}}^{2}+m_{h^{0}}^{2}\sim 2m_{A^{0}}^{2}$.
 \item $F_{II}$-model: Stable zones are in almost all parameter space.  This a scenario of non-alignment since $H^{0}$ emulates the couplings among SM-Higgs with fermions and bosons. To avoid non-perturbative scenarios for quartic couplings, $h^{0}$ and $A^{0}$ should be almost degenerate in mass. 
 \end{itemize}

Finally, we note the strong dependence and sensitivity of a unique global solution with the $\tan\beta$ when $\beta\rightarrow0$ and $\beta\to\pi/2$. Therefore, in these zones and models, a more carefully analysis must be done to describe the real behavior of metastable states and global minimum.

\section{One loop analysis for quartic couplings for \texorpdfstring{$B-E$}{lg} and \texorpdfstring{$C-F$}{lg} like models}
\label{sec:NLO}
\begin{figure}[htp]
 \centering
 \includegraphics[scale=0.2]{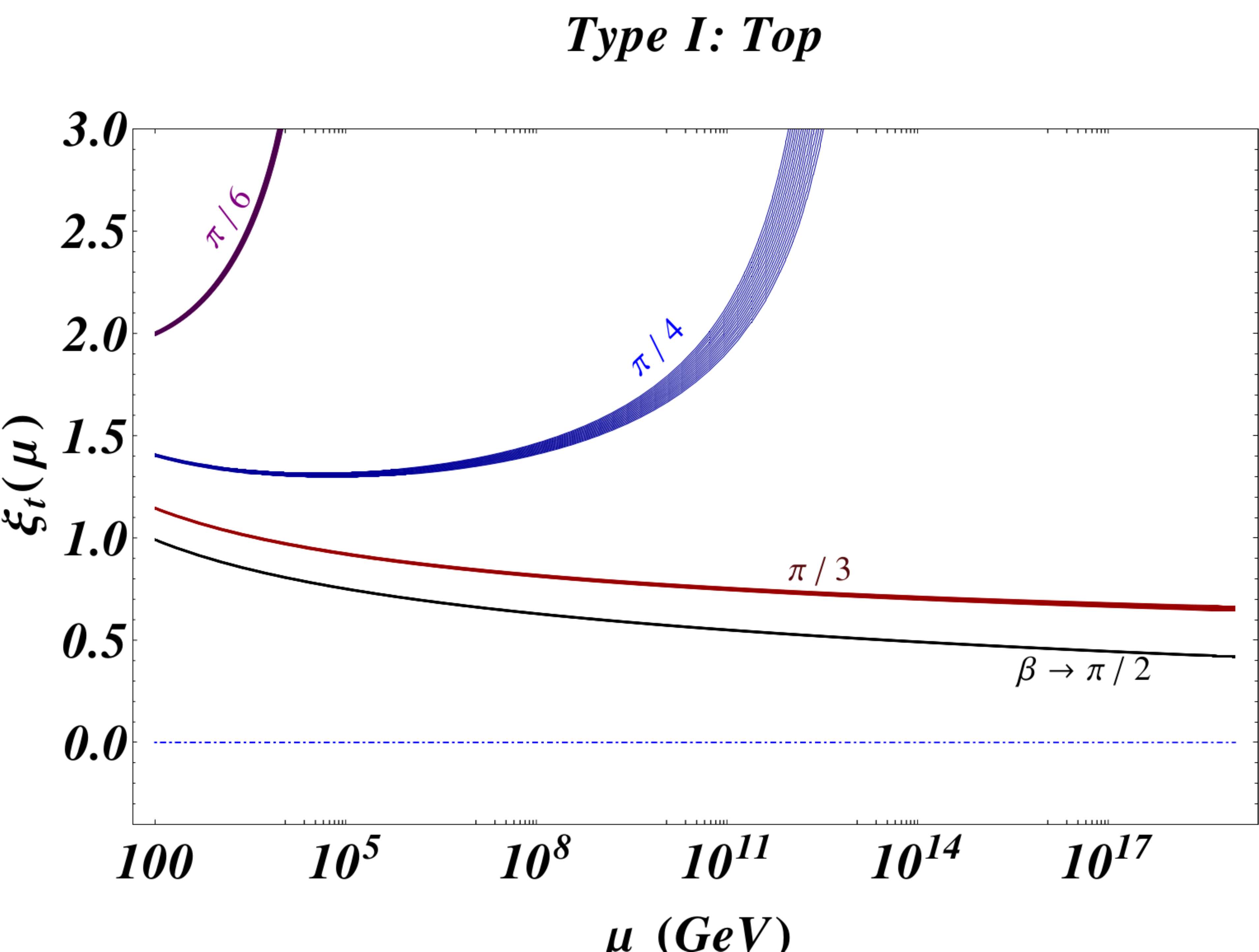}
 \includegraphics[scale=0.2]{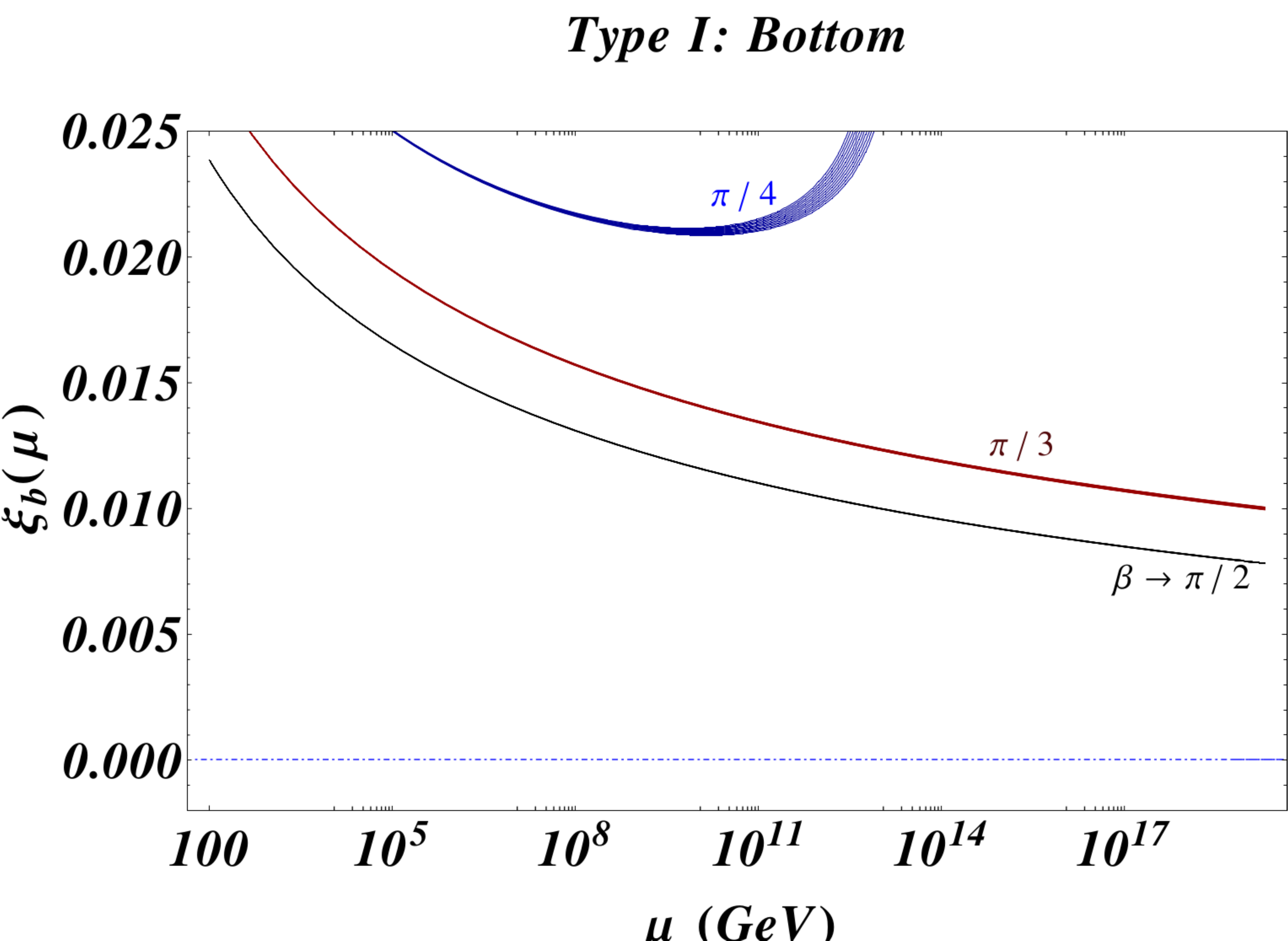}
 \vspace{0.4cm}
 \caption{Top and bottom Yukawa couplings evolution varying mixing angle $\beta$ in the initial condition \ref{etatop}. Width in each evolution is due to top mass (pole) uncertainty $m_{t}=(173.34\pm0.76)$ GeV \cite{Top}.}
 \label{fig:topyuk}
\end{figure}

\begin{figure}[tph]
\centering 
\includegraphics[scale=0.18]{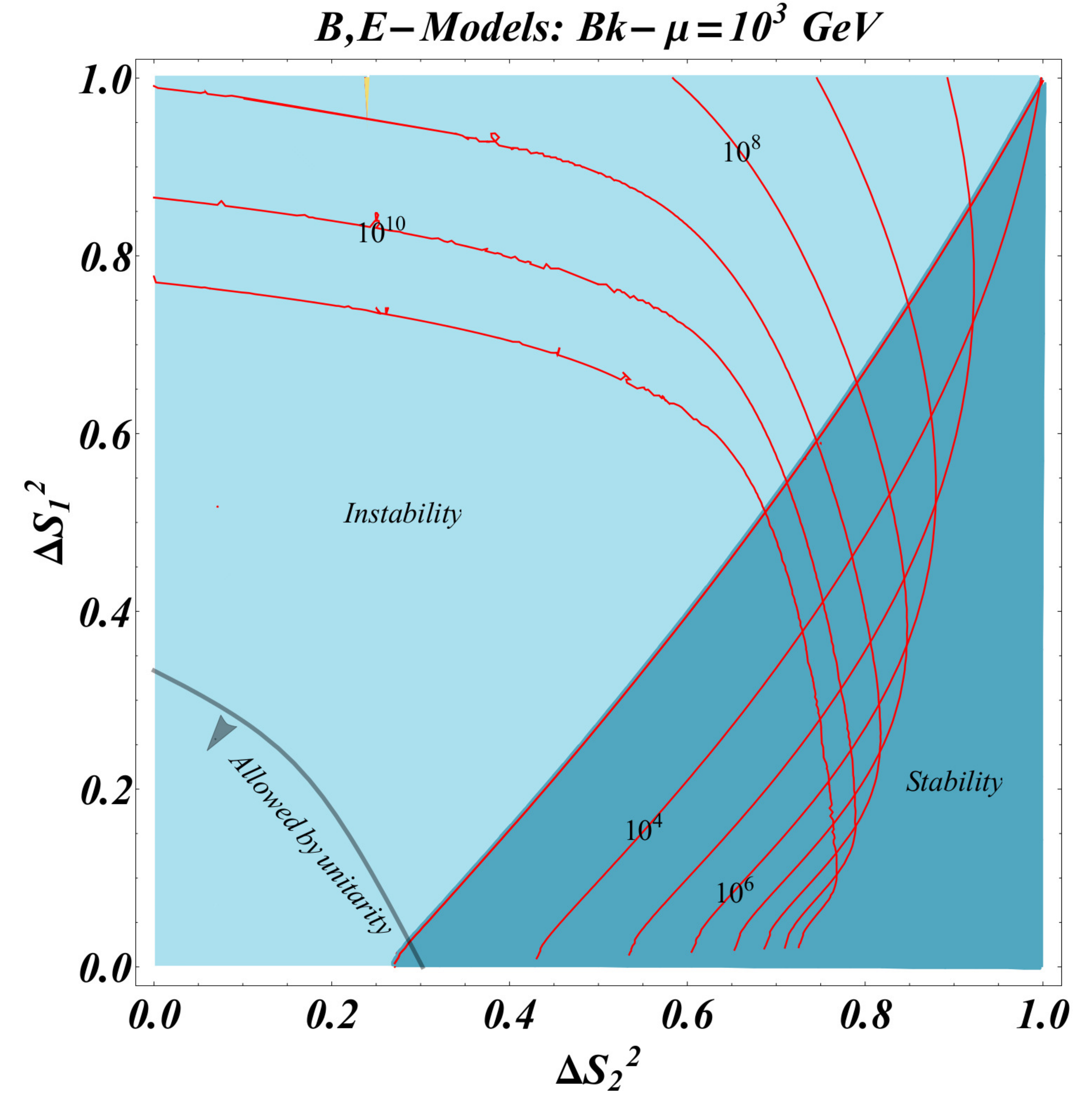} %
\includegraphics[scale=0.18]{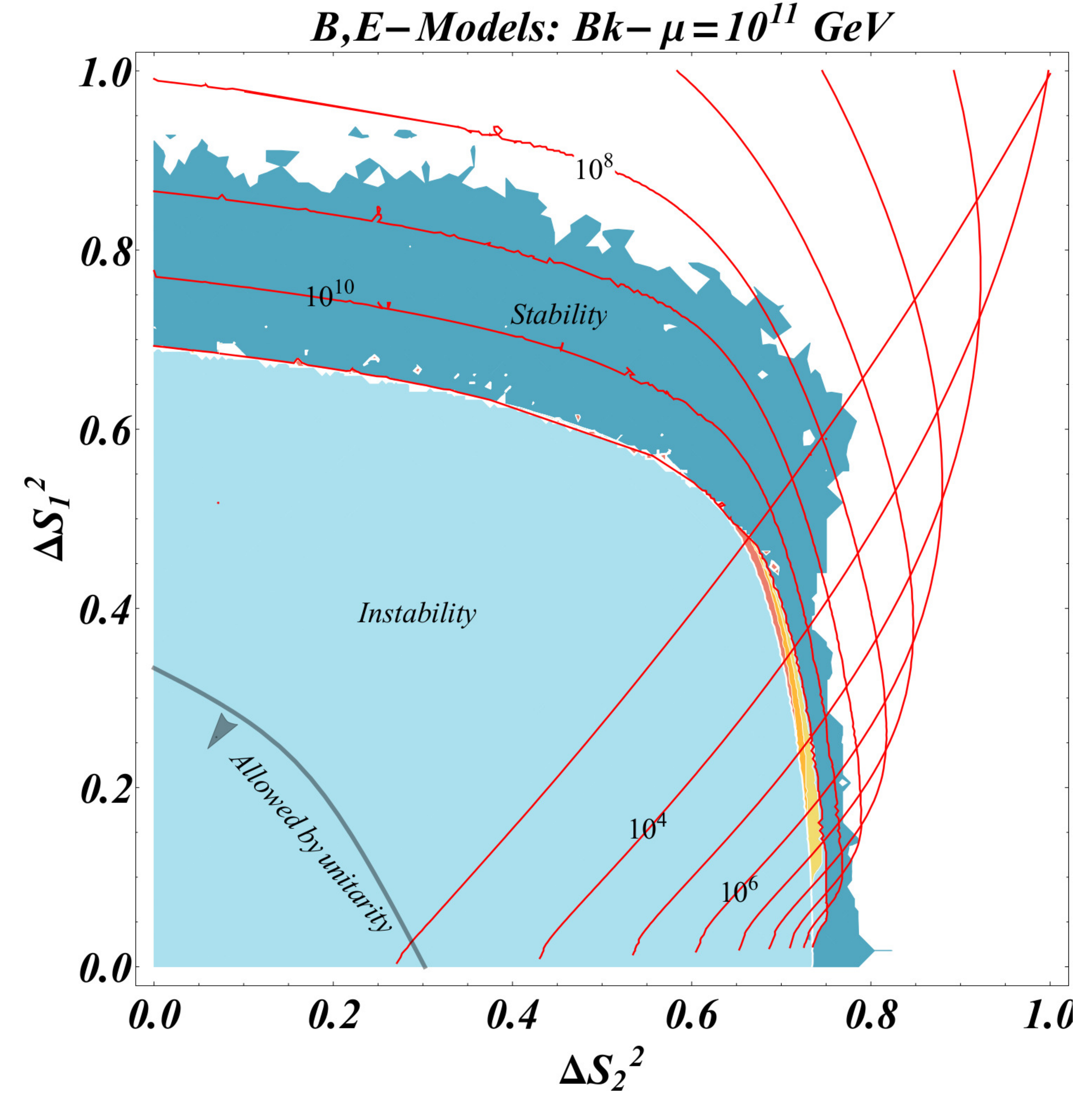} 
\includegraphics[scale=0.18]{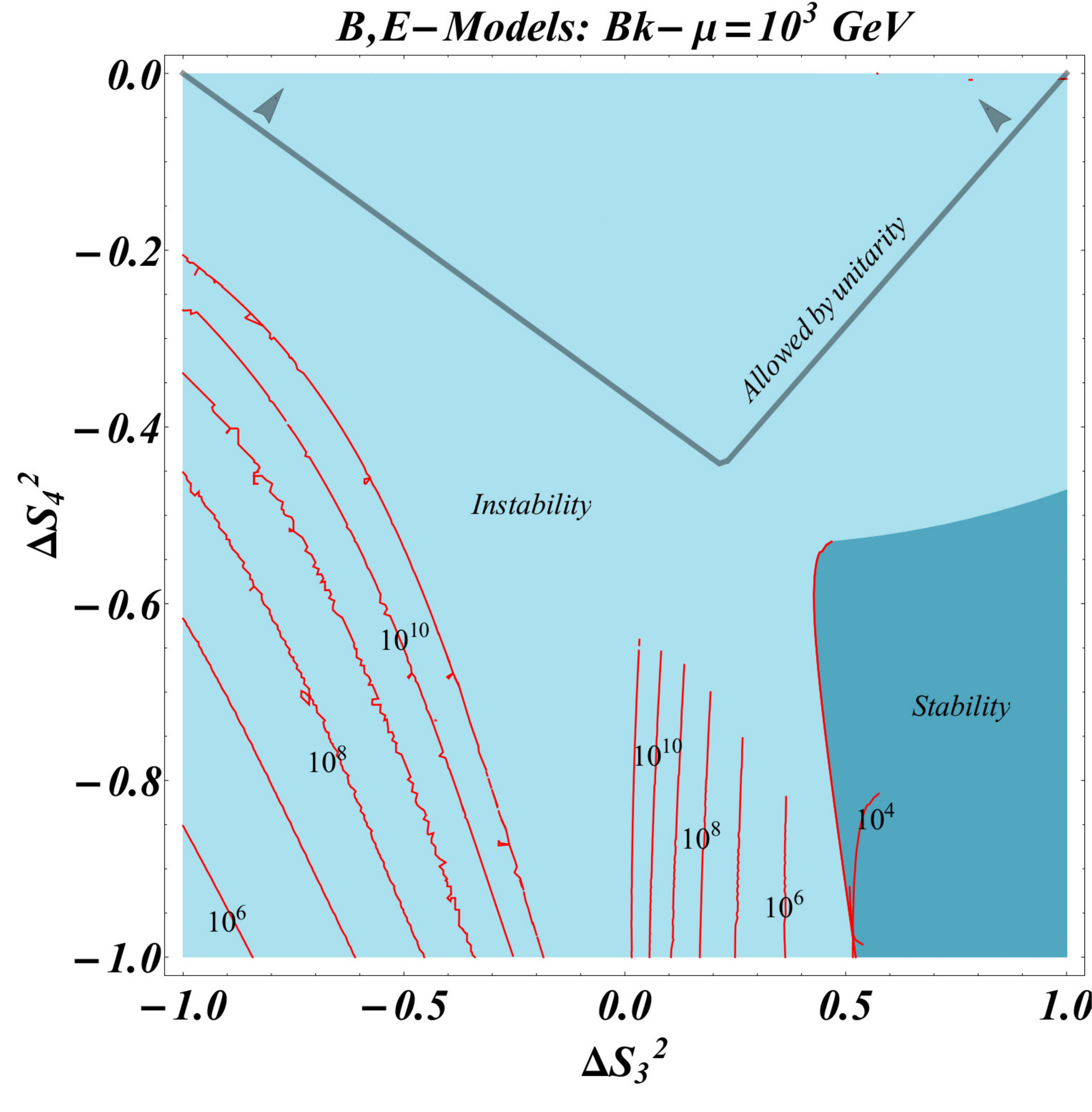} %
\includegraphics[scale=0.18]{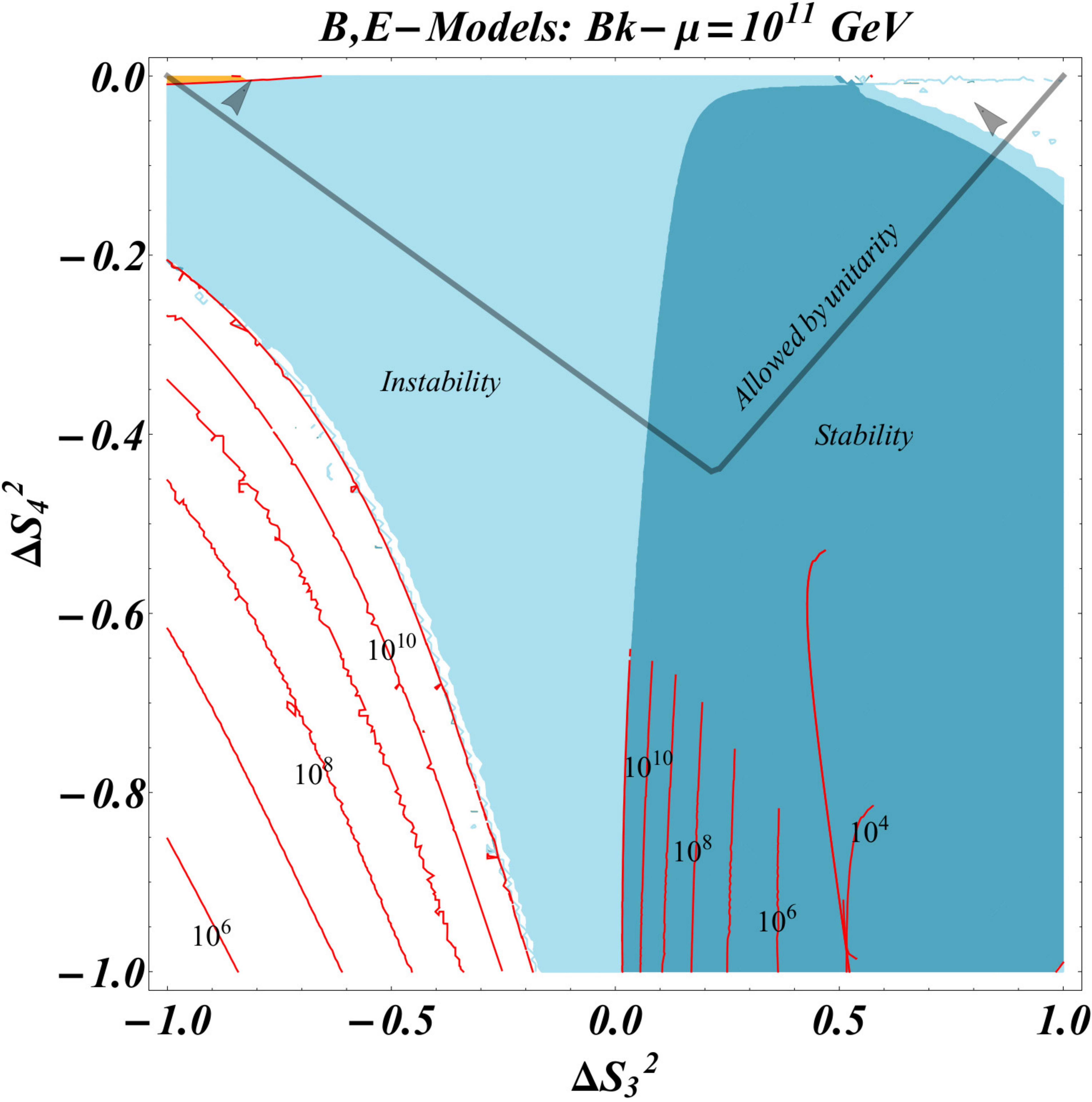}
\vspace{0.4cm}
\caption{(\textbf{Up}) Phase diagrams with the evolution of contours from $\protect\mu %
=10^{3}$ GeV (\textbf{Background-Left}) up to $\protect\mu =10^{11}$ GeV
(\textbf{Background-Right}) in the $\Delta S_{1}^{2}$-$\Delta S_{2}^{2}$ plane for B-models. Here $%
-0.25\leq\protect\lambda_{3,4}(m_{Z})\leq0$ and starting with $\protect%
\lambda _{3,4}(m_{Z})=-\protect\lambda _{2}(m_{Z})/2$. Red lines are the
remaining contours between $\protect\mu =10^{3}$ and $10^{11}$ GeV. (\textbf{Down}) Phase diagrams with the evolution of contours from $\protect\mu %
=10^{3}$ GeV (\textbf{Background-Left}) up to $\protect\mu =10^{11}$ GeV (\textbf{Background-Right}) in the $\Delta
S_{4}^{2}$-$\Delta S_{3}^{2}$ plane for B-models. Here $0\leq\protect\lambda%
_{1,2}(m_{Z})\leq0.25$ and starting with $\protect\lambda _{1,2}(m_{Z})=%
\protect\lambda _{3}(m_{Z})$. Red lines are the remaining contours between $%
\protect\mu =10^{3}$ and $10^{11}$ GeV. Higher values of $\mu$ are incompatible with perturbativity and RGEs convergence. We have taking into account quark top mass in the pole for Yukawa evolution.}
\label{fig:L4L3L1L2B}
\end{figure}

From the possible presence of two non-degenerate minima in the Higgs potential at tree level, it is feasible to extract the following consequences. Firstly, metastable states are strongly dependent on  $\alpha$ and $\beta$ angles, and secondly, the alignment regime, where is reliable, is the boundary between absolute stable zones and metastable zones. 
By the form of Yukawa couplings evolution $A$-$D$ models could be ruled highly constrained from vacuum stability analyses. Non-perturbative zones exclude models with $\tan\beta<1$ since compatible areas with a bounded from below Higgs potential are highly reduced. This argument comes from top-Yukawa couplings given in Fig. \ref{fig:topyuk} evolution and structure of stability contours in $B$-$E$ models shown in Fig. \ref{fig:L4L3L1L2B}, which are inconsistent with perturbative unitarity in the $\lambda_{1}-\lambda_{2}$ plane. 

One relevant scenario emerges when $\tan \beta \rightarrow 1$ and $\alpha \rightarrow \pi/4, -\pi /4$, where the improvement under metastability
forbids every zone allowed by stability at one loop level; being restricted by perturbativity and unitarity analyses as well.  In addition, when $\alpha=\pi/4$ (non-alignment) this zone is also forbidden by divergent solutions for metastability discriminant (\ref{MetastabilityDiscriminant}). Alignment scenario is present when $\alpha\to\pi/4$, where metastability does not allow zones where $\Delta S_{1}^{2}<\Delta S_{2}^{2}$.
Despite perturbative unitarity is consistent with stable zones at $10^{11}$ GeV in $\Delta S_{4}^{2}-\Delta S_{3}^{2}$, at $10^{3}$ GeV unitarity exclude all possible stable zones in the $\Phi_{1}-\Phi_{2}$ plane.

\begin{figure}[tph]
\centering \includegraphics[scale=0.18]{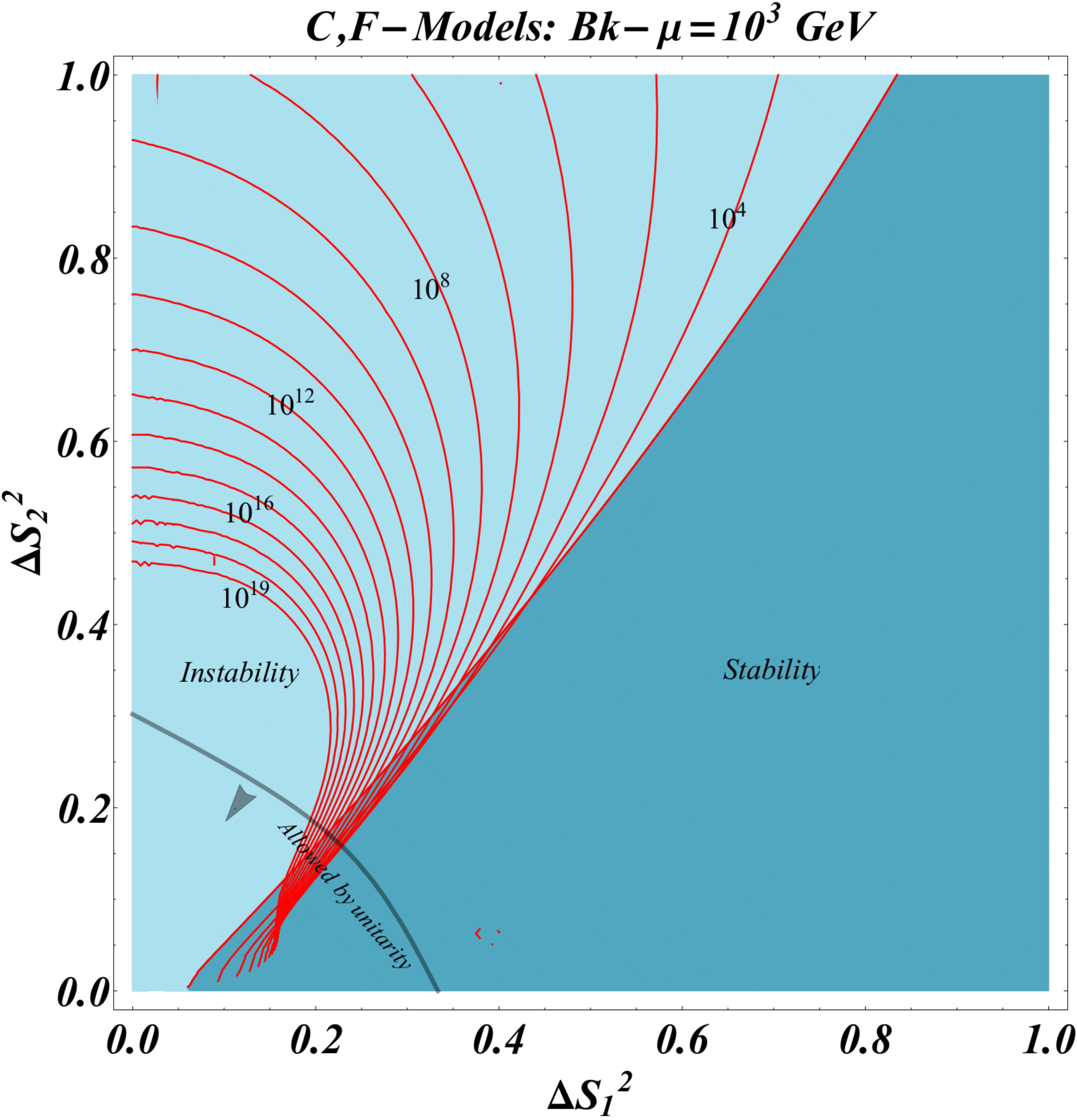} %
\includegraphics[scale=0.18]{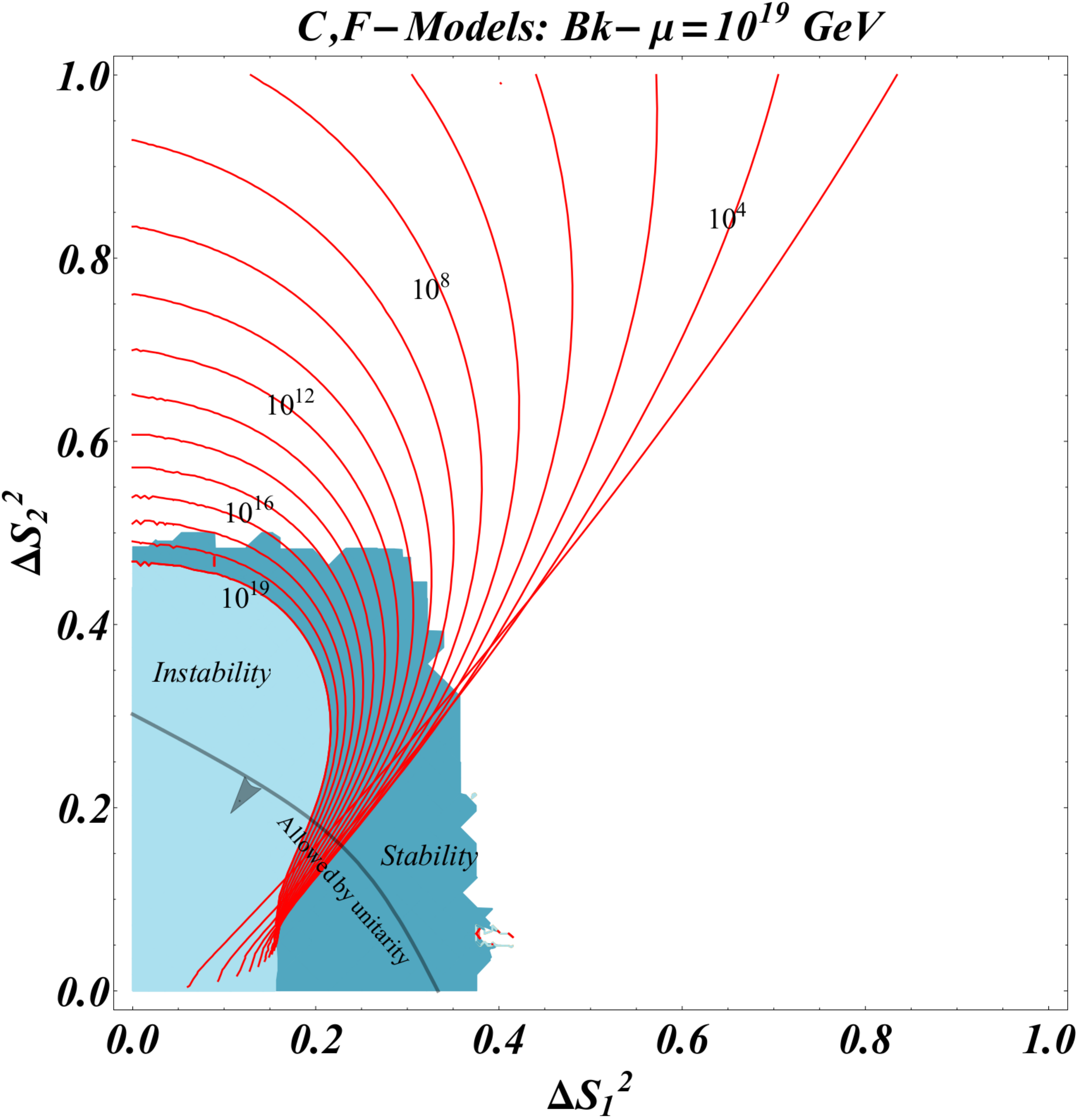}
\centering \includegraphics[scale=0.18]{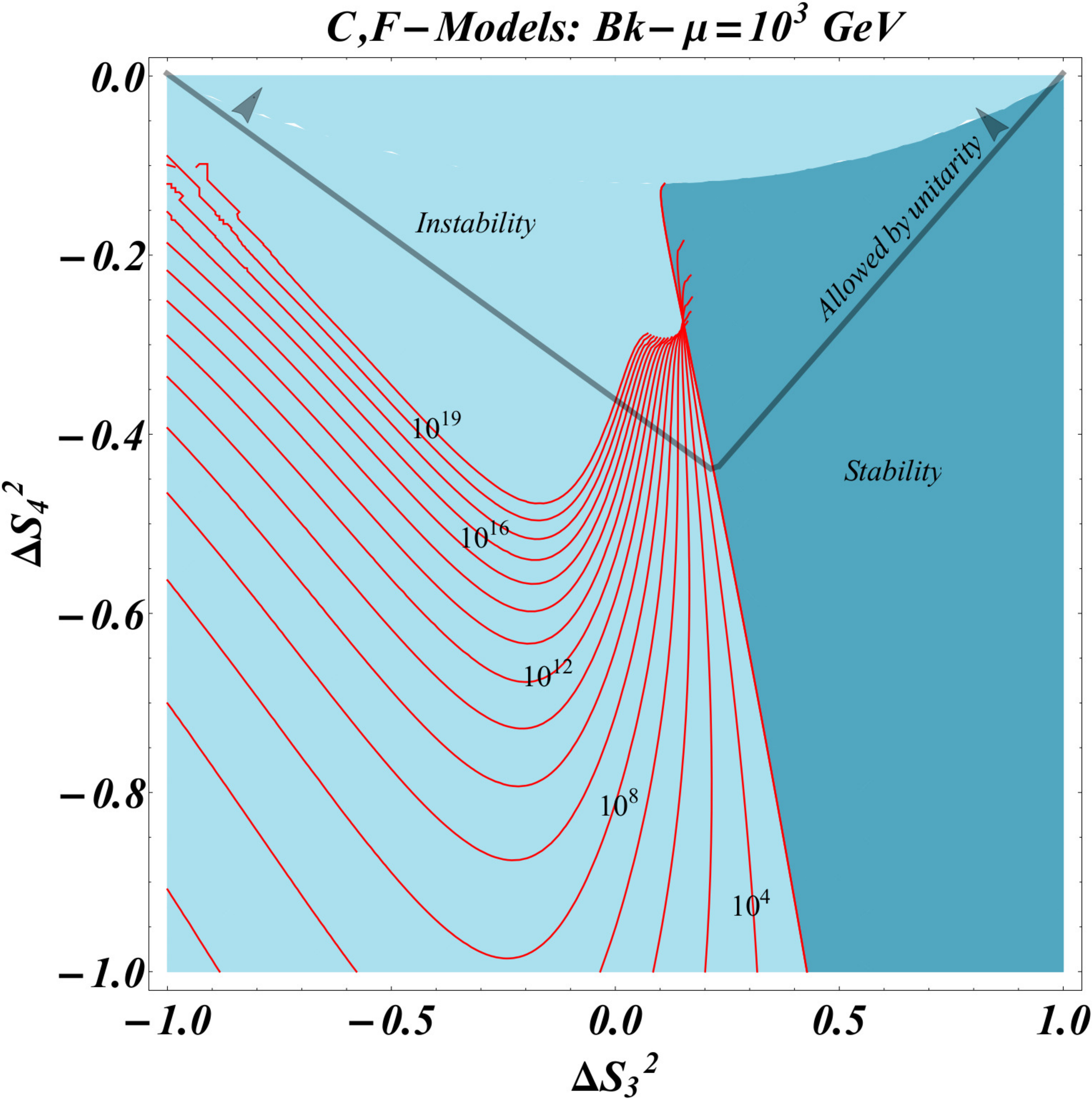} %
\includegraphics[scale=0.18]{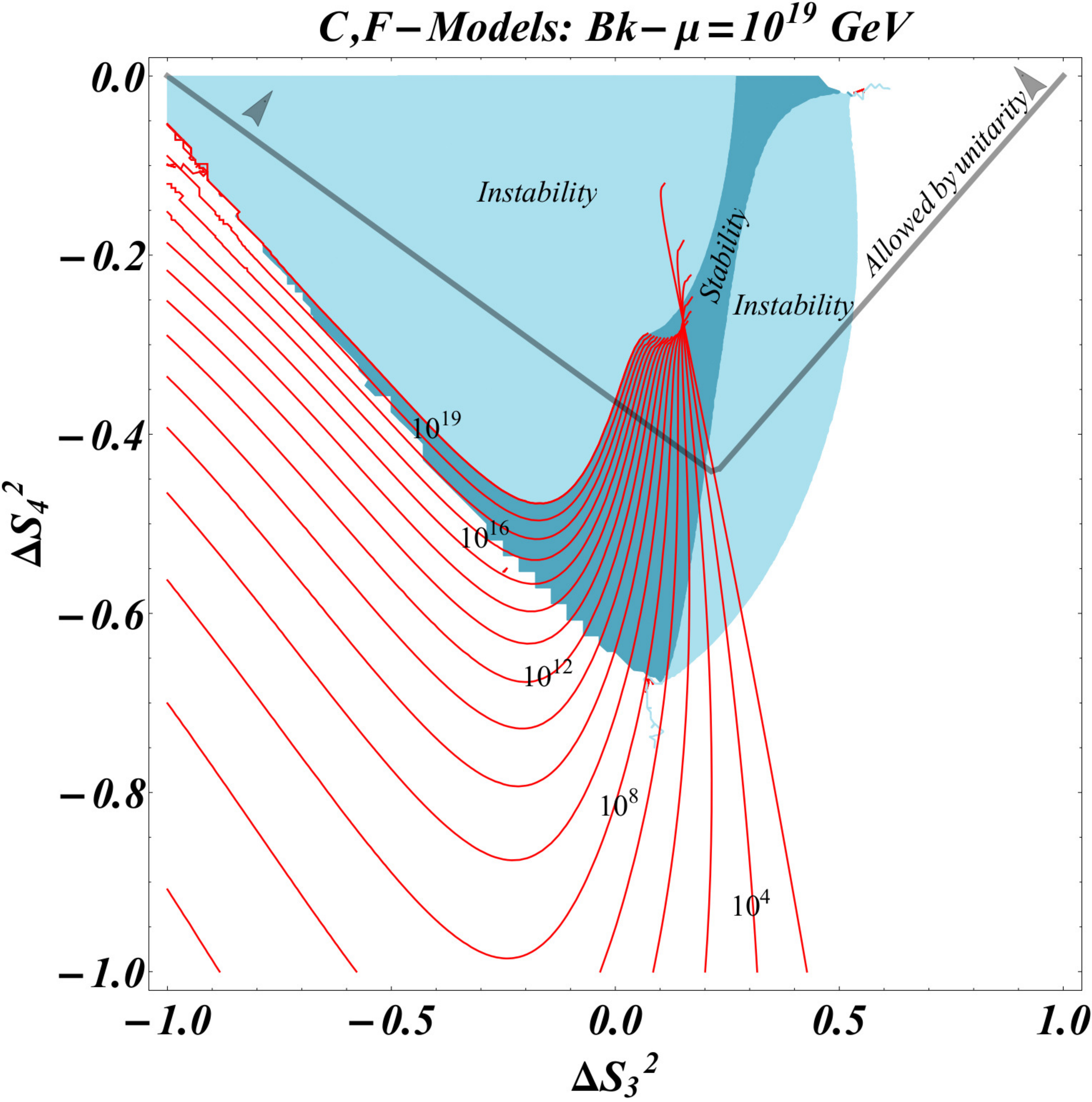}
\vspace{0.4cm}
\caption{(\textbf{Up}) Phase diagrams with the evolution of contours from $\protect\mu %
=10^{3}$ GeV (\textbf{Background-Left}) up to $\protect\mu =10^{19}$ GeV
(\textbf{Background-Right}) in the $\Delta S_{1}^{2}$-$\Delta S_{2}^{2}$ plane for C-models. Here $%
-0.25\leq\protect\lambda_{3,4}(m_{Z})\leq0$ and starting with $\protect%
\lambda _{3,4}(m_{Z})=-\protect\lambda _{2}(m_{Z})/2$. Red lines are the
remaining contours between $\protect\mu =10^{3}$ and $10^{19}$ GeV. (\textbf{Down}) Phase diagrams with the evolution of contours from $\protect\mu %
=10^{3}$ GeV (\textbf{Background-Left}) up to $\protect\mu =10^{19}$ GeV (\textbf{Background-Right}) in the $\Delta
S_{4}^{2}$-$\Delta S_{3}^{2}$ plane for C-models. Here $0\leq\protect\lambda%
_{1,2}(m_{Z})\leq0.25$ and starting with $\protect\lambda _{1,2}(m_{Z})=%
\protect\lambda _{3}(m_{Z})$. Red lines are the remaining contours between $%
\protect\mu =10^{3}$ and $10^{19}$ GeV.}
\label{fig:L4L3L1L2C}
\end{figure}

 At one loop level, instabilities in the 2HDM type I in $C$-$E$ models are present among in intermediate energies and GUT and Planck scales. Hence we focused on $%
10^{3}\leq \mu (\text{ GeV })\leq 10^{19}$ scenario. By crossing stability and metastability analyses, it is possible to find out stronger exclusion regions. This procedure might be an improvement of analysis at NLO for Higgs potential since it avoids the possibility of having two minima in the EW scale.

On the other hand, $C$-$F$ models are broadly compatible with perturbative unitarity and vacuum stability analysis. For instance in the plane $\Delta S_{1}^{2}-\Delta S_{2}^{2}$, stable zones shown in Fig. \ref{fig:L4L3L1L2C} are also consistent with unitarity analysis for lower values of $\Delta S_{1}^{2}$. In the $\Phi_{1}$ direction, values beyond $0.4$ are non-perturbative. This zone is compatible with alignment regime given by the $C_{I}$ model. In the $\Delta S_{4}^{2}-\Delta S_{3}^{2}$ exist zones compatible with stability and perturbative unitarity for considered energy scales. However they are broadly suppressed and have set in $\Delta S_{3}^{2}>0$ and $-0.35<\Delta S_{4}^{2}<-0.1$; which can enter in conflict with $ST$ oblique parameters \cite{Castillo}. By $\Delta S_{3}^{2}$ results, a new hierarchical structure appears in mass eigenstates for $F_{I}$ and $C_{II}$ models: $2m_{H^{\pm}}^{2}>m_{A^{0}}^{2}+m_{H^{0}}^{2}-m_{h^{0}}^{2}$. Stability analyses in the $\lambda_{3}$ coupling for $F_{II}$ and $C_{II}$ models are also consistent with hypothesis avoiding charge breaking minima i.e. $m_{A^{0}}<m_{H^{\pm}}$.

One reference point to study is the limit of a quasi-inert model, where one vacuum expectation value emulates the unique VEV of SM vacuum, and the remaining one is equal to zero. This case recovers when $\tan\beta >>1$ and $\alpha \rightarrow 0$; identifying highly compatibility from metastability at tree level. It is worthwhile to say that this scenario is perturbatively reliable, and RGEs can be solved if and only if there is a degeneracy between $H^{0}$ and $A^{0}$. This quasi-inert limit can easily be identified with the scalar sector for neutrino-specific 2HDMs considered broadly in \cite{Logan1,Logan2}; which are motivated to introduce naturally neutrino masses if one VEV acquires a small enough value in the scale of eV compatible with cosmological and experimental constraints.

\section{Phenomenological aspects in the alignment regime}
\label{sec:Likelihood}

\begin{figure}[htp]
 \centering    
 \includegraphics[height=7.4cm,width=10cm]{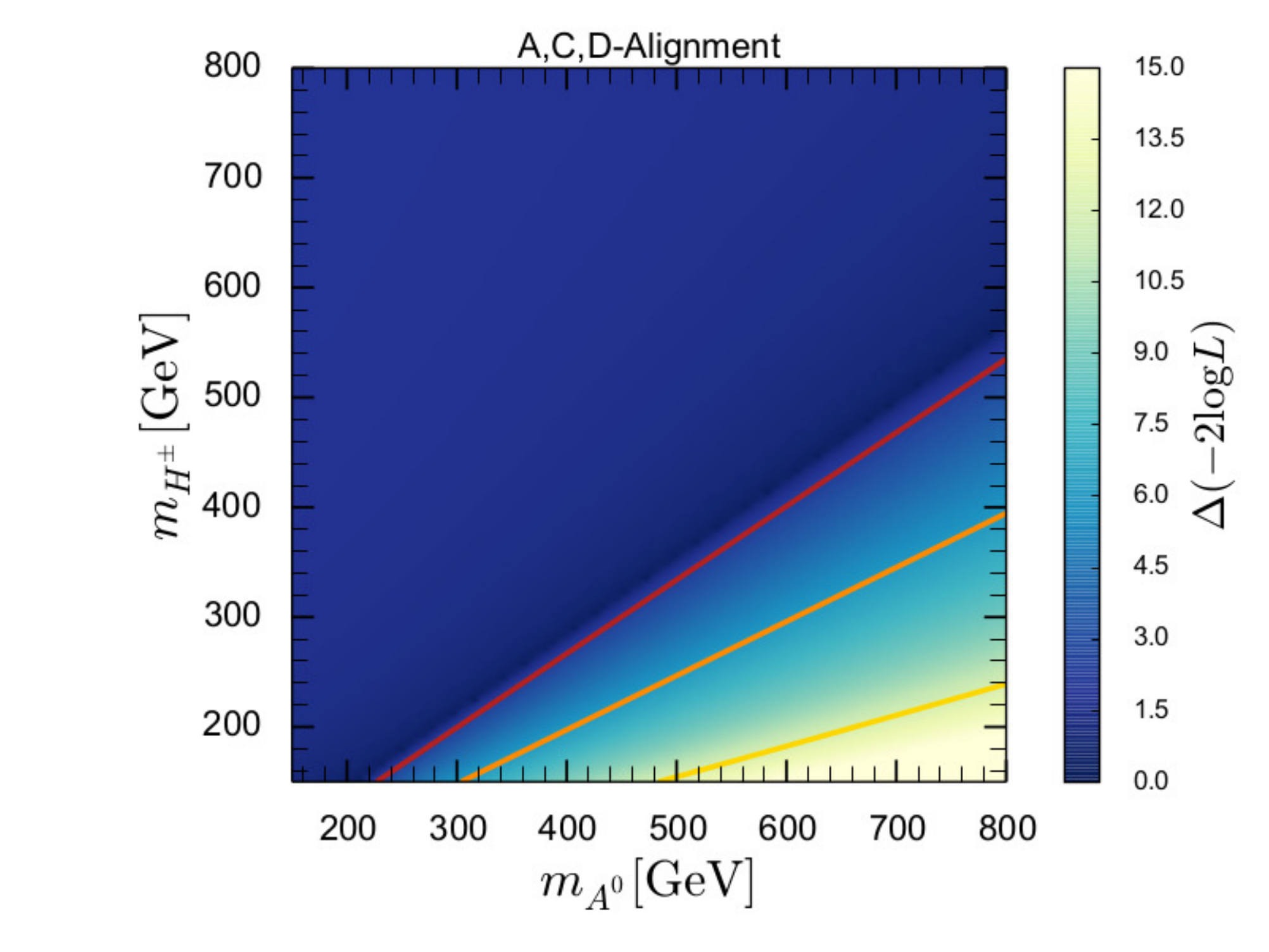}
 \vspace{0.4cm}
 \caption{Likelihood analysis in the alignment regime described by the $A_{III}, C_{I}$ and $D_{II}$ models in the $m_{H^{\pm}}-m_{A^{0}}$ plane in the diphotonic channel.  Here $h^{0}$, the SM-like Higgs has a mass of 125 GeV. The red, orange and yellow
lines correspond to the allowed boundaries of $68\%, 95\%$ and $99.7\%$ CL regions, respectively. We take $k_{S}=1$, by the perturbativity argument given in Tab. \ref{tab:Models}.}
 \label{fig:likelihoodI}
 \end{figure}

\begin{figure}[htp]
 \centering
 \includegraphics[scale=0.45]{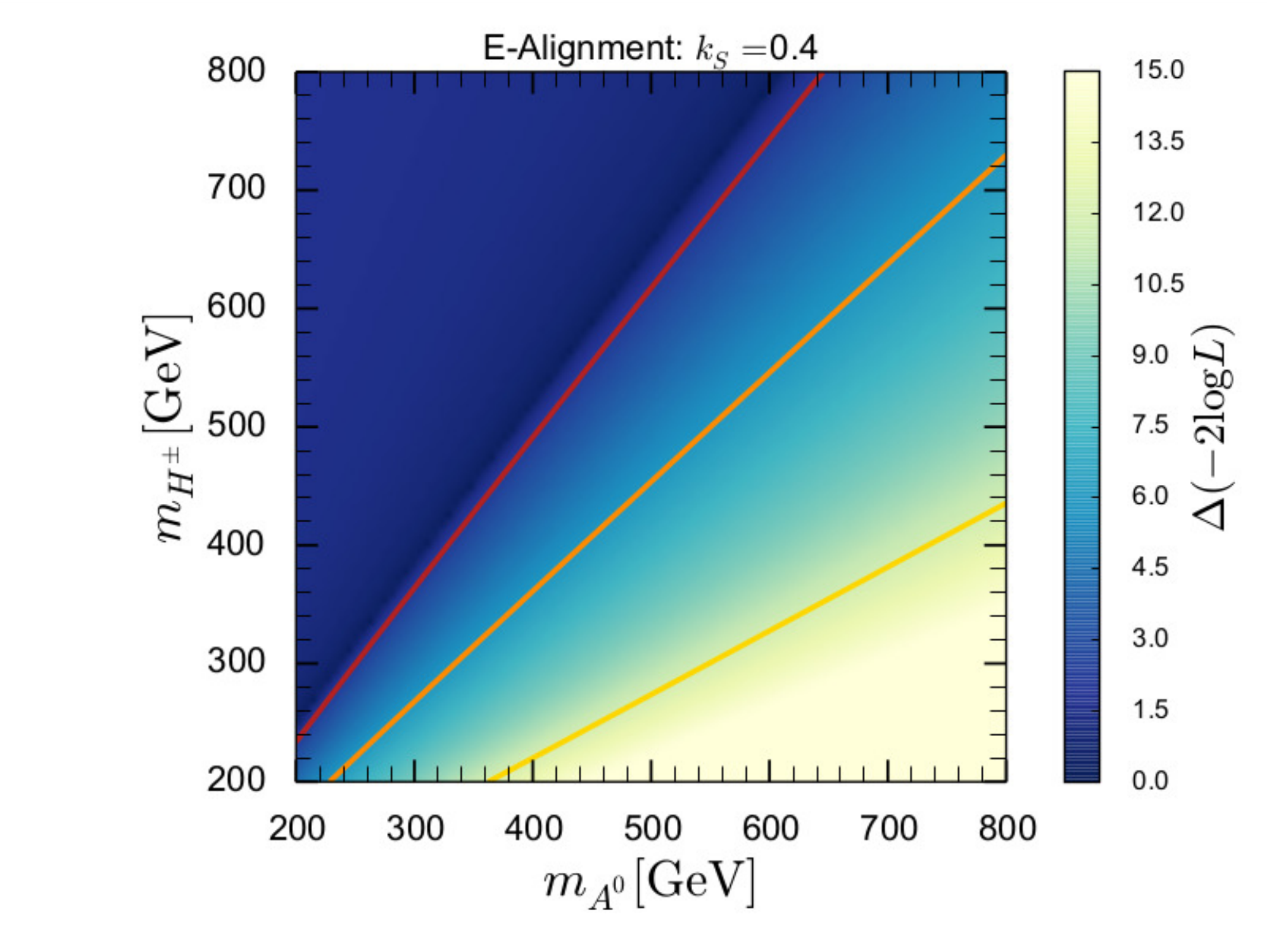}\hspace*{-0.3cm}
 \includegraphics[scale=0.45]{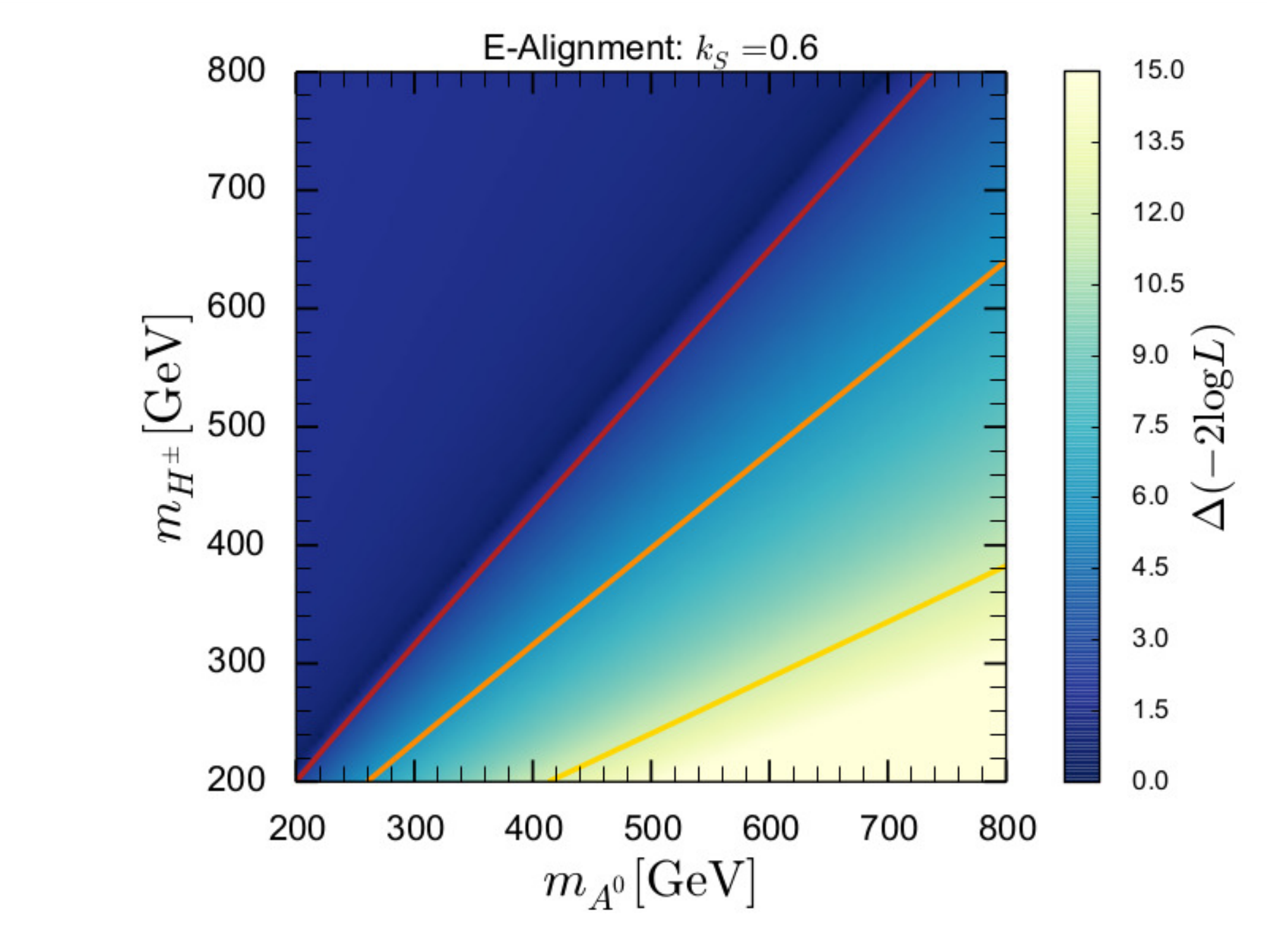}
 \includegraphics[scale=0.45]{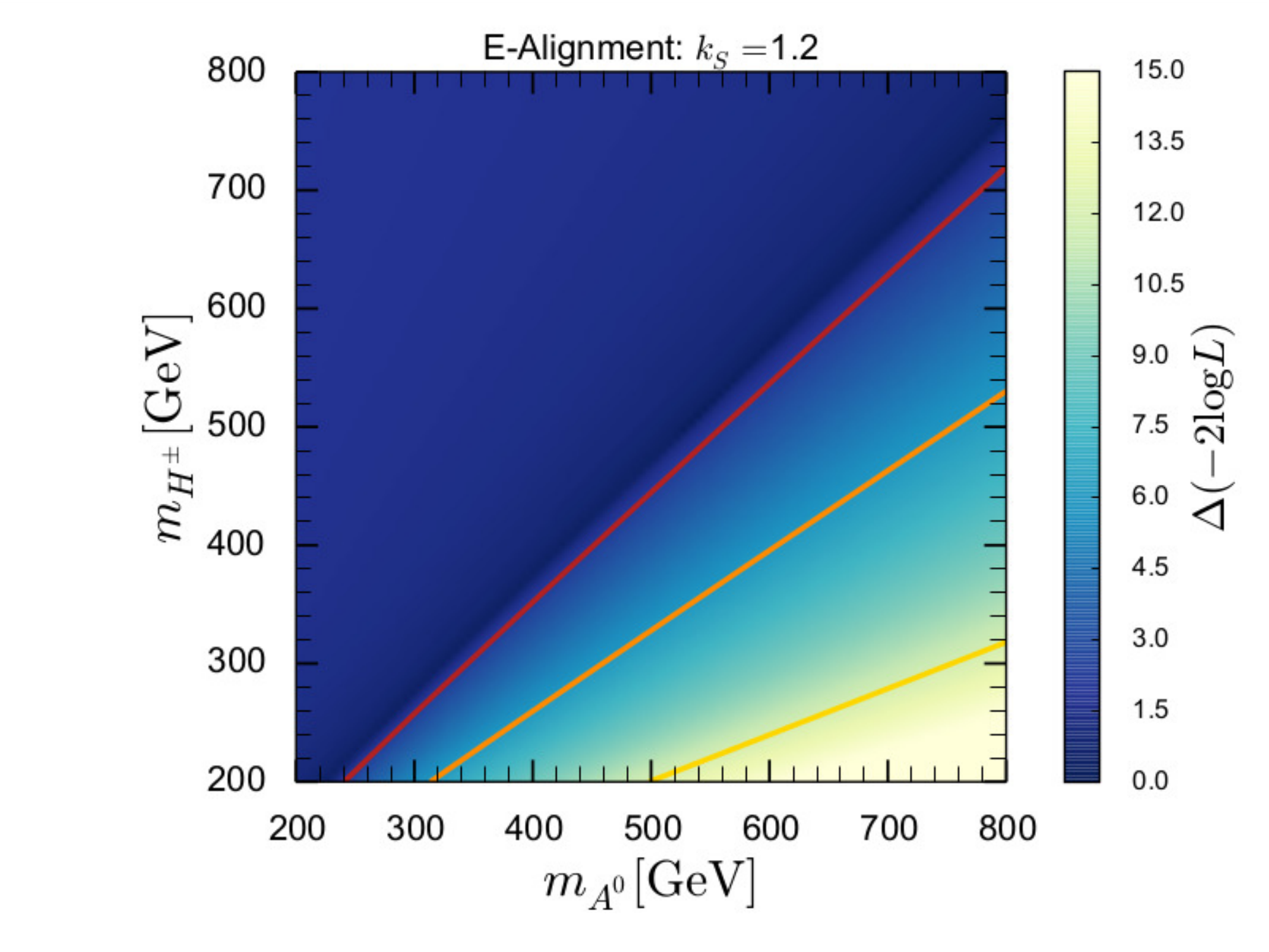}\hspace*{-0.3cm}
 \includegraphics[scale=0.45]{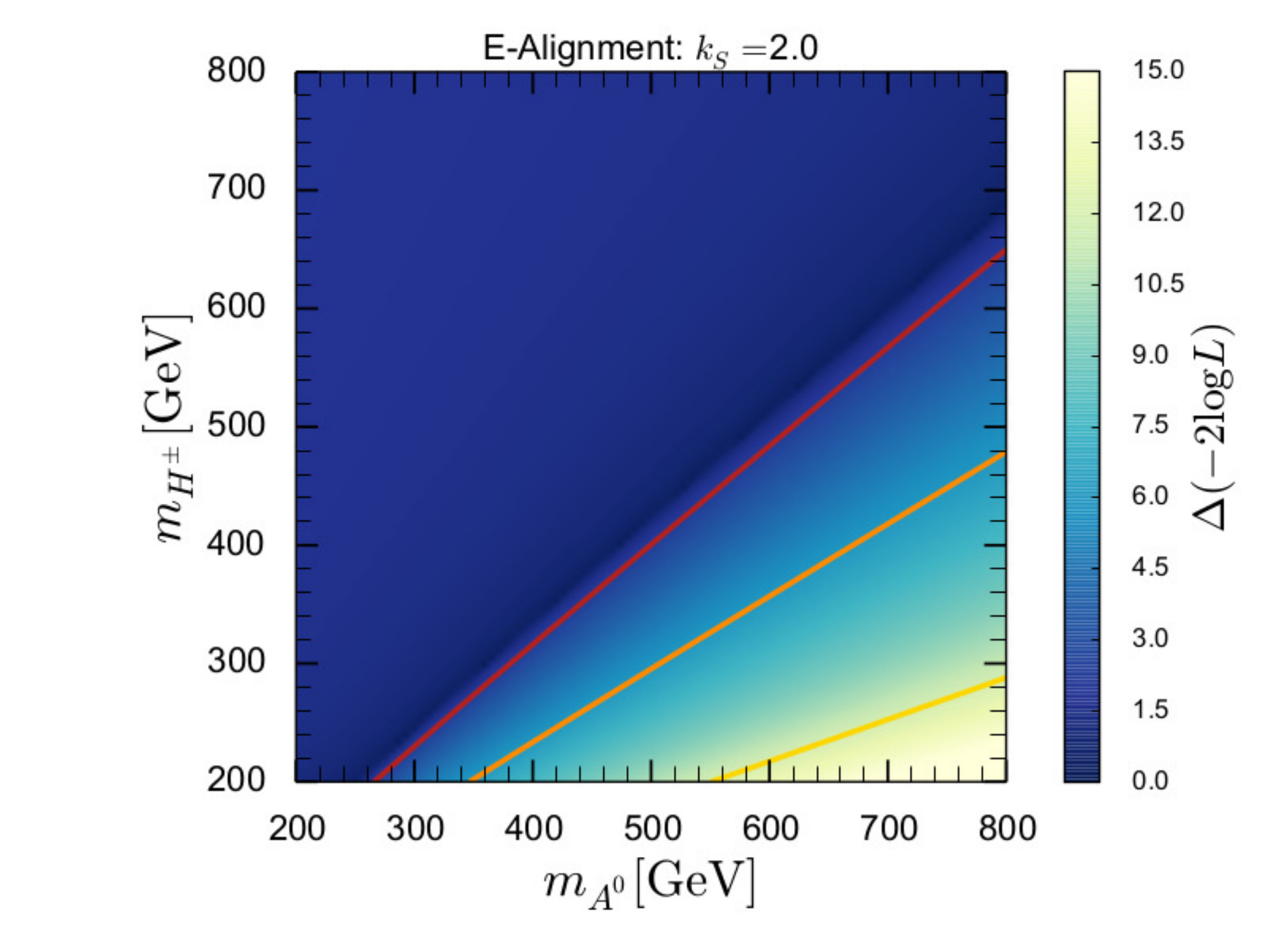}
 \vspace{0.4cm}
  \caption{Likelihood analysis in the alignment scenario described by $E_{I}$ model in the $m_{H^{\pm}}-m_{A^{0}}$ plane in the diphotonic channel and varying $k_{S}=m_{A^{0}}^{2}/m_{H^{0}}^{2}$ ratio.  Here $h^{0}$-SM like Higgs has a mass of 125 GeV. The red, orange and yellow
lines correspond to the allowed boundaries of $68\%, 95\%$ and $99.7\%$ CL regions, respectively. }
 \label{fig:likelihoodII}
 \end{figure}

We analyze phenomenological compatibility in the alignment regimen with the likelihood proof for two photons decay and the oblique parameters realization in the $ST$ plane.
In the former case, contours built from \texttt{Lilith} in Fig. \ref{fig:likelihoodI} operates with the method: From definitions of effective coupling $C_{\gamma}$ among two photons and one Higgs $h^{0}$ at LO of (\ref{ECoupling}), likewise of $-2\log (C_{\gamma})$ relation, we scan the 2-dimensional parameter space in the $m_{H^{\pm}}-m_{A^{0}}$ plane fixing $\beta\to 0,\pi/4,\pi/2$ in contours. We study particularly an alignment scenario where $\cos(\beta-\alpha)\approx0$ ($A_{III}$, $C_{I}$, $D_{II}$ and $E_{I}$ models). In addition for $E_{I}$ model, in each contour of Fig. \ref{fig:likelihoodII}  we are varying $k_{S}=m_{A^{0}}^{2}/m_{H^{0}}^{2}$ ratio. The 2-dimensional $68\%, 95\%, 99.7\%$ CL regions in the plane ($m_{H^{\pm}}-m_{A^{0}}$) are obtained with $\Delta(-2 log L) < 2.3, 5.99, 11.83$, respectively. 

On the other hand, the systematic with oblique parameters in the $S, T$ plane works in the following way: Taking the experimental constraints $S=(0.05\pm0.11)$ and $T=(0.09\pm0.13)$, we examine $99\%$ CL contours for model predictions in splittings and direct masses. The $ST$ formulas for 2HDMs have been extracted from \cite{2HDMC,Report}.

In the alignment regime for $ACD$ models,  perturbativity analysis presented in the section \ref{sec:LO}, demands that $A^{0}$ and $H^{0}$ to have small splittings ($k_{S}\approx 1$). Likelihood proof shows as states where $m_{H^{\pm}}>m_{H^{0}}$ are compatible with a Gaussian distribution of the measurements in the diphotonic channel. At this level, hypothesis to avoid charge vacua where $m_{H^{\pm}}>m_{A^{0}}$ becomes consistent with phenomenological approach of this two photon decay for a SM like Higgs. However, regions with $m_{H^{\pm}}<m_{A^{0}}$ splittings are still compatible with measurements for $h\to \gamma\gamma$ decay. Finally, zones with $m_{A^{0},H^{0}}>500$ GeV and $m_{H^{\pm}}<150$ GeV are excluded at 99.7 $\%$ of CL.  

\begin{figure}[htp]
 \centering
 \includegraphics[height=7.4cm]{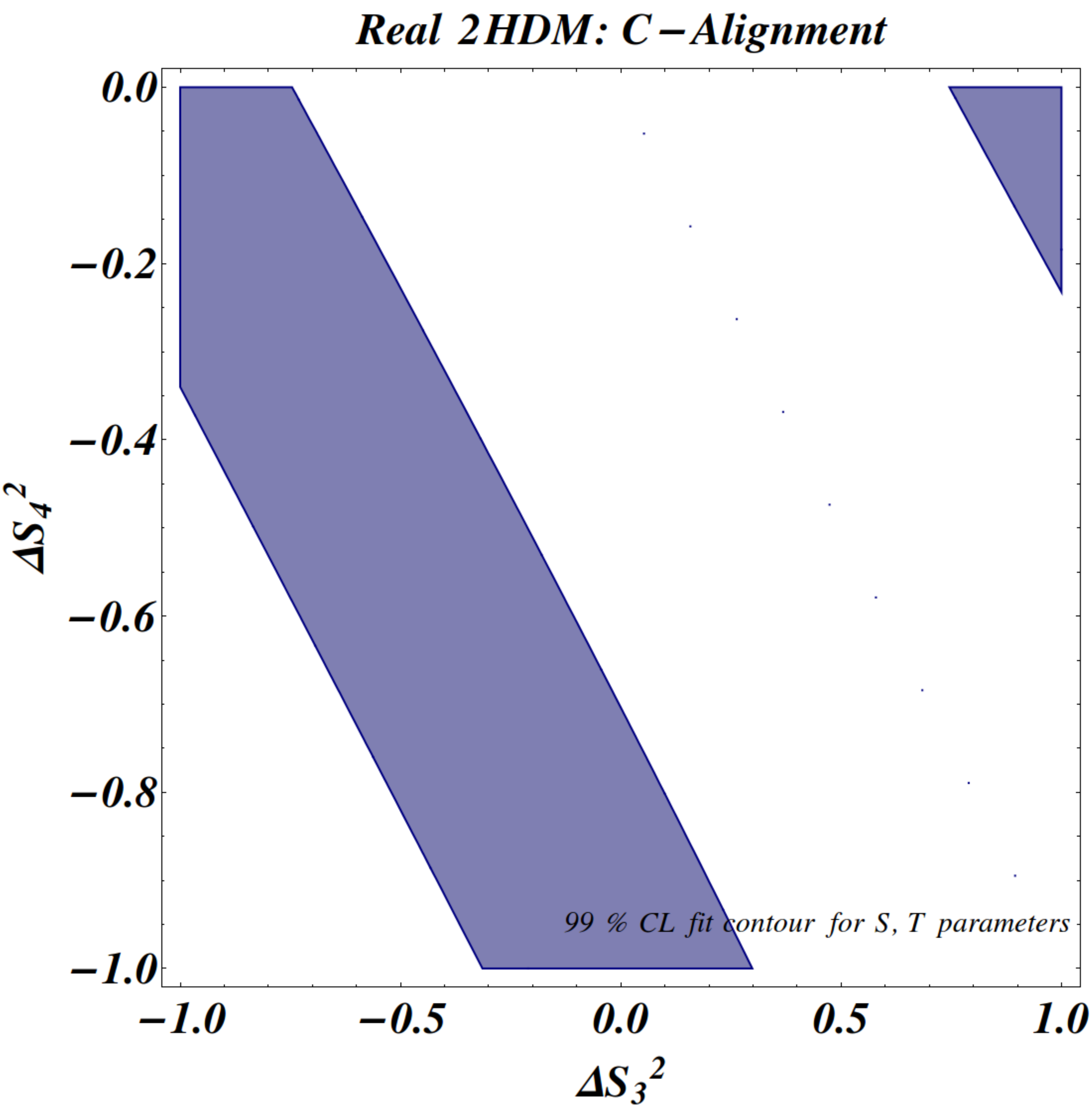}
 \includegraphics[height=7.4cm]{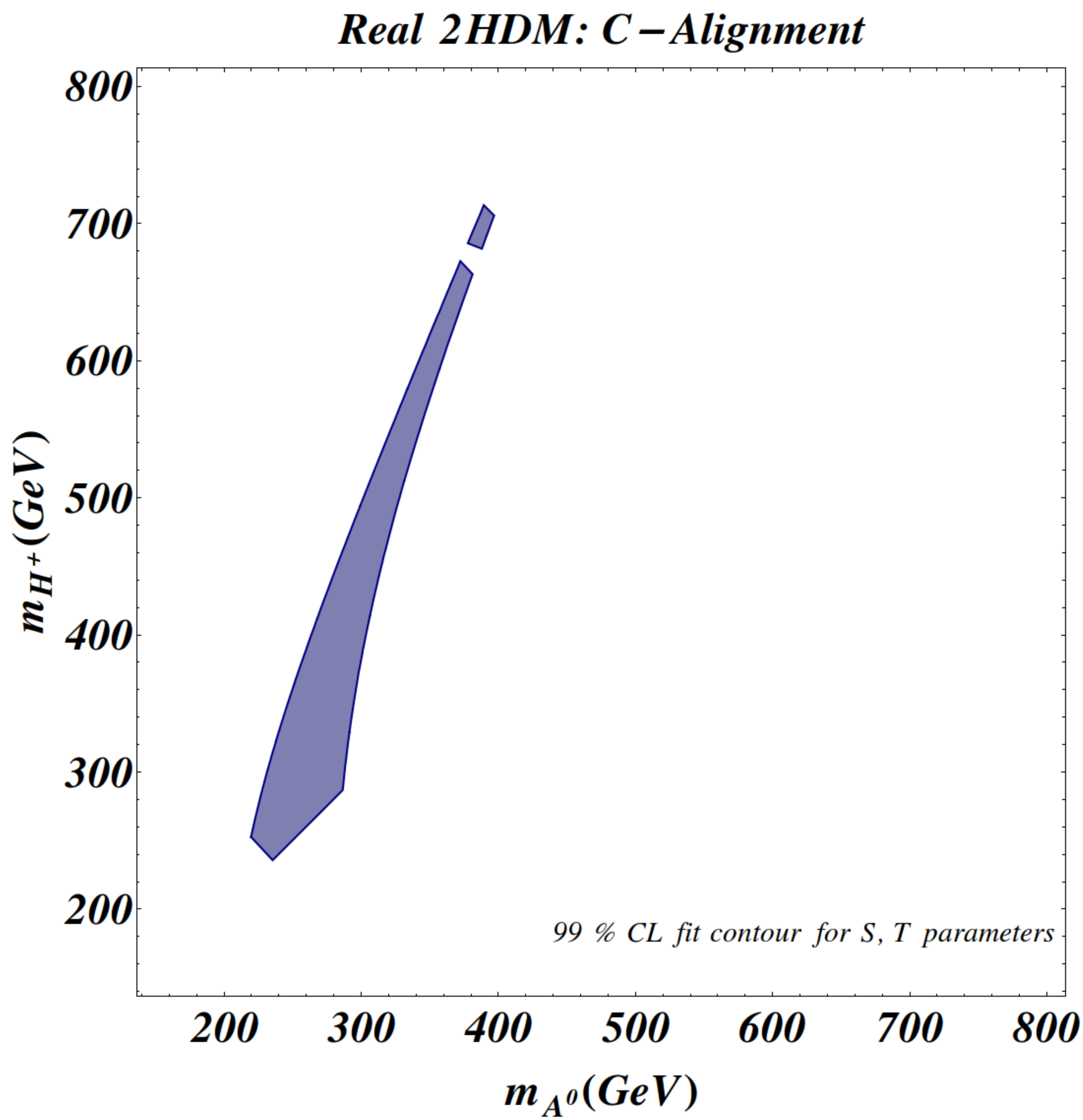}
 \vspace{0.4cm}
 \caption{Oblique parameters in the 2HDM-$U(1)$ with the \emph{S%
}, \emph{T} fit results for softly breaking of $U(1)$ symmetry in the alignment regime given by the $C_{I}$ model; in terms of splittings (\textbf{Left}) and in terms of masses for scalar states (\textbf{Right}). Computations over $ST$ plane have used \texttt{Mathematica} module described in \cite{Deva}.}
 \label{fig:obliqueC}
 \end{figure}

The oblique parameters in the $C_{I}$-model are depicted in Fig \ref{fig:obliqueC} in terms of compatible contours at $99\%$ for splittings and masses (taking $m_{H^{\pm}}>m_{A^{0},H^{0}}$). These analyses shows how masses for charged Higgs are highly constrained for $m_{H^{\pm}}>700$ GeV. In the same way, at this level, pseudoscalar masses are excluded for $m_{A^{0}}>400$ GeV.  

From vacuum stability and perturbativity analyses, the most stable model is the $C_{I}$ one, requiring additionally a degeneracy between $m_{A^{0}}$ and $m_{H^{0}}$ yielding $k_{S}\approx1$; thus alignment regime requires more information to probe $\lambda_{4}<0$ constraint. On the other hand, ST-oblique parameters plane at $99\%$ CL show in Fig. \ref{fig:obliqueC} yields two zones of compatibility for the $C_{I}$ model, where we have taken the hypothesis $m_{H^{\pm}}>m_{A^{0}}$. These zones are consistent with the stable regions in the plane $\Delta S_{4}^{2}-\Delta S_{3}^{2}$ from scales of $\mu=10^{5}$ up to scales $10^{19}$ GeV in values of $\Delta S_{4}^{2}<-0.4$ (Fig \ref{fig:L4L3L1L2C}). However, these zones are outside of unitarity behavior and perturbativity regime for $\lambda_{4}$ and $\lambda_{3}$ couplings at high energy scales. Oblique parameters $ST$ at $99\%$ CL, locate pseudoscalar and $H^{0}$ Higgs with masses in $200<m_{A^{0},H^{0}}(\text{GeV})<400 $, meanwhile the charged Higgs mass satisfies $200<m_{H^{\pm}}(\text{GeV})<700$.

On the other hand, a most constrained parameter space comes from likelihood analysis for the alignment regime given in the $E_{I}$ model as is shown in Fig. \ref{fig:likelihoodII}. For $k_{S}>1$ choice and at least at 68$\%$ C.L. hypothesis where $m_{H^{\pm}}<m_{H^{0}}$ is excluded from the compatibility of the diphotonic decay for a SM-like Higgs boson. Nevertheless, this model has a stringent zone for stability and no-metastability at tree level. Therefore, even though the model is highly compatible with measurements and likelihood hypothesis, stability can be ruled out broadly zones of the respective parameter space.

\begin{figure}[htp]
 \centering
 \includegraphics[height=7.4cm]{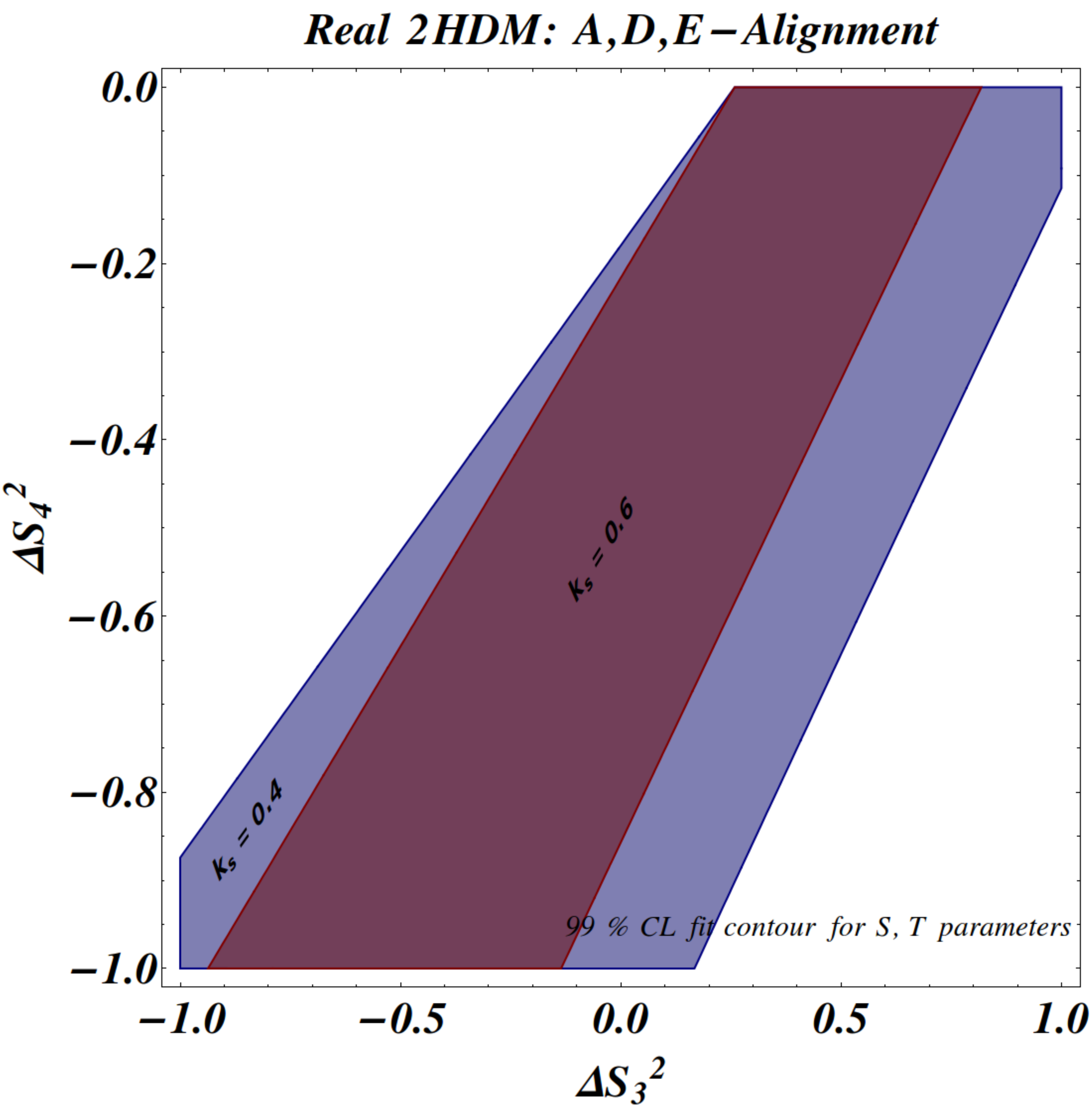}
 \includegraphics[height=7.4cm]{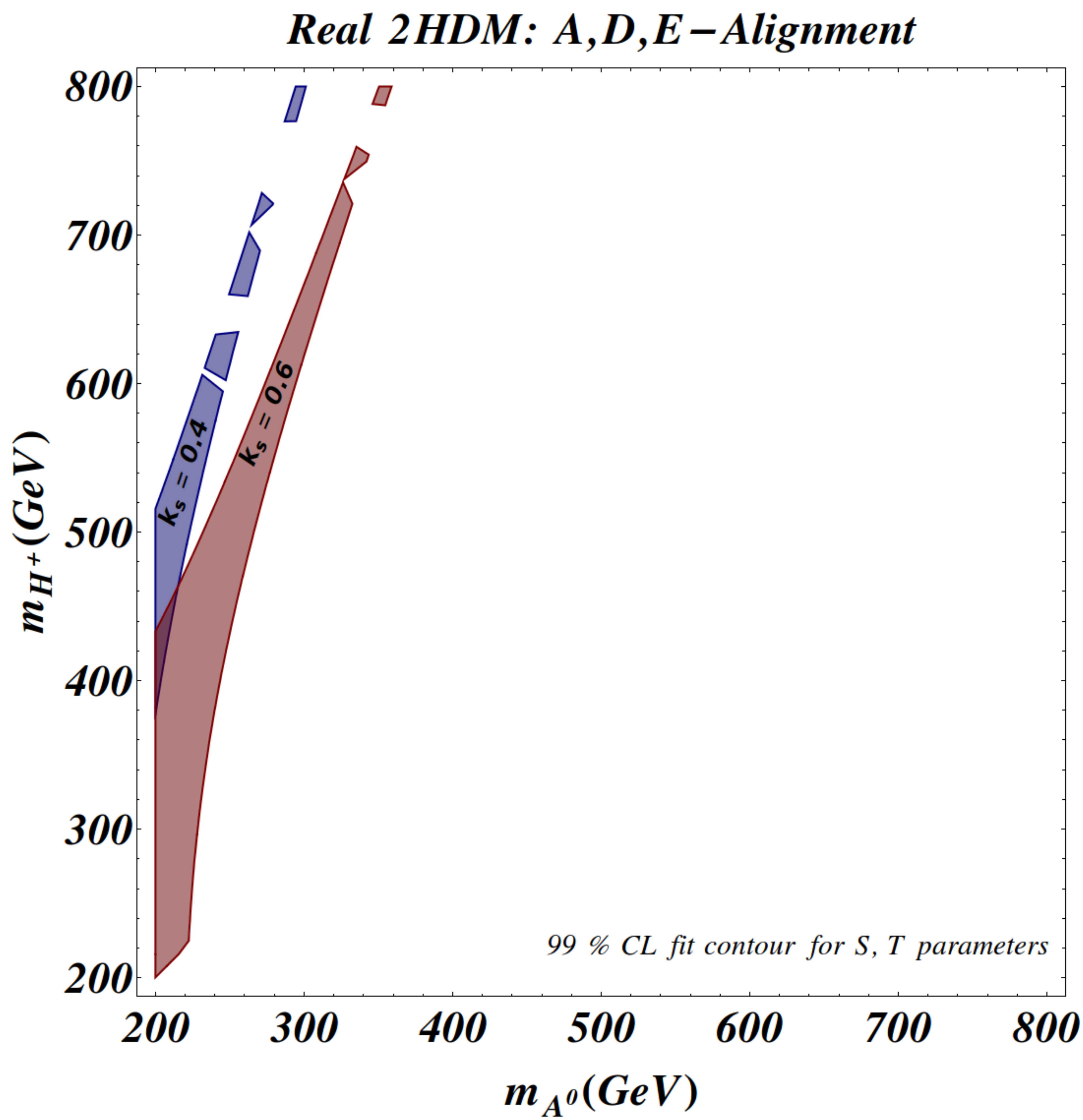}\\
 \includegraphics[height=7.4cm]{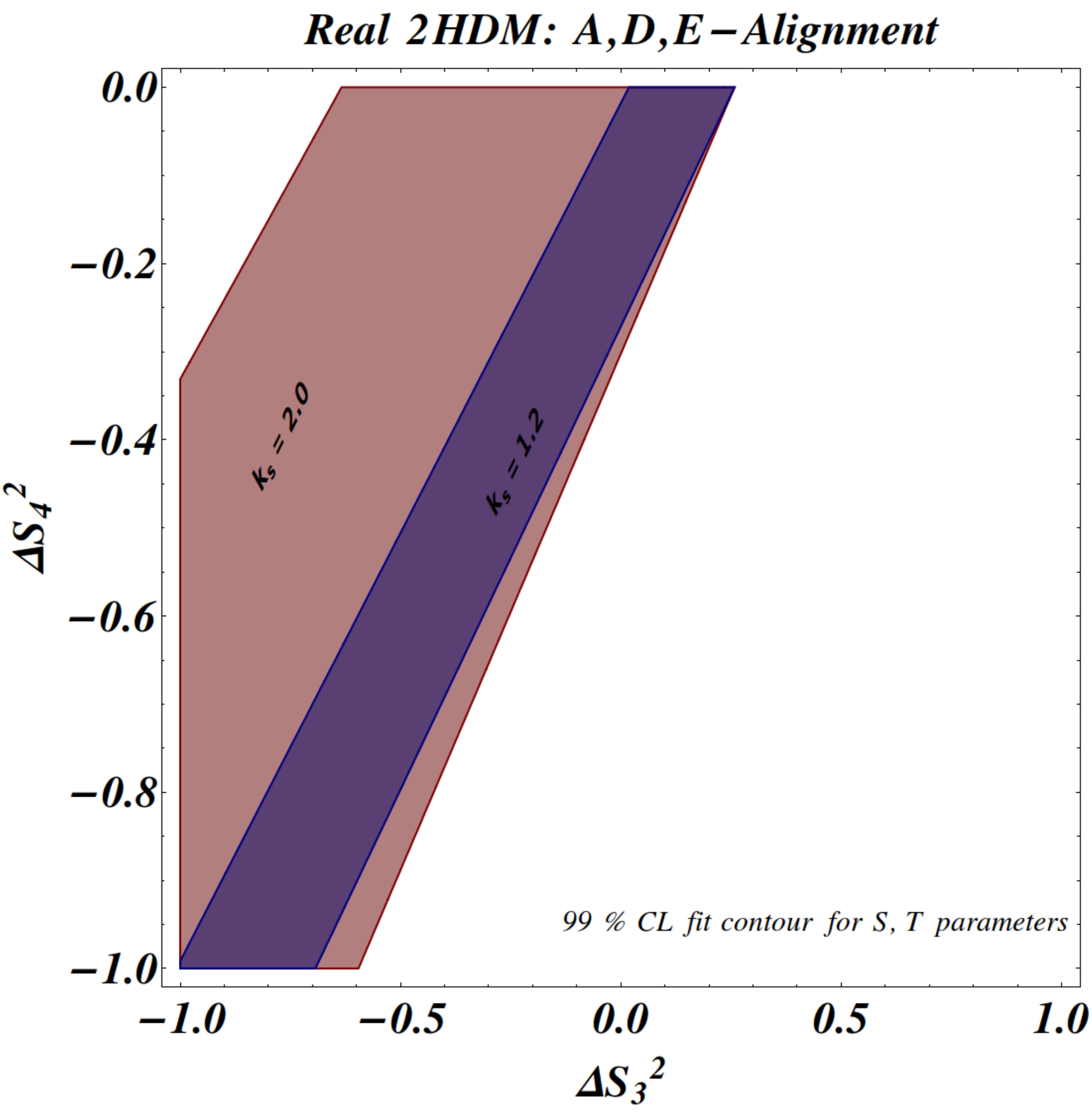}
 \includegraphics[height=7.4cm]{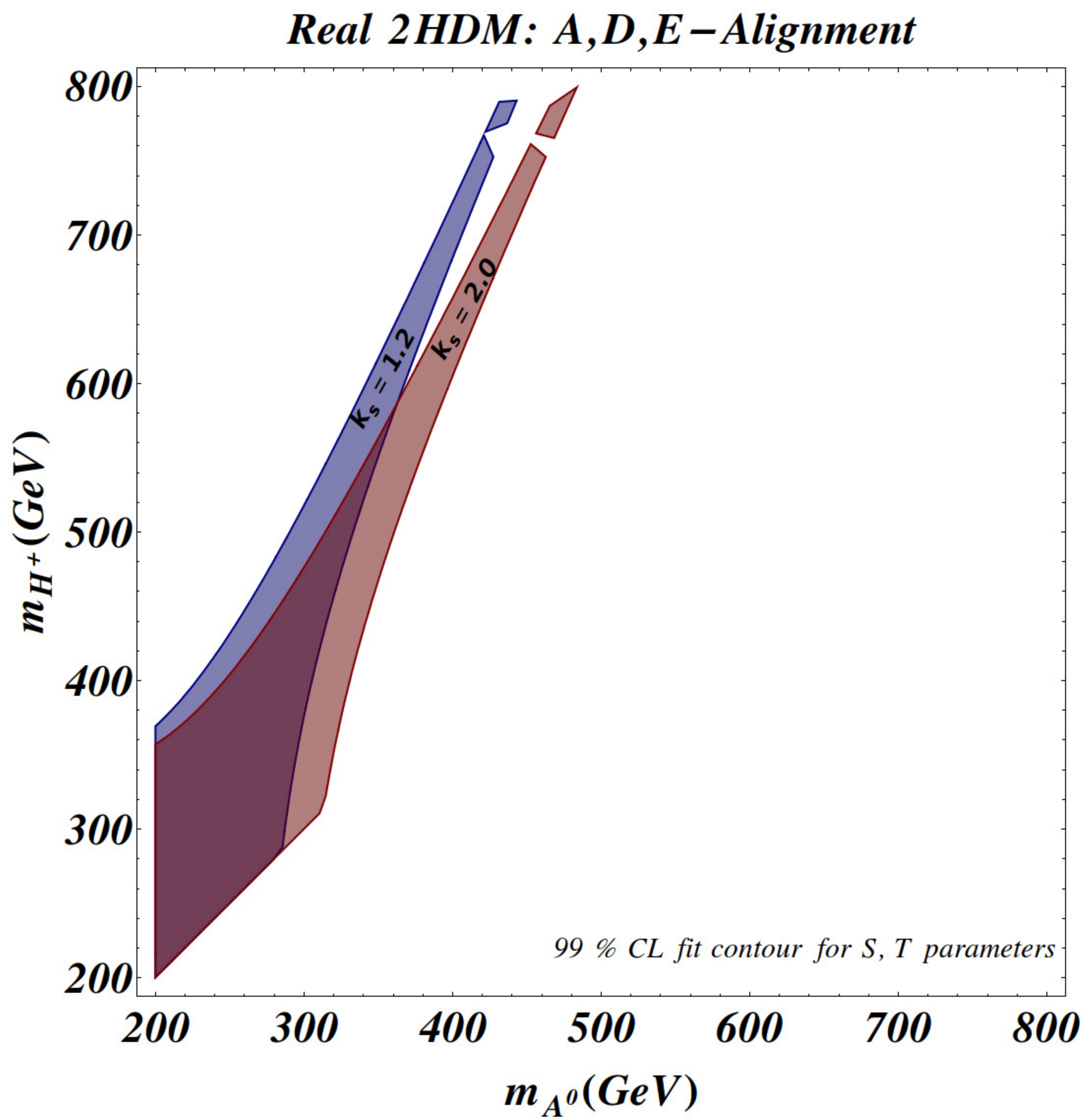}
 \vspace{0.4cm}
 \caption{Oblique parameters in the 2HDM-$U(1)$ with the \emph{S%
}, \emph{T} fit results for softly breaking of $U(1)$ symmetry in the alignment regime given by the $A_{III},D_{II}$ and $E_{I}$ models (varying $k_{s}$);  in terms of splittings (\textbf{Left}) and in terms of masses for scalar states (\textbf{Right}). Computations over $ST$ plane have used \texttt{Mathematica} module described in \cite{Deva}.}
 \label{fig:ADEoblique}
 \end{figure}

 In the aligned $A_{III}, D_{II}$ and $E_{I}$ models, oblique parameters at $99\%$ C.L. exclude masses for pseudoscalar Higgs of $m_{A^{0}}>400$ GeV when $k_{s}=0.4,0.6$ (Fig. \ref{fig:ADEoblique}).  In that regimen, charged Higgs mass could get up values up to of $800$ GeV. When $k_{s}=1.2,2.0$, pseudoscalar Higgs boson has a maximum mass close to $480$ GeV, with a mass for charged Higgs close to $800$ GeV.

\section{Remarks and conclusions}
\label{sec:Remarks}

The Standard Model having a light Higgs boson in a mass around of 125 GeV provides an accurate description of a significant quantity of experimental data associated with the Spontaneous Symmetry Breaking mechanism. The consistency of the precision electroweak data gives predictions of SM suggest that, if new physics is present at the EW scale, it is most probably weakly interacting and consistent with the presence of a light Higgs boson in the spectrum. Extension with these features is the Two Higgs Doublet Model with a softly broken global symmetry and implementing an additional hypothesis of alignment for scalar states.

Based on the vacuum behavior of SM, we comprehensively study metastable and stable states in the model with softly breaking of a $U(1)$ global symmetry in the 2HDM. Initially, softly terms are implemented to forbid massless axion-like particles. Besides, these parameters are related to metastable states at tree level in the field space. If theses states are not long-lived enough, dramatical consequences forbid a well-grounded theory. Thus, we consider this fact as possible exclusions for different configurations of parameter space. Metastability searches arise on Minkowskian formalism of reparameterization group of the Higgs potential to search one global minimum in the theory, which is strongly dependent on $\tan\beta$ and $\alpha$ mixing. There is a high sensitivity of discriminant in zones approaching to $\beta\to0$ and $\pi/2$, and when $\beta\pi/4$ and $\alpha\pi/4$. The alignment scenarios present in the parameters sweeping define the boundary between stable and metastable zones, being important to characterize possible phase transitions due to formation the multiple local minima in the Higgs potential.

Once studied possible exclusion regions for metastability behavior, we describe vacuum analyses at one loop level for the models with $\tan\beta=1$ and $\tan\beta>>1$, for different crucial values of mixing angle $\alpha$. Regimes for $\tan\beta<1$ are highly nonperturbative and drive out rapidly to instabilities in the Higgs potential; being a consequence of initial conditions for Yukawa dynamics for type I 2HDM. Besides of this model with $v_{1}=v_{2}$ present most stringent regions for parameter space.

Likewise, the hypothesis to avoid electromagnetic-charged vacua are proved in a phenomenological point of view, employing likelihood proof of charged Higgs boson influence in the triangle loop corrections of the diphotonic decay of SM Higgs. In this direction, run 1 data of LHC strongly favored scenarios where $m_{H^{\pm}}>m_{A^{0}}$, which is also a necessary exigency from couplings in the Higgs potential to get one neutral minimum consistent with the EW symmetry breaking. Notwithstanding stringent behavior from vacuum analyses, the alignment scenario with $v_{2}=v_{1}$ is phenomenological compatible by likelihood proofs with the normal vacuum hypothesis for charged Higgs boson with the highest mass value in the scalar spectrum. 
\appendix

\section{Renormalization Group Equations for \texorpdfstring{$U(1)$}{lg}-2HDM}
\label{ap:RGEsection}
The behavior of the parameters and relations among arises through the Renormalization Group Equations (RGE). At
higher levels in perturbation theory, exist a dependence on the energy scale $\mu $ as a result of quartic couplings renormalization. Regularization of ultraviolet divergences in loop integrals for radiative corrections introduces the $\mu$-dependence. With the aim to evaluate the presence of instabilities in all field space, energy scale dependent couplings accomplish with the same constraints at tree level (for different regions in the field area) described in section \ref{sec:TL}, ensuring a Higgs potential bounded from below.

Besides, the RGE are a powerful tool to determine the triviality
behavior energy limits of the parameters and perturbative validity of the
theory. To numerically evaluate the energy dependence of the quartic
couplings one loop level, it is also necessary considering RGEs of all remaining couplings, i.e., the gauge group couplings $g^{\prime },g,$ $g_{s}$ of the symmetry groups $U(1)$, $SU(2)$, $SU(3)$, and the Yukawa couplings of the top and the down quark sectors $\xi_{t},\xi _{b}$ and $\xi_{\tau}$ (coupled to $\Phi _{2}$ doublet) and $\eta _{t},\eta _{b}$ and $\eta _{\tau }$ (couplings to $%
\Phi _{2}$ doublet) respectively are computed in Refs \cite{Das,Piard}. The one loop RGEs for a general gauge theory are presented in \cite%
{ChengLi, Machacek} and for NHDM theory with gauge group $SU(2)_{L}\times U(1)_{Y}$ are demonstrated in \cite{RGE21}.

We summarize the RGEs for 2HDM with SM gauge group through the following
settlement of equations for $U\left( 1\right) $ global invariant Higgs
potential. The RGEs for gauge sector at one loop level are %
\begin{eqnarray}
\frac{dg}{dt} &=&\frac{1}{16\pi ^{2}}\left( \frac{4}{3}n_{f}+\frac{1}{6}%
n_{H}-\frac{22}{3}\right) g^{3}=-3g^{3}, \\
\frac{dg^{\prime }}{dt} &=&\frac{1}{16\pi ^{2}}\left( \frac{20}{9}n_{F}+%
\frac{1}{6}n_{H}\right) g^{^{\prime }3}=7g^{^{\prime }3}, \\
\frac{dg_{s}}{dt} &=&\frac{1}{16\pi ^{2}}\left( \frac{4}{3}n_{f}-11\right)
g_{s}^{3}=-7g_{s}^{3}.
\end{eqnarray}%
In the 2HDM, $n_{H}=2$ and $n_{f}=3$.
In all equations $t=\log \mu .$ For type I-2HDM, the RGEs
for Yukawa couplings in the third family ($\tau ,b,t$) are

\begin{align}
16\pi ^{2}\frac{d\xi_{\tau}}{dt}& =-\left( \frac{9}{4}g^{2}+\frac{15%
}{4}g^{^{\prime }2}\right) \xi_{\tau}+T_{11}\xi_{\tau}+\frac{%
3}{2}\xi_{\tau}^{3}, \\
16\pi ^{2}\frac{d\xi_{b}}{dt}& =-\left( 8g_{s}^{2}+\frac{9}{4}g^{2}+\frac{%
5}{12}g^{^{\prime }2}\right) \xi _{b}+T_{11}\xi _{b}+\frac{3}{2}\xi_{b}^{3}-\frac{3}{2}\xi_{t}^{2}\xi _{b}, \\
16\pi ^{2}\frac{d\xi_{t}}{dt}& =-\left( 8g_{s}^{2}+\frac{9}{4}g^{2}+\frac{%
17}{12}g^{^{\prime }2}\right) \xi_{t}+T_{11}\xi_{t}+\frac{3}{2}\xi_{t}^{3}-\frac{3}{2}\xi_{b}^{2}\xi_{t}.
\end{align}
By virtue of Yukawa matrices are diagonal, $T_{11}=3\left( \xi_{t}^{2}+\xi_{b}^{2}\right) +\xi_{\tau }^{2}.$ Here $\Phi _{2}$ decouple from all fermions; top quark, bottom quark and $\tau $ lepton only couple to $\Phi
_{2}$ doublet.  Initial conditions for type I Yukawa couplings are the following (without QCD-infrared corrections)
\begin{subequations}
\begin{align}
 \xi_{t}(m_{Z})&=\frac{1}{\sin\beta}\frac{\sqrt{2}m_{t}}{v} \label{etatop}\\
 \xi_{b}(m_{Z})&=\frac{1}{\sin\beta}\frac{\sqrt{2}m_{b}}{v}\\
 \xi_{\tau}(m_{Z})&=\frac{1}{\sin\beta}\frac{\sqrt{2}m_{\tau}}{v}
\end{align}
\end{subequations}

The RGE for scalar couplings ($U\left( 1\right) $ global
invariant Higgs potential) at one loop in a 2HDM type I are described by

\begin{align}
16\pi ^{2}\frac{d\lambda _{1}}{dt}& =12\lambda _{1}^{2}+4\lambda
_{3}^{2}+4\lambda _{3}\lambda _{4}+2\lambda _{4}^{2}+\frac{%
9g^{4}+6g^{2}g^{^{\prime }2}+3g^{^{\prime }4}}{4}-\left( 9g^{2}+3g^{^{\prime
}2}\right) \lambda _{1}, \\
16\pi ^{2}\frac{d\lambda _{2}}{dt}& =12\lambda _{2}^{2}+4\lambda
_{3}^{2}+4\lambda _{3}\lambda _{4}+2\lambda _{4}^{2}+\frac{%
9g^{4}+6g^{2}g^{^{\prime }2}+3g^{^{\prime }4}}{4}-\left( 9g^{2}+3g^{^{\prime
}2}\right) \lambda _{2}  \notag \\
& -6\left( \eta _{bb}^{4}+\eta _{tt}^{4}\right) -2\eta _{\tau \tau
}^{4}+4\left( 3\left( \eta _{tt}^{2}+\eta _{bb}^{2}\right) +\eta _{\tau \tau
}^{2}\right) \lambda _{2}, \\
16\pi ^{2}\frac{d\lambda _{3}}{dt}& =2\left( \lambda _{1}+\lambda
_{2}\right) \left( 3\lambda _{3}+\lambda _{4}\right) +4\lambda
_{3}^{2}+2\lambda _{4}^{2}+\frac{9g^{4}-6g^{2}g^{^{\prime }2}+3g^{^{\prime
}4}}{4}-\left( 9g^{2}+3g^{^{\prime }2}\right) \lambda _{3}  \notag \\
& +2\left( 3\left( \eta _{tt}^{2}+\eta _{bb}^{2}\right) +\eta _{\tau \tau
}^{2}\right) \lambda _{3}, \\
16\pi ^{2}\frac{d\lambda _{4}}{dt}& =2\lambda _{4}\left( \lambda
_{1}+\lambda _{2}\right) +2\left( 2\lambda _{4}^{2}+4\lambda _{3}\lambda
_{4}\right) +3g^{2}g^{^{\prime }2}-\left( 9g^{2}+3g^{^{\prime }2}\right)
\lambda _{4}  \notag \\
& +2\left( 3\left( \eta _{tt}^{2}+\eta _{bb}^{2}\right) +\eta _{\tau \tau
}^{2}\right) \lambda _{4}.
\end{align}

Since fermions are coupled to one and only one doublet $\Phi_{2}$, we might expect many
contributions to unstable zones from $\lambda _{2}$ ($\Delta S_{2}^{2}$) coupling associated to the quartic coupling of $\left( \Phi _{2}^{\dagger }\Phi _{2}\right) ^{2}$ dimension fourth operator. Besides, for high values of all couplings in
their initial conditions might exist additional sources for nonperturbative
scenarios \cite{Kanemura2}.

\section{Higgs decay in two photons and likelihood proof}

From 2HDM fundamentals, we know that charged Higgs might have a substantial contribution to the $h\rightarrow \gamma \gamma $ decay rate. Since this channel has been an important situation for Higgs-like scalar detection in LHC-experiments, limits on the parameters controlling this new physics contribution can be obtained from the Higgs precision measurements. Hence we consider a 2HDM with a softly broken $U(1)$ symmetry in the Higgs potential and Yukawa Lagrangian and assume that the only deviation from a SM-like Higgs behavior is due to the contribution of charged Higgs to the loop-induced process $H\rightarrow \gamma \gamma $.
More precisely, in this particular model other scalar states are decoupled, and only the charged scalar contribution is present. In this case, the input from status to the $h\rightarrow \gamma \gamma $ decay width is parameterized by charged Higgs mass. The corresponding amplitude at LO reads

\begin{equation}
\mathcal{M}_{h\gamma \gamma }^{H^{\pm }}=\frac{v^{2}g\left( m_{H^{\pm
}}\right) }{2m_{H^{\pm }}^{2}}A_{0}^{h}\left( \tau _{H^{\pm }}\right) ,\text{
}\tau _{H^{\pm }}=\frac{m_{h}^{2}}{4m_{H^{\pm }}^{2}}. \label{Higgsphotonphoton}
\end{equation}

where $g\left( m_{H^{\pm }}\right) $ is the $h,H^{+}H^{-}$ coupling
and $A_{0}^{h}\left( m_{h}^{2}/4m_{H^{\pm }}^{2}\right) $ form factor. For
lower values in $\cos \left( \beta -\alpha \right) \approx 0$ (alignment
scenario)$,$ $g\left( H^{\pm }\right) $ coupling is given by \cite{HaberLilith}
\begin{equation}
g\left(H^{\pm }\right) =%
-\frac{1}{4}\sin ^{2}2\beta \left( \lambda _{1}+\lambda _{2}-2\lambda
_{34}\right) -\lambda _{3}
\end{equation}

Therefore in the alignment limit, $g(m_{H^{\pm}})$ coincides for $A,C,D$ models. In models with alignment, discrepancy comes from $E_{I}$ since $\beta\to\pi/4$. In all cases, the effective Higgs-$\gamma \gamma $ coupling can therefore be expressed by%
\begin{equation}
C_{\gamma }=\frac{|\mathcal{M}_{h\gamma \gamma }^{SM}+\mathcal{M}_{h\gamma
\gamma }^{H^{\pm }}|}{|\mathcal{M}_{h\gamma \gamma }^{SM}|}.
\label{ECoupling}
\end{equation}

Note that the SM amplitude $h^{0}\gamma \gamma $ appears both in the numerator and denominator of Eq. (\ref{ECoupling}) because of  SM tree-level couplings-like are assumed. For SM, this contribution has the following terms

\begin{equation}
\mathcal{M}_{h\gamma \gamma
}^{SM}=\sum_{f}N_{c}Q_{f}^{2}g_{hff}A_{1/2}^{h}\left( \tau _{f}\right)
+g_{hVV}A_{1}^{h}\left( \tau _{w}\right) .
\end{equation}

$g_{hff}$ and $g_{hVV}$ are the reduced couplings among Higgs and
fermion or vector boson respectively. The form factors in Eq. (\ref{Higgsphotonphoton}) are

\begin{eqnarray}
A_{0}^{h}\left( \tau _{H^{\pm }}\right) &=&-\left[ \tau _{H^{\pm }}-f\left(
\tau _{H^{\pm }}\right) \right] \tau _{H^{\pm }}^{-1}. \\
A_{1/2}^{h}\left( \tau _{f}\right) &=&2\left[ \tau _{f}+\left( \tau_{f}-1\right) f\left( \tau _{f}\right) \right] \tau _{f}^{-2}. \\
A_{1}^{h}\left( \tau _{w}\right) &=&-\left[ 2\tau _{w}^{2}+3\tau
_{w}+3\left( 2\tau _{w}-1\right) f\left( \tau _{w}\right) \right] \tau
_{w}^{-2}.
\end{eqnarray}

where equivalently to charged Higgs%
\begin{equation}
\tau _{f}=\frac{m_{h}^{2}}{4m_{f}^{2}}\text{ and }\tau _{w}=\frac{m_{h^{0}}^{2}}{%
4m_{W}^{2}}
\end{equation}
Finally the complex functions $f\left( \tau \right) $ is itself given by the
integral%
\begin{equation}
f\left( \tau \right) =-\frac{1}{2}\int_{0}^{1}\frac{dy}{y}\ln \left[ 1-4\tau
y\left( 1-y\right) \right] =\left\{ 
\begin{array}{c}
\arcsin ^{2}\left( \sqrt{\tau }\right) ;\hspace{2.8cm} \tau \leq 1 \\ 
-\frac{1}{4}\left[ \ln \left( \frac{\sqrt{\tau }+\sqrt{\tau -1}}{\sqrt{\tau }%
-\sqrt{\tau -1}}\right) -i\pi \right] ^{2}\hspace{1.cm}\tau >1%
\end{array}%
\right.
\end{equation}

\section*{Acknowledgments}

We are indebted to financial support from Colciencias and DIB-Universidad Nacional de Colombia. In particular, Andr\'es Castillo, Rodolfo Diaz, and John Morales would like to thank the \emph{Programa Nacional Doctoral of Colciencias-\textbf{567}} for its academic and financial support. Carlos G. Tarazona is also indebted to the financial aid from DIB-Project with Q-number \textbf{110165843163} (Universidad Nacional de Colombia).

\bibliographystyle{ieeetr}
\bibliography{Biblipro12}

\begin{thebibliography}{10}

\bibitem{PDG2}
K.~A. Olive {\em et~al.}, ``{\textit{Review of Particle Physics}},'' {\em Chin.
  Phys.}, vol.~C38, p.~090001, 2014.

\bibitem{Atlas}
G.~Aad and et~al, ``\textit{Observation of a new particle in the search for the
  Standard Model {H}iggs boson with the {{ATLAS}} detector at the {{LHC}}},''
  {\em Phys. Lett. B}, vol.~716, no.~1, pp.~1--29, 2012.

\bibitem{CMS}
S.~Chatrchyan and et~al., ``\textit{Observation of a new boson at a mass of 125
  GeV with the {CMS} experiment at the {LHC}},'' {\em Phys. Lett. B}, vol.~716,
  no.~1, pp.~30--61, 2012.

\bibitem{CMSspin}
V.~Khachatryan {\em et~al.}, ``{\textit{Constraints on the spin-parity and
  anomalous HVV couplings of the Higgs boson in proton collisions at 7 and 8
  TeV}},'' {\em Phys. Rev.}, vol.~D92, no.~1, p.~012004, 2015.
\newblock arXiv:hep-ph/1411.3441.

\bibitem{Carena2}
M.~Carena, I.~Low, N.~R. Shah, and C.~E.~M. Wagner, ``{\textit{Impersonating
  the Standard Model Higgs Boson: Alignment without Decoupling}},'' {\em JHEP},
  vol.~04, p.~015, 2014.
\newblock arXiv:hep-ph/1310.2248.

\bibitem{Degrassifin}
J.~Elias-Miro~J., Espinosa, G.~Giudice, F.~Isidori, A.~Riotto, and A.~Strumia,
  ``\textit{{H}iggs mass and vacuum stability in the {S}tandard {M}odel at
  {NNLO}},'' {\em JHEP}, vol.~08, p.~098, 2012.
\newblock CERN-PH-TH/2012-134, RM3-TH/12-9. arxiv:1205497 [hep-ph].

\bibitem{Elias}
J.~Elias-Miro, J.~R. Espinosa, G.~F. Giudice, G.~Isidori, A.~Riotto, and
  A.~Strumia, ``{Higgs mass implications on the stability of the electroweak
  vacuum},'' {\em Phys. Lett.}, vol.~B709, pp.~222--228, 2012.
\newblock arXiv:hep-ph/1112.3022.

\bibitem{ArkaniHamed}
N.~Arkani-Hamed, S.~Dubovsky, L.~Senatore, and G.~Villadoro, ``{\textit{(No)
  Eternal Inflation and Precision Higgs Physics}},'' {\em JHEP}, vol.~03,
  p.~075, 2008.
\newblock arXiv:hep-ph/0801.2399.

\bibitem{Bezrukov}
F.~Bezrukov and M.~Shaposhnikov, ``{\textit{Standard Model Higgs boson mass
  from inflation: Two loop analysis}},'' {\em JHEP}, vol.~07, p.~089, 2009.
\newblock arXiv:hep-ph/0904.1537.

\bibitem{Burgess}
D.~C.~V. Burgess~C. and J.~Espinosa, ``\textit{Effective operators and vacuum
  instability as heralds of new physics},'' {\em JHEP}, vol.~0201, p.~041,
  2002.
\newblock arXiv:hep-ph/0201160.

\bibitem{JCasas}
J.~Casas, J.~Espinosa, and M.~Quiros, ``\textit{Improved {H}iggs mass stability
  bound in the standard model and implications for supersymmetry },'' {\em
  Phys. Lett. B}, vol.~342, pp.~171--179, 1995.
\newblock arXiv:hep-ph/9409458.

\bibitem{Report}
G.~C. Branco, P.~M. Ferreira, L.~Lavoura, M.~N. Rebelo, M.~Sher, and J.~P.
  Silva, ``\textit{Theory and phenomenology of two-{H}iggs-doublet models},''
  {\em Physics Reports}, vol.~516, no.~1-2, pp.~1--102, 2012.
\newblock arXiv:hep-ph/1106.0034.

\bibitem{IvanovSimetrias}
I.~Ivanov, ``{\textit{Minkowski space structure of the {H}iggs potential in
  {2HDM}}},'' {\em Phys.Rev.}, vol.~D75, p.~035001, 2007.
\newblock arXiv:hep-ph/0609018.

\bibitem{FerreiraPer}
P.~M. Ferreira and D.~R.~T. Jones, ``{\textit{Bounds on scalar masses in two
  Higgs doublet models}},'' {\em JHEP}, vol.~08, p.~069, 2009.
\newblock arXiv:hep-ph/0903.2856.

\bibitem{Ferreira1}
P.~M. Ferreira, R.~Santos, and A.~Barroso, ``\textit{Stability of the
  tree-level vacuum in two {H}iggs doublet models against charge or {CP}
  spontaneous violation},'' {\em Phys. Lett. B}, vol.~603, no.~3-4,
  pp.~219--229, 2004.
\newblock arXiv:hep-ph/0406231.

\bibitem{Ivanov}
I.~F. Ginzburg and I.~P. Ivanov, ``\textit{Tree-level unitarity constraints in
  the most general two {H}iggs doublet model},'' {\em Phys. Rev. D.}, vol.~72,
  no.~11, 2005.
\newblock arXiv:hep-ph/0508020.

\bibitem{8Ferreira1}
J.~Velhinho, R.~Santos, and A.~Barroso, ``{\textit{Tree level vacuum stability
  in two {H}iggs doublet models}},'' {\em Phys.Lett.}, vol.~B322, pp.~213--218,
  1994.

\bibitem{Barroso2}
A.~Barroso, P.~M. Ferreira, R.~Santos, and J.~P. Silva, ``\textit{Stability of
  the normal vacuum in multi-{H}iggs-doublet models},'' {\em Phys. Rev. D.},
  vol.~74, no.~8, 2006.
\newblock arXiv:hep-ph/0608282.

\bibitem{Ivanovnuevo}
I.~P. Ivanov and J.~P. Silva, ``{Tree-level metastability bounds for the most
  general two Higgs doublet model},'' {\em Phys. Rev.}, vol.~D92, no.~5,
  p.~055017, 2015.
\newblock arXiv:hep-ph/1507.05100.

\bibitem{Barroso:2007rr}
A.~Barroso, P.~M. Ferreira, and R.~Santos, ``{\textit{Neutral minima in
  two-Higgs doublet models}},'' {\em Phys. Lett.}, vol.~B652, pp.~181--193,
  2007.
\newblock arXiv: hep-ph/0702098.

\bibitem{Met1}
A.~Barroso, P.~Ferreira, I.~Ivanov, R.~Santos, and J.~P. Silva,
  ``{\textit{Avoiding Death by Vacuum}},'' {\em J.Phys.Conf.Ser.}, vol.~447,
  p.~012051, 2013.
\newblock arXiv:hep-ph/1305.1906.

\bibitem{Met2}
A.~Barroso, P.~Ferreira, I.~Ivanov, and R.~Santos, ``{\textit{Metastability
  bounds on the two {H}iggs doublet model}},'' {\em JHEP}, vol.~1306, p.~045,
  2013.
\newblock arXiv:hep-ph/1303.5098.

\bibitem{Dipankar}
D.~Das and I.~Saha, ``{\textit{Search for a stable alignment limit in
  two-Higgs-doublet models}},'' {\em Phys. Rev.}, vol.~D91, no.~9, p.~095024,
  2015.
\newblock arXiv:hep-ph/1503.02135.

\bibitem{Ferreira2015}
P.~M. Ferreira and B.~Swiezewska, ``{One-loop contributions to neutral minima
  in the inert doublet model},'' {\em JHEP}, vol.~04, p.~099, 2016.
\newblock arXiv:hep-ph/1511.02879.

\bibitem{Cline}
J.~M. Cline, K.~Kainulainen, and A.~P. Vischer, ``{\textit{Dynamics of two
  {H}iggs doublet {CP} violation and baryogenesis at the electroweak phase
  transition}},'' {\em Phys.Rev.}, vol.~D54, pp.~2451--2472, 1996.
\newblock arXiv:hep-ph/9506284.

\bibitem{Lilith}
J.~Bernon and B.~Dumont, ``{\textit{Lilith: a tool for constraining new physics
  from Higgs measurements}},'' {\em Eur. Phys. J.}, vol.~C75, no.~9, p.~440,
  2015.
\newblock arXiv:hep-ph/1502.04138.

\bibitem{Ginzburg2}
I.~Ginzburg and K.~Kanishev, ``{\textit{Different vacua in {2HDM}}},'' {\em
  Phys.Rev.}, vol.~D76, p.~095013, 2007.
\newblock arXiv:hep-ph/0704.3664.

\bibitem{Kanishev2}
I.~F. Ginzburg, I.~P. Ivanov, and K.~A. Kanishev, ``{\textit{The Evolution of
  vacuum states and phase transitions in 2HDM during cooling of Universe}},''
  {\em Phys. Rev.}, vol.~D81, p.~085031, 2010.
\newblock arXiv:hep-ph/0911.2383.

\bibitem{Kanishev3}
I.~F. Ginzburg, K.~A. Kanishev, M.~Krawczyk, and D.~Sokolowska,
  ``{\textit{Evolution of Universe to the present inert phase}},'' {\em Phys.
  Rev.}, vol.~D82, p.~123533, 2010.
\newblock arXiv:hep-ph/1009.4593.

\bibitem{Ivanov2}
I.~P. Ivanov, ``{\textit{Minkowski space structure of the {H}iggs potential in
  {2HDM}. II. Minima, symmetries, and topology}},'' {\em Phys.Rev.}, vol.~D77,
  p.~015017, 2008.
\newblock arXiv:hep-ph/0710.3490.

\bibitem{Borah}
D.~Borah and J.~M. Cline, ``{\textit{Inert Doublet Dark Matter with Strong
  Electroweak Phase Transition}},'' {\em Phys.Rev.}, vol.~D86, p.~055001, 2012.
\newblock arXiv:hep-ph/1204.4722.

\bibitem{MorseT}
P.~Gabor, ``\textit{Morse Theory},'' in {\em \textit{Morse Theory}}
  (U.~Cambridge, ed.), Lecture Notes, pp.~1--51, 2001.

\bibitem{Peccei1}
R.~D. Peccei and H.~R. Quinn, ``\textit{Constraints imposed by "{CP}"
  conservation in the presence of pseudoparticles},'' {\em Phys. Rev. D.},
  vol.~16, no.~6, pp.~1791--1797, 1977.

\bibitem{Peccei2}
R.~D. Peccei and H.~R. Quinn, ``\textit{"{CP}" conservation in the presence of
  pseudoparticles},'' {\em PRL}, vol.~38, no.~25, pp.~1440--1443, 1977.

\bibitem{Peccei4}
R.~Peccei, ``\textit{The Strong {CP} Problem and Axions},'' in {\em Axions}
  (M.~Kuster, G.~Raffelt, and B.~Beltran, eds.), vol.~741 of {\em Lecture Notes
  in Physics}, pp.~3--17, Springer Berlin Heidelberg, 2008.

\bibitem{Kim}
J.~E. Kim, ``{\textit{A Review on axions and the strong {CP} problem}},'' {\em
  AIP Conf. Proc.}, vol.~1200, pp.~83--92, 2010.

\bibitem{Kim2}
J.~E. Kim and G.~Carosi, ``\textit{Axions and the strong {$CP$} problem},''
  {\em Rev. Mod. Phys.}, vol.~82, pp.~557--601, Mar 2010.

\bibitem{Logan1}
S.~M. Davidson and H.~E. Logan, ``{\textit{Dirac neutrinos from a second Higgs
  doublet}},'' {\em Phys. Rev.}, vol.~D80, p.~095008, 2009.
\newblock arXiv:hep-ph/0906.335.

\bibitem{Logan2}
S.~M. Davidson and H.~E. Logan, ``{\textit{LHC phenomenology of a
  two-Higgs-doublet neutrino mass model}},'' {\em Phys. Rev.}, vol.~D82,
  p.~115031, 2010.
\newblock arXiv:hep-ph/1009.4413.

\bibitem{Ferreira2}
P.~M. Ferreira, R.~Santos, and A.~Barroso, ``\textit{Erratum: Stability of the
  tree-level vacuum in two {H}iggs doublet models against charge or {CP}
  spontaneous violation (Physics Letters B (2004) 603 (219)},'' {\em PLB},
  vol.~629, no.~2-4, p.~114, 2005.
\newblock arxiv:hep-ph/0406231.

\bibitem{Top}
``{\textit{{F}irst {C}ombination of {T}evatron and {LHC} {M}easurements of the
  {T}op-{Q}uark {M}ass}},'' 2014.
\newblock {arXiv:hep-ph/1403.4427}.

\bibitem{Castillo}
A.~Castillo, R.~A. Diaz, J.~Morales, and C.~G. Tarazona, ``{\textit{Study of
  vacuum behavior for inert models with discrete $Z_{2}$-like and abelian
  $U(1)$ symmetries}},'' 2015.
\newblock arXiv:hep-ph/1510.00494.

\bibitem{2HDMC}
D.~Eriksson, J.~Rathsman, and O.~Stal, ``{2HDMC} - two-{H}iggs-doublet model
  calculator,'' {\em Computer Physics Communications}, vol.~181, no.~1,
  pp.~189--205, 2010.
\newblock arXiv:hep-ph/0902.0851.

\bibitem{Deva}
G.~Funk, D.~O'Neil, and R.~M. Winters, ``{\textit{What the Oblique Parameters
  S, T, and U and Their Extensions Reveal About the 2HDM: A Numerical
  Analysis}},'' {\em Int. J. Mod. Phys.}, vol.~A27, p.~1250021, 2012.
\newblock arXiv:Hep-ph/1110.3812.

\bibitem{Das}
C.~R. Das and M.~K. Parida, ``{\textit{New formulas and predictions for running
  fermion masses at higher scales in SM, 2HDM, and MSSM}},'' {\em Eur. Phys.
  J.}, vol.~C20, pp.~121--137, 2001.
\newblock arXiv:hep-ph/0010004.

\bibitem{Piard}
H.~Arason, D.~J. Casta\~no, B.~Kesthelyi, S.~Mikaelian, E.~J. Piard, P.~Ramond,
  and B.~D. Wright, ``\textit{Renormalization-group study of the standard model
  and its extensions: The standard model},'' {\em Phys. Rev. D}, vol.~46,
  pp.~3945--3965, Nov 1992.

\bibitem{ChengLi}
T.~P. Cheng, E.~Eichten, and L.-F. Li, ``\textit{Higgs phenomena in
  asymptotically free gauge theories},'' {\em Phys. Rev. D}, vol.~9,
  pp.~2259--2273, Apr 1974.

\bibitem{Machacek}
M.~E. Machacek and M.~T. Vaughn, ``\textit{Two-loop renormalization group
  equations in a general quantum field theory (II). Yukawa couplings },'' {\em
  Nuclear Physics B}, vol.~236, no.~1, pp.~221 -- 232, 1984.

\bibitem{RGE21}
H.~E. Haber and R.~Hempfling, ``{\textit{The Renormalization group improved
  Higgs sector of the minimal supersymmetric model}},'' {\em Phys. Rev.},
  vol.~D48, pp.~4280--4309, 1993.
\newblock arXiv:hep-ph/9307201.

\bibitem{Kanemura2}
S.~Kanemura, T.~Kasai, and Y.~Okada, ``{\textit{Upper bounds and lower bounds
  of the lightest {CP} even {H}iggs boson in the two {H}iggs doublet model}},''
  1999.
\newblock arXiv:hep-ph/9911312.

\bibitem{HaberLilith}
J.~Bernon, J.~F. Gunion, H.~E. Haber, Y.~Jiang, and S.~Kraml,
  ``{\textit{Scrutinizing the alignment limit in two-Higgs-doublet models:
  m$_h$=125 GeV}},'' {\em Phys. Rev.}, vol.~D92, no.~7, p.~075004, 2015.
\newblock arXiv:hep-ph/1507.00933.

\end{thebibliography}

\end{document}